\documentclass[12pt]{report}
\usepackage[letterpaper,top=1in,bottom=1in,right=1in,left=1in]{geometry}

\usepackage{graphicx}
\usepackage{amsmath}
\usepackage{cite}
\usepackage{siunitx}
\usepackage[hidelinks]{hyperref}
\usepackage{enumerate}
\usepackage{listings}
\usepackage{xcolor}
\usepackage{setspace}
\usepackage[numbers]{natbib}
\usepackage{doi}
\usepackage[utf8]{inputenc}
\usepackage{feynmp-auto}
\usepackage{scalerel}
\usepackage{subcaption}
\usepackage{amssymb}
\usepackage{amsmath}
\usepackage{multirow}
\usepackage{notoccite}

\usepackage[nottoc]{tocbibind}

\usepackage{titlesec}
\titleformat{\chapter}[display]{\normalfont\bfseries}{\centering CHAPTER \thechapter}{0pt}{\centering}
\titleformat{\section}[block]{\normalfont\bfseries}{\thesection}{12pt}{}
\titleformat{\subsection}[block]{\normalfont}{\thesubsection}{12pt}{}

\newcommand{\marrow}[5]{%
    \fmfcmd{style_def marrow#1
    expr p = drawarrow subpath (1/4, 3/4) of p shifted 6 #2 withpen pencircle scaled 0.4;
    label.#3(btex #4 etex, point 0.5 of p shifted 6 #2);
    enddef;}
    \fmf{marrow#1,tension=0}{#5}}
    
\newcommand{\marrowz}[5]{%
    \fmfcmd{style_def marrow#1
    expr p = drawarrow subpath (1/4, 3/4) of p shifted 12 #2 withpen pencircle scaled 0.4;
    label.#3(btex #4 etex, point 0.5 of p shifted 12 #2);
    enddef;}
    \fmf{marrow#1,tension=0}{#5}}
    
    \newcommand{\marrowzz}[5]{%
    \fmfcmd{style_def marrow#1
    expr p = drawarrow subpath (1/4, 3/4) of p shifted 18 #2 withpen pencircle scaled 0.4;
    label.#3(btex #4 etex, point 0.5 of p shifted 18 #2);
    enddef;}
    \fmf{marrow#1,tension=0}{#5}}
    
\newcommand{\myrbrace}[2]{\vspace{#2pt}\scaleleftright[\dimexpr5pt+#1\dimexpr0.06pt]{.}{\rule[\dimexpr2pt-#1\dimexpr0.5pt]{-4pt}{#1pt}}{\rbrace}\hspace{6pt}}

\begin{document}

\doublespacing
\pagenumbering{roman}

\chapter*{A Strong-QCD Regime Measurement of the Proton's Spin Structure}
\thispagestyle{empty}

\begin{center}

BY

\vfill

DAVID RUTH

B.A. in Physics, McDaniel College, 2015

\vfill

DISSERTATION

\vfill

Submitted to the University of New Hampshire

in Partial Fulfillment of

the Requirements for the Degree of

\vfill

Doctor of Philosophy

in

Physics

\vfill

December, 2022

\end{center}

\pagebreak
\null

\vfill

\begin{center}

ALL RIGHTS RESERVED

\copyright 2022

David Ruth

\end{center}

\pagebreak
\noindent This dissertation has been examined and approved in partial fulfillment of the requirements for the degree of Doctor of Philosophy in Physics by:

\vspace{0.5in}

\hfill
\begin{minipage}{0.7\textwidth}

Dissertation Director, Karl Slifer \\Professor of Physics
\vspace{0.3in}

Mark McConnell\\Professor of Physics
\vspace{0.3in}

William Brooks\\Associate Professor of Physics\\Universidad Tecnica Federico Santa Maria
\vspace{0.3in}

David Mattingly\\Associate Professor of Physics
\vspace{0.3in}

Elena Long\\Assistant Professor of Physics
\vspace{0.3in}

\hfill On November 16th, 2022

\end{minipage}

\vspace{0.5in}

\noindent Original approval signatures are on file with the University of New Hampshire Graduate School.

\pagebreak
\chapter*{DEDICATION}
\addcontentsline{toc}{chapter}{DEDICATION}


\begin{center}

This thesis is dedicated to my granddad, Ray Somerlock, who helped me to become interested in science very early with his tireless curiosity and enthusiasm for engineering. I know he would have loved to read this thesis and I strive to always carry his passion with me.

\end{center}

\vfill

\pagebreak
\chapter*{ACKNOWLEDGEMENTS}
\addcontentsline{toc}{chapter}{ACKNOWLEDGEMENTS}

Above all else, I want to thank my amazing wife Michelle. You make this all feel worthwhile, and I couldn't ask for a better partner to be on this journey with. Thank you sunshine, for all of your support and encouragement. I couldn't have gotten through this without you.

I want to thank my advisor Karl Slifer, who I am honored to say is both a mentor and a true friend. Karl has helped me to become a better researcher and created one of the best environments to learn in that I could imagine. Thank you for sharing your experience with me over the last seven years.

I'm also eternally grateful for my parents, Nick and Sheila, who have supported me on every step of this journey. I couldn't ask for a more caring family and I will never forget that you've been there for me every time I needed you, in every way possible. Thank you for always coming through for us, and for setting such an amazing example with your own lives that I can aspire to reach.

Thank you also to all my friends in the polarized target group, especially Elena Long and Nathaly Santiesteban, who have both helped me to mature as a scientist and been some of the best friends I could ask for at UNH. Some of my fondest grad school memories are working on cooldowns with all of you, despite the hard work and long hours!

Thank you to my family who has always supported and encouraged me, my cousins Rachel, Jessie, Nicole, and Kasey, my grandparents Dolores, John, Edna, and Ray, my aunts and uncles Ann, George, Joan, Bill, Robin, Dave, Carl, and Christy, and my lovely in-laws Robin, Frank, and Matt.

Thank you also to Jian-Ping Chen, whose experience and keen eye for detail have helped me learn a lot and become a more rigorous scientist, and without whom the E08-027 analysis would never have been completed.

Thank you as well to my best man and friend Sam, who has helped me stay sane and avoid becoming completely antisocial, and the rest of the 203 crew at McDaniel, who I hope I will stay lifelong friends with.

\pagebreak
\tableofcontents
\listoftables
\listoffigures
\chapter*{ABSTRACT}
\addcontentsline{toc}{chapter}{ABSTRACT}









\vspace{0.2in}

\noindent The theory of the strong force, Quantum Chromodynamics (QCD) remains one of the most important ways to understand the fundamental properties of ordinary matter. However, at low momentum transfer $Q^2$, in the regime where the strong force becomes extremely strong, our understanding of QCD for ordinary nucleons becomes hazy. Several cutting edge theories such as Chiral Perturbation Theory ($\chi$PT) and Lattice QCD have provided valuable predictions in this regime, but Lattice QCD has not yet extended predictions of many important quantities to this kinematic region, and Chiral Perturbation Theory has faced several important disagreements with experimental data for the neutron over the last several decades. It is therefore of extreme importance to have a benchmark of experimental data in the low energy regime for the proton's behavior, as a test of leading theories for the behavior of QCD in this regime.

The E08-027 (g2p) experiment ran at Jefferson Lab in 2012 with the goal of collecting this valuable data, and though I was still completing my undergraduate studies at the time, I became involved in the analysis in 2015 and built on the previous work to complete it and analyze the exciting results. This experiment achieved a high precision measurement of the spin structure functions $g_1$ and $g_2$ for the proton, quantities which describe the internal spin structure of the proton. These measurements were taken in the valuable low $Q^2$ region described above, and used to extract several moments of these spin structure functions which can be directly compared to the cutting-edge predictions of Chiral Perturbation Theory. 

Though the experiment's timeline was such that I didn't have a chance to work directly on the experimental setup, I had the opportunity to acquire hands-on experience working on a polarized target at UNH which is very similar to the crucial polarized target used in the g2p experiment.

Full details of the g2p experiment and my experimental work at the University of New Hampshire are presented in this thesis, as well as a detailed description of the analysis process and the exciting benchmark results, which serve as a direct test of all current and future theories of QCD in the low-$Q^2$ regime.

\pagebreak

\pagenumbering{arabic}

\chapter{Background}

\begin{figure}[htb]
\centering
\includegraphics[width=0.7\textwidth]{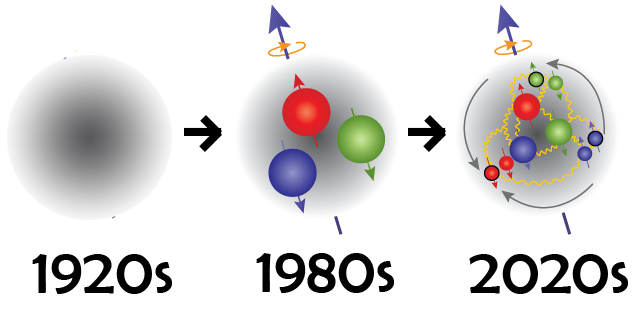}
\caption{Model of the Proton over time. Adapted from a figure in \cite{proton_struc}}
\label{fig:proton}
\end{figure}

It is currently estimated that the proton accounts for upwards of 90\% of ordinary matter in the visible universe~\cite{proton_number}. Protons make up the nuclei of atoms which compose humans and just about everything else we see every day, and their charge dictates most electromagnetic interactions we can study. Accordingly, it is surprising that we are still without a robust understanding of this essential particle. In 1920 when Ernest Rutherford proposed the proton's existence, he believed it was a fundamental particle, one of the most basic building blocks of the universe~\cite{rutherford}. In 1964, it was theorized that the proton was itself composed of three resonances Gell-Mann would come to call ``quarks''~\cite{gellmann}. Over time, this picture too has evolved, revealing a tremendously complex internal structure, where the three valence quarks of the proton are bound with gluons to each other and to a shifting ``quark sea'' of quarks and antiquarks, as in Fig.~\ref{fig:proton}.

This complex structure comes with a number of ongoing mysteries about the proton which still require resolution. In 2010, the radius of the proton was measured to a value about 4\% less than was observed for the last century~\cite{pradius}, a value well outside of statistical error bars which raised many ongoing questions about the proton's true size, a problem which is now known as the ``Proton Radius Puzzle.'' Similar to the mystery of its exact size, and relevant to the topic of this dissertation, the proton's spin is also not yet well understood, and the source of several puzzles. Spin is a fundamental property of particles which acts as an intrinsic angular momentum that exceeds the size which relativistic physics would allow in a real angular momentum, and which has great impact on the particle's electromagnetic behavior~\cite{griffiths}. Most famous among the mysteries about the proton's spin is the fact that though the proton is known to be a composite particle, adding the spins of its known constituents fails to produce the measured spin of the proton. This is known as the ``Proton Spin Puzzle'', and current research focuses on investigating the idea that the proton's spin may in part derive from the real internal angular momentum of its constituents\cite{spinquest}.

\begin{figure}
\centering
\begin{subfigure}{.5\textwidth}
  \centering
  \includegraphics[width=1.0\linewidth]{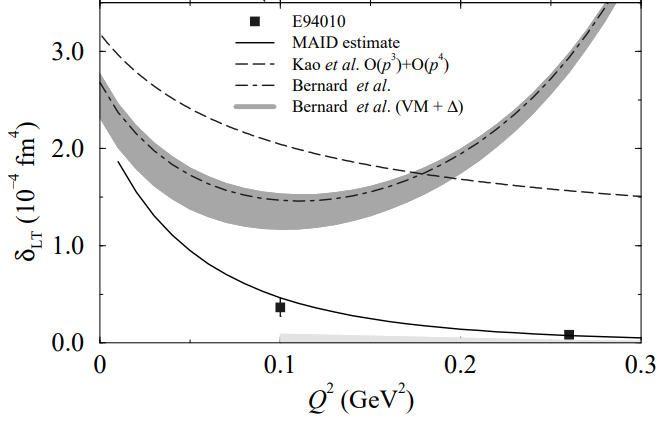}
  \label{fig:sub1}
\end{subfigure}%
\begin{subfigure}{.5\textwidth}
  \centering
  \includegraphics[width=1.0\linewidth]{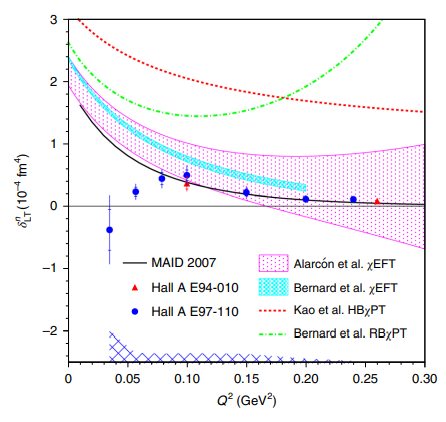}
  \label{fig:sub2}
\end{subfigure}
\caption{The ``$\delta_{LT}$ Puzzle'': (Left) The results of Jefferson Lab E94-010 showing the initial disagreement~\cite{E94010}. (Right) The results of Jefferson Lab E97-110 showing a new disagreement in 2021~\cite{saGDH}.}
\label{fig:dltpuzzle}
\end{figure}

Additional difficulties arise when trying to determine how the proton's spin is distributed within this complex particle. The proton's spin distribution can be studied with a quantity known as a structure function, the moments of these quantities can likewise be directly compared to theoretical predictions to test the theory of the strong nuclear force, quantum chromodynamics (QCD). It was expected that one of these moments, $\delta_{LT}$, would be a reliable test due to its insensitivity to the contribution of the $\Delta$-resonance~\cite{g2pProposal}. However, initial measurements of this quantity showed sizable disagreement with the predictions of chiral perturbation theory ($\chi$PT) for the neutron~\cite{E94010}. Chiral perturbation theory is to date one of the best theoretical frameworks for predicting the behavior of QCD in low momentum collisions. Newer predictions of $\chi$PT seem to resolve this initial disagreement, but recent data for the neutron show a new disagreement at even lower $Q^2$, as is shown in Fig.~\ref{fig:dltpuzzle}. The theoretical difficulty in reproducing data at low momentum transfer for the neutron implies the dire need for proton data in this region.

The low momentum transfer ($Q^2$) region is important, as it is in this regime that the strong nuclear force becomes ``truly strong''; while high $Q^2$ is useful for mapping out the partonic constituents of the nucleon, at low $Q^2$ it is possible to study the behavior of the nucleon as a whole and how its properties can derive continuously from those of its component particles. There is some important theoretical work done in this region, most notably chiral perturbation theory and lattice QCD, which can both reproduce some quantities at low-to-medium $Q^2$. However, there is a severe lack of experimental data, as taking low $Q^2$ data, especially for the proton, is very difficult. This thesis will detail the results of the Jefferson Lab E08-027 Experiment (g2p) which attempts to remedy this with a new high precision measurement of the proton's spin structure in this low $Q^2$, strong-QCD regime.


\chapter{Theory}

\section{Inclusive Electron Scattering}

To study the structure of the nucleon, modern experiments rely on the experimental technique of particle scattering, a technique which helped to pioneer the field of nuclear physics when Ernest Rutherford used it to discover the existence of the nucleus by scattering helium nuclei ($\alpha$-particles) through a gold foil. Rutherford and his students Hans Geiger and Ernest Marsden discovered that most of these positively charged particles passed straight through the foil, while a small number deflected at dramatic angles. The study of the deflected particles led Rutherford to theorize that the mass of the atom is compressed into a tiny volume known as a nucleus, with most of the atom composed of empty space, and likewise, modern scattering experiments can glean ever more detailed information about the nucleus by studying the scattered particles~\cite{GoldFoil}.

The quantity that is usually measured and studied in these experiments is known as a scattering cross section. This quantity describes the normalized interaction rate over a solid angle area of the target, and varies with the energy and momentum of the measured particles. The nucleon's structure can be probed by measuring the cross sections which result from scattering a beam of electrons off of a fixed proton target ($ep$ scattering). Depending on the energy of the electron beam, a number of different particles may result from this as the proton's constituents may be knocked out and form new hadrons, but here we are concerned with inclusive electron scattering, in which only the scattered electron is measured, ignoring other products which may result.

\begin{figure}[htb]
\centering


\begin{fmffile}{feyngraph}
  \begin{fmfgraph*}(320,160)
    \fmfleft{ip,il}
    \fmfright{x1,x2,K,x3,x4,o2,o3,o4,ol}
    \fmfset{arrow_len}{10}
    \fmf{fermion}{il,vl,ol}
    \marrow{ea}{ up }{top}{$k^\mu$}{il,vl}
    \marrow{eb}{ up }{top}{$k'^\mu$}{vl,ol}
    \fmflabel{\large $e^-$}{il}
    \fmflabel{\large $e^-$}{ol}
    \fmf{photon,tension=1,label=$\gamma^*$}{vl,vp}
    \marrow{ec}{ right }{ lrt }{$q^\mu$}{vl,vp}
    \fmf{phantom}{ip,vp,x1}
    \fmffreeze
    \fmf{phantom}{ip,vp}
    \fmfi{fermion}{vpath (__ip,__vp) scaled 1.01 shifted (-1.8, 6)}
    \marrowz{ed}{ up }{top}{$p^\mu$}{ip,vp}
    \fmfi{fermion}{vpath (__ip,__vp) scaled 1.01}
    \fmfi{fermion}{vpath (__ip,__vp) scaled 1.01 shifted ( 1.8,-6)}
    \fmfshift{15 left}{x1,x2,x3,x4}
    \fmf{phantom}{vp,x1} 
    \fmf{phantom}{vp,x2} 
    \fmf{phantom}{vp,x3} 
    \fmf{phantom}{vp,x4} 
    \fmfi{fermion}{vpath (__vp,__x1) scaled 1.02 shifted ( 0.0,-6.0)}
    \fmfi{fermion}{vpath (__vp,__x2) scaled 1.00 shifted ( 0.0,-2.0)}
    \fmfi{fermion}{vpath (__vp,__x3) scaled 0.98 shifted ( 0.0, 2.0)}
    \fmfi{fermion}{vpath (__vp,__x4) scaled 0.96 shifted ( 0.0, 6.0)}
    \fmfv{l.d=0,l.a=0,l={\myrbrace{96}{0}$\text{\large X}$}}{K}
    \fmflabel{\large $p$}{ip}
    \fmfblob{50}{vp}
  \end{fmfgraph*}
\end{fmffile}

\caption{$ep$ Scattering interaction}
\label{fig:epscattering}
\end{figure}

A Feynman diagram for the electron-proton scattering interaction is shown in Figure~\ref{fig:epscattering}. In this process, an electron enters with four-momentum (in natural units) $k^\mu$ = (E,\textbf{k}), and exchanges a virtual photon with a proton in the target, which has initial four-momentum $p^\mu$ = ($\epsilon$,\textbf{p}). This virtual photon carries a four momentum of $q^\mu$ = ($\nu$,\textbf{q}), where $\nu$ is the energy transfer from the electron to the struck proton, and \textbf{q} is the momentum exchanged in the interaction. The electron then scatters at an angle $\theta$ from its initial path, with a four-momentum of $k'^\mu$ = (E$'$,\textbf{k$'$}), and some unknown products X are produced by the proton~\cite{Povh}. Since the electron is comparatively well understood, and we are only measuring the scattered electrons in inclusive scattering, we define everything in the rest frame of the proton where $\epsilon$ = $M_p$, and in terms of the initial and final state of the electron. The energy transfer $\nu$ is then defined simply as:
\begin{equation}
\label{eqn:nudef}
\nu = E - E'
\end{equation}
with E and E$'$ the initial and final energies of the electron, respectively. We can also use these quantities in combination with the angle  at which the scattered electron is measured to define the momentum transfer $Q^2$ of the interaction:
\begin{equation}
\label{eqn:q2def}
Q^2 = -\textbf{q}^2 = 4 E E' \sin^2{\frac{\theta}{2}}
\end{equation}
This is an especially important kinematic variable because it defines the distance scale which the scattering interaction is probing. The de Broglie wavelength of a particle is considered to be~\cite{DeBroglie}
\begin{equation}
\label{eqn:debroglie}
\lambda = \frac{h}{\textbf{q}}
\end{equation}
where h is the Planck constant. So it is plain to see that the distance scale associated with our interaction is inversely proportional to the momentum transfer. A high $Q^2$ probes the small-distance region of the proton where we are looking at the nucleon's individual constituents, while a low $Q^2$ conversely probes the large distance region and reveals information about how the proton's constituents work together to give the proton its observed properties. If we wish to test the aforementioned theories of QCD such as chiral perturbation theory, and to understand how the proton obtains its spin from the quarks and gluons that make it up, the latter is our kinematic region of interest~\cite{KarlThesis}.

We must also define the invariant mass of the hadron, a frame-independent kinematic variable which quantifies the total mass-energy of the proton and its unknown products after scattering:
\begin{equation}
\label{eqn:invariantmass}
W^2 = (p^\mu + q^\mu)^2 = M_p^2 + 2 M_p \nu - Q^2
\end{equation}
Wherein $M_p$ $\approx 938.3$ MeV is the mass of the proton. Using the kinematic variables already defined, we can also define the \textit{Bjorken-x}, a quantity which can be identified with the fraction of the proton's momentum carried by the struck quark:
\begin{equation}
\label{eqn:bjorkenx}
x_{bj} = \frac{Q^2}{2 M_p \nu}
\end{equation}
This quantity can also be considered to track the elasticity of the interaction, as discussed in the next section. Finally, we define the fractional energy loss of the electron simply as:
\begin{equation}
\label{eqn:fracloss}
y = \frac{\nu}{E}
\end{equation}
These invariant quantities will be used to map out the response of the nucleon over the tested kinematic range.

\section{Interaction Rates \& Cross Sections}
To define the measurable cross section mentioned in the previous section, we will first need to discuss the interaction rate of the $ep$ scattering process. The interaction rate $W_{fi}$ describes the number of transitions from the initial state of $e + p$ to the final state of $e' + X$ per unit time. Fermi's Golden Rule states~\cite{HalzenMartin}:
\begin{equation}
\label{eqn:goldenrule1}
W_{fi} = 2 \pi |V_{fi}|^2 \delta(k_f - k_i)
\end{equation}
where $V_{fi}$ is the interaction potential and the delta function enforces energy and momentum conservation. By adopting a covariant normalization, we can rewrite per unit volume in four dimensions for a generic scattering interaction A+B $\rightarrow$ C+D~\cite{HalzenMartin}:
\begin{equation}
\label{eqn:goldenrule2}
W_{fi} = \bigg (\frac{2 \pi}{V}\bigg)^4 |\mathcal{T}|^2 \delta^{(4)} \big ([p_C + p_D] - [p_A + p_B] \big)
\end{equation}
Here, $\mathcal{T}$ is the probability amplitude of the interaction process, V is the volume over which the interactions take place, and $p_{A,B,C,D}$ are the four-momenta of the particles involved in scattering. The probability amplitude will be a combination of the electron part, which is easy to compute using the Feynman rules because Quantum Electrodynamics (QED) is well understood, and the hadronic part, which comprises both the initial proton state and the final unobserved products X~\cite{Srednicki}. Using the Feynman rules for spinor electrodynamics, we have one incoming and outgoing electron, one internal photon, and two vertices. Though the exact vertex on the hadronic side of the virtual photon is not known, we know it must be a QED vertex, and so add the associated vertex factor. We then find:
\begin{equation}
\label{eqn:spinorfeynman}
\mathcal{T}_{Lepton} = i e^2 \gamma^\mu \gamma^\nu \frac{g^{\mu \nu}}{\textbf{q}^2 - i \epsilon} \overline{u}_{s'}(\textbf{k$'$}) u_{s}(\textbf{k})
\end{equation}
Where $\gamma$ is the gamma-matrix, $u_s$ and $\overline{u}_s$ are the spinors, and e is the fundamental charge. Because we don't know the exact structure of the hadronic part of the process, we must instead write it generally as a transition amplitude from the initial state of a proton with momentum P and spin S to a final unknown state X~\cite{Zielinski:2017gwp}:
\begin{equation}
\label{eqn:hadronamplitude}
\mathcal{T}_{Hadron} = <P,S|J^\mu|X>
\end{equation}

Plugging into Equation~\ref{eqn:goldenrule2} and simplifying by making use of gamma matrix identities to contract the metric with two of our gamma matrices we obtain:
\begin{equation}
\begin{split}
\label{eqn:goldenrule3}
W_{fi} = \bigg (\frac{2 \pi}{V}\bigg)^4 \bigg (\frac{e^4}{Q^4}\bigg)  \overline{u}_{s'}(\textbf{k$'$}) u_{s}(\textbf{k}) \gamma^\mu \overline{u}_{s}(\textbf{k}) u_{s'}(\textbf{k$'$}) \gamma^\nu <P,S|J^\mu|X>\\
<X|J^\nu|P,S>\delta^{(4)} \big ([p + p_x] - [k + k'] \big)
\end{split}
\end{equation}
We can define the cross section which can be experimentally measured in terms of the interaction rate~\cite{HalzenMartin}:
\begin{equation}
\label{eqn:xs}
d\sigma = \frac{W_{fi}}{\phi}N_f
\end{equation}
Here $\phi$ is the initial flux of particles, and $N_f$ is the number of available final states. We will later associate these quantities with the number of incident electrons and the density of scattering centers in the target, respectively, to form an experimental cross section~\cite{Povh}.

For now, the initial flux will be a product of the number of beam particles passing through unit area per unit time, $\frac{2E_e}{V}$ and the number of target particles per unit volume, $\frac{2E_p}{V}$~\cite{HalzenMartin}. Since the electron mass $m_e$ is much smaller than $M_p$, we neglect it, and in our lab frame, the protons are at rest, so we can simplify the initial flux as:
\begin{equation}
\label{eqn:initflux}
\phi = \frac{4 E M_p}{V^2}
\end{equation}
The number of available final states for a given particle A in a volume V is limited by quantum theory to be~\cite{HalzenMartin}:
\begin{equation}
\label{eqn:nfgeneric}
N_f = \frac{V d^3 p_A}{(2 \pi)^3 2 E_A}
\end{equation}
Our final states are the scattered electron with momentum $k'$ energy $E'$ and some unknown products with momenta $p_x$ and energy $E_x$. We will need to sum over all possible X states to obtain the full cross section, so combining equations~\ref{eqn:nfgeneric},~\ref{eqn:initflux} and~\ref{eqn:xs} we obtain the differential cross section:
\begin{equation}
\begin{split}
\label{eqn:xs_full}
d\sigma = \frac{2 \pi}{8 E E' M_p} \bigg (\frac{e^4}{Q^4}\bigg) d^3 k'  \overline{u}_{s'}(\textbf{k$'$}) u_{s}(\textbf{k}) \gamma^\mu \overline{u}_{s}(\textbf{k}) u_{s'}(\textbf{k$'$}) \gamma^\nu \times \\
\displaystyle\sum_{X}\frac{d^3 p_x}{(2 \pi)^3 2 E_x}<P,S|J^\mu|X><X|J^\nu|P,S>\delta^{(4)} \big ([p + p_x] - [k + k'] \big)
\end{split}
\end{equation}

To simplify this, in natural units the fine structure constant $\alpha$ = $\frac{e^2}{4 \pi}$, and we can define the lepton tensor:
\begin{equation}
\label{eqn:leptontensor}
L^{\mu \nu} = \overline{u}_{s'}(\textbf{k$'$}) u_{s}(\textbf{k}) \gamma^\mu \overline{u}_{s}(\textbf{k}) u_{s'}(\textbf{k$'$}) \gamma^\nu
\end{equation}
and the hadron tensor:
\begin{equation}
\label{eqn:hadrontensor}
W_{\mu \nu} = \frac{1}{2 M_p} \displaystyle\sum_{X}\frac{d^3 p_x}{2 E_x}<P,S|J^\mu|X><X|J^\nu|P,S>\delta^{(4)} \big ([p + p_x] - [k + k'] \big)
\end{equation}

This tensor can also be written in a simpler form by employing completeness:
\begin{equation}
\label{eqn:hadron_tensor_simpler}
W_{\mu \nu} = i \int e^{i q x}<P,S|J^\mu(x)J^\nu(0)|P,S>d^4x
 \end{equation}
The meaning of these two quantities will be elaborated on in the following section, but for now we use them to write our cross section as:
\begin{equation}
\label{eqn:xs_compact}
d\sigma = \frac{\alpha^2}{Q^4} \frac{d^3 k'}{E'} \frac{1}{E} L^{\mu \nu} W_{\mu \nu}
\end{equation}
Finally, we will transform into spherical coordinates~\cite{Srednicki}:
\begin{equation}
\label{eqn:differential}
d^3 k' = k'^2 dk' d\Omega
\end{equation}
Where $d\Omega$ = $\sin\theta d\theta d\phi$ is the differential solid angle which will be measured in the detector. Since the mass of the electron is small enough to be negligible, k$'$ $\simeq$ E$'$. This gives us a doubly differential cross section definition of:
\begin{equation}
\label{eqn:xs_pretty}
\frac{d^2\sigma}{d E' d \Omega} = \frac{\alpha^2}{Q^4} \frac{E'}{E} L^{\mu \nu} W_{\mu \nu}
\end{equation}
How this quantity is derived experimentally is detailed in Chapter 6. But we can interpret it most readily with equation~\ref{eqn:xs} as being the fraction of incoming electrons which interact with a proton per unit time, scaled by the number of possible scattering centers in the target.

For the coming sections, it will be useful to make a distinction for an elastic measurement of this quantity as compared to an inelastic measurement. In an elastic measurement, the energy transfer is absorbed entirely by the recoil proton, such that the final invariant mass of the hadronic system is simply the rest mass of the proton: $W_{Elastic}$ = $M_p$. For elastic reactions, the proton behaves as a point particle, corresponding to a Bjorken-x of $x_{bj}$ = 1, and the final energy of the scattered electron can be easily calculated. For reactions with a higher W, the final hadronic system is more complex, and we are probing more the partonic behavior of the proton's constituents, corresponding with lower values of x. If we go to a very high W, we reach the Deep Inelastic Scattering (DIS) region where we can probe the individual quarks and gluons, but between this region and the elastic response is known as the Resonance Region, this is our area of interest because it concerns how the proton's constituents contribute to its behavior on a hadronic level.
\begin{figure}[htb]
\centering
\includegraphics[width=1.0\textwidth]{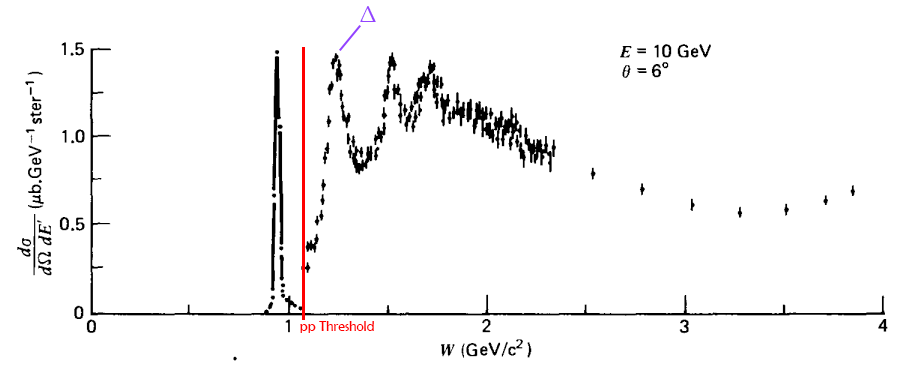}
\caption{ep $\rightarrow$ eX scattering cross section as a function of the invariant mass W. Figure edited from~\cite{HalzenMartin} and produced with SLAC data.}
\label{fig:xs_elastic}
\end{figure}

The behavior of the cross section as a function of W is shown in figure~\ref{fig:xs_elastic}. The elastic response is shown in the peak at the proton mass of $M_p$ $\approx 938.3$ MeV. The resonance region then begins at the pion production (pp) threshold, which is the minimum energy transfer necessary to create a pion, W $\approx 1038.2$ MeV. Several excited states of the proton form resonances in the following region, but the most important resonance is the first one, corresponding to the $\Delta$-baryon at W $\approx 1232$ MeV. Our experiment will focus on measurements which begin at the pion production threshold, and extend to the highest W which is able to be measured at the relevant kinematics. These inelastic measurements require a slightly more complex treatment than an elastic measurement would.

\section{Lepton and Hadron Tensors}

The leptonic and hadronic tensors of the previous section encode the nonperturbative behavior of the electron and hadronic system involved in the scattering, respectively. What we have defined so far can be considered a generic form used in unpolarized scattering. But for polarized scattering in the inelastic regime, we will need to define both tensors a bit differently. Since the leptonic tensor is a bit simpler, we will start by looking at the unpolarized version, which is averaged over the initial and final spin of the scattering electron. To average the T-Matrix we write~\cite{Thomas}:
\begin{equation}
\label{eqn:tmatrix}
<|\mathcal{T}|^2> = \frac{1}{2} \displaystyle\sum_{s,s'}|\mathcal{T}_{Lepton}|^2
\end{equation}
For an unpolarized initial electron, we can adopt the normalization condition~\cite{Thomas}:
\begin{equation}
\label{eqn:norm1}
\displaystyle\sum_{s=\pm \frac{1}{2}}u_s(k)\overline{u}_s(k) = \gamma^{\mu}k_{\mu} + m_{e}
\end{equation}

To average the unpolarized version of the lepton tensor of~\ref{eqn:leptontensor} we must take the trace of the T-matrix, so using this normalization condition:
\begin{equation}
\label{eqn:lepton_unpolarized}
L^{\mu \nu}_U = \frac{1}{2}Tr\bigg ((\gamma^{\mu}k_{\mu} + m_{e})\gamma^{\mu}(\gamma^{\nu}k'_{\nu} + m_{e})\gamma^{\nu} \bigg)
\end{equation}
Multiplying out these terms and performing the trace we find:
\begin{equation}
\label{eqn:lepton_unpolarized2}
L^{\mu \nu}_U = 2\big[k^{\mu}k'^{\nu}+k'^{\mu}k^{\nu} - g^{\mu \nu}(\textbf{k}\cdot \textbf{k$'$} - m_{e}^2)\big]
\end{equation}
In the case that the electron in question has an initial polarization, we will need to pick a different normalization, yielding an extra term~\cite{Zielinski:2017gwp,Thomas}:
\begin{equation}
\label{eqn:lepton_polarized}
L^{\mu \nu}_P = 2\big[k^{\mu}k'^{\nu}+k'^{\mu}k^{\nu} - g^{\mu \nu}(\textbf{k}\cdot \textbf{k$'$} - m_{e}^2) + i \epsilon^{\mu \nu \alpha \beta}s_{\alpha}q_{\beta}\big]
\end{equation}
Here $\epsilon$ is the Levi-Civita tensor, which enforces $\epsilon_{0123} = +1$, $q_{\beta}$ is the momentum of the virtual photon, and $s_{\alpha}$ is the lepton spin vector, which points along the spin quantization vector z with magnitude $s_{\mu} = \overline{u}\gamma_{\mu}\gamma_{5}u$ and sign determined by whether the electron helicity is forward (along z) or backward (antiparallel to z). Note that the polarization dependent term makes the lepton tensor asymmetric, but the asymmetric term cancels when averaged over the initial spins.

The lepton tensor characterizes the scattering electron part of the diagram in Fig ~\ref{fig:epscattering}. However, we are far more interested in the inelastic behavior of the other part, since it contains information on our proton target. This part is characterized by the hadronic tensor of~(\ref{eqn:hadrontensor}). This can be split into a symmetric unpolarized part and an antisymmetric polarized part, much like the lepton tensor. We can first get the most general form of the unpolarized part by expanding its previous form considering Lorentz and gauge invariance~\cite{Thomas}:
\begin{equation}
\label{eqn:hadron_unpolarized}
W_{\mu \nu}^U = W_{1}(\nu,Q^2)\bigg(\frac{q_{\mu} q_{\nu}}{Q^2}-g_{\mu \nu}\bigg) +W_{2}(\nu,Q^2)\frac{1}{M_p^2}\bigg(p_{\mu} - \frac{\textbf{p}\cdot\textbf{q}}{Q^2}q_{\mu}\bigg)\bigg(p_{\nu} - \frac{\textbf{p}\cdot\textbf{q}}{Q^2}q_{\nu}\bigg) \end{equation}
Here, $W_{1}$ and $W_{2}$ are inelastic \textit{unpolarized} structure functions which will be discussed in the next section. For now, as in the lepton tensor, we add an antisymmetric term to get the polarized variant of the hadronic tensor~\cite{Thomas}:
\begin{equation}
\label{eqn:hadron_polarized}
W_{\mu \nu}^P = W_{\mu \nu}^U + i \epsilon_{\mu \nu \alpha \beta}q^{\alpha}\bigg(G_{1}(\nu,Q^2)M_{p}S^{\beta} + G_{2}(\nu,Q^2)\frac{1}{M_{p}}(S^{\beta}\textbf{p}\cdot\textbf{q} - p^{\beta}\textbf{S}\cdot\textbf{q})\bigg) \end{equation}

Where $S^{\beta}$ corresponds to the proton's spin vector, similar to the electron's in the prior discussion. $G_1$ and $G_2$ are polarized structure functions which will be discussed fully in the next section. To continue our understanding of an unpolarized part with additional terms that represent the polarized physics, let us first contract the unpolarized leptonic tensor of (\ref{eqn:lepton_unpolarized2}) and the unpolarized hadronic tensor of (\ref{eqn:hadron_unpolarized}) to get a fully symmetric, unpolarized result:
\begin{equation}
\label{eqn:unpol_contraction}
L^{\mu \nu}W_{\mu \nu} = 4 W_{1}(\nu,Q^2)(\textbf{k}\cdot\textbf{k$'$}) + 2 W_{2}(\nu,Q^2)\bigg[\frac{2}{M_p^2}(\textbf{p}\cdot\textbf{k})(\textbf{p}\cdot\textbf{k$'$}) - (\textbf{k}\cdot\textbf{k$'$})\bigg] \end{equation}

In the lab reference frame, the proton is at rest, and the electron only has energy in the direction of propagation along the z-axis. It scatters from the central axis by an angle $\theta$ along a vector with azimuthal angle $\phi$. This allows us to write in full the relevant four-vectors~\cite{Zielinski:2017gwp}:
\begin{equation}
\label{eqn:fourv_k}
k_{\mu} = (E,\textbf{k}) = (E,0,0,E)
 \end{equation}
 \begin{equation}
\label{eqn:fourv_kprime}
k'_{\mu} = (E',\textbf{k$'$}) = (E',E'\sin\theta\cos\phi,E'\sin\theta\sin\phi,E'\cos\theta)
 \end{equation}
  \begin{equation}
\label{eqn:fourv_p}
p_{\mu} = (M_p,\textbf{p}) = (M_p,0,0,0)
 \end{equation}
 Plugging these definitions and~(\ref{eqn:unpol_contraction}) into our equation for the cross section in~(\ref{eqn:xs_pretty}) we can write our unpolarized cross section as:
 \begin{equation}
\label{eqn:xs_unpol}
\frac{d^2\sigma}{d E' d \Omega} = \frac{\alpha^2 E'}{4 E^3} \bigg(2W_{1}(\nu,Q^2)\frac{1}{\sin^2\frac{\theta}{2}}+W_{2}(\nu,Q^2)\frac{\cos^2\frac{\theta}{2}}{\sin^4\frac{\theta}{2}}\bigg)
\end{equation}
To simplify our understanding of this quantity, it is helpful to factor out the Mott cross section, which describes scattering off of a simple point particle~\cite{HalzenMartin}:
\begin{equation}
\label{eqn:mott}
\bigg(\frac{d\sigma}{d \Omega}\bigg)_{Mott} = \frac{\alpha^2 \cos^2\frac{\theta}{2}}{4 E^2 \sin^4\frac{\theta}{2}}\frac{E'}{E}
\end{equation}
 \begin{equation}
\label{eqn:xs_unpol2}
\frac{d^2\sigma}{d E' d \Omega} = \bigg(\frac{d\sigma}{d \Omega}\bigg)_{Mott} \bigg(2W_{1}(\nu,Q^2)\tan^2\frac{\theta}{2}+W_{2}(\nu,Q^2)\bigg)
\end{equation}
This allows us to interpret the structure functions as quantities which characterize the difference in behavior from a pointlike particle. If we add in the asymmetric terms from the unpolarized lepton and hadron tensors of~(\ref{eqn:lepton_polarized}) and~(\ref{eqn:hadron_polarized}) to our contraction, we get several additional terms in this cross section expansion:
\begin{equation}
\begin{split}
\label{eqn:xs_pol}
\frac{d^2\sigma}{d E' d \Omega} = \bigg(\frac{d\sigma}{d \Omega}\bigg)_{Mott}\frac{E^2}{Q^2}\tan^2\frac{\theta}{2}\sin^2\frac{\theta}{2} \bigg(2W_{1}(\nu,Q^2)\tan^2\frac{\theta}{2}+W_{2}(\nu,Q^2)\frac{1}{\tan^2\frac{\theta}{2}} +\\
4 M_{p} G_{1}(\nu,Q^2)\big[(\textbf{s}\cdot\textbf{S})+\frac{1}{Q^2}(\textbf{s}\cdot\textbf{q})(\textbf{q}\cdot\textbf{S})\big]+\\
\frac{4}{M_{p}} G_{2}(\nu,Q^2)\big[(\textbf{p}\cdot\textbf{q})(\textbf{s}\cdot\textbf{S})-(\textbf{S}\cdot\textbf{q})(\textbf{p}\cdot\textbf{s})\big]\bigg)
\end{split}
\end{equation}
To simplify this, we will define the spin vectors as we did the others in the lab frame. The electron's spin vector \textbf{s} points along the initial momentum \textbf{k} and thus differs only by a factor of the electron mass. The proton's spin vector is situated at some angle $\beta$ from the scattering plane at an azimuthal angle of $\rho$:
\begin{equation}
\label{eqn:fourv_s}
s_{\mu} = (\frac{E}{m_e},\textbf{s}) = \frac{1}{m_e}(E,0,0,E)
 \end{equation}
 \begin{equation}
\label{eqn:fourv_S}
S_{\mu} = (0,\textbf{S}) = (0,\sin\beta\cos\rho,\sin\beta\sin\rho,\cos\beta)
 \end{equation}
 Using these and our earlier four vector definitions we can write, making use a few trigonometric identities:
\begin{equation}
\begin{split}
\label{eqn:xs_pol2}
\frac{d^2\sigma}{d E' d \Omega} = \bigg(\frac{d\sigma}{d \Omega}\bigg)_{Mott}\frac{E^2}{Q^2}\tan^2\frac{\theta}{2}\sin^2\frac{\theta}{2} \bigg(2W_{1}(\nu,Q^2)\tan^2\frac{\theta}{2}+W_{2}(\nu,Q^2)\frac{1}{\tan^2\frac{\theta}{2}} +\\
4 M_{p} G_{1}(\nu,Q^2)\big[(E\cos\beta+E'\sin\theta\sin\beta\cos(\phi-\rho)+E'\cos\theta\cos\beta\big]+\\
4 G_{2}(\nu,Q^2)\big[2E\sin\theta\sin\beta\cos(\phi-\rho)-Q^2\cos\beta\big]\bigg)
\end{split}
\end{equation}
Though this seems messy, this construction of the cross section will be our basic template for extracting the structure functions, and will make significantly more sense with smart choices of angle for the proton polarization vector $S_{\mu}$.
\section{Structure Functions}
We have already identified four structure functions $W_1$, $W_2$, $G_1$, and $G_2$, though we have not attempted to explain them properly. Before we try to do so, it is necessary to consider how we can simplify the cross section above to extract the structure functions we care about. Firstly, we consider the cross section where the electron has forward helicity, to one where it has backward helicity. The latter will flip the sign of the latter two terms by inverting the sign of $s_{\mu}$, so if we subtract one from the other we will cancel the unpolarized terms and double the magnitude of the polarized terms, leaving us with:
\begin{equation}
\begin{split}
\label{eqn:xs_poldiff1}
\frac{d^2\sigma^\uparrow}{d E' d \Omega} - \frac{d^2\sigma^\downarrow}{d E' d \Omega} = \bigg(\frac{d\sigma}{d \Omega}\bigg)_{Mott}\frac{8 E^2}{Q^2}\tan^2\frac{\theta}{2}\sin^2\frac{\theta}{2} \times \\
\bigg(M_{p} G_{1}(\nu,Q^2)\big[(E\cos\beta+E'\sin\theta\sin\beta\cos(\phi-\rho)+E'\cos\theta\cos\beta\big]+\\
G_{2}(\nu,Q^2)\big[2EE'\sin\theta\sin\beta\cos(\phi-\rho)-Q^2\cos\beta\big]\bigg)
\end{split}
\end{equation}
Let us now consider the proton polarization. We will look at two specific cases, one where the proton is polarized longitudinally, or in the same direction as k, in which case $\beta$ = 0 or $\pi$.
\begin{equation}
\begin{split}
\label{eqn:xs_poldiff_long}
\Delta\sigma_\parallel = \frac{d^2\sigma^{\uparrow\Uparrow}}{d E' d \Omega} - \frac{d^2\sigma^{\downarrow\Uparrow}}{d E' d \Omega} =  \bigg(\frac{d\sigma}{d \Omega}\bigg)_{Mott}\frac{4 E^2}{Q^2}\tan^2\frac{\theta}{2}\sin^2\frac{\theta}{2}\times \\ \bigg(M_{p} G_{1}(\nu,Q^2)\big[E+E'\cos\theta\big]-Q^2 G_{2}(\nu,Q^2)\bigg)
\end{split}
\end{equation}
This is known as the longitudinal cross section difference. We can also define a transverse cross section difference where $\beta$ = $\frac{\pi}{2}$ or $\frac{3 \pi}{2}$:
\begin{equation}
\begin{split}
\label{eqn:xs_poldiff_trans}
\Delta\sigma_\perp = \frac{d^2\sigma^{\uparrow\Rightarrow}}{d E' d \Omega} - \frac{d^2\sigma^{\downarrow\Rightarrow}}{d E' d \Omega} =  \bigg(\frac{d\sigma}{d \Omega}\bigg)_{Mott}\frac{4 E^2}{Q^2}\tan^2\frac{\theta}{2}\sin^2\frac{\theta}{2}\times \\ E'\sin\theta\cos(\phi-\rho)\bigg(M_{p} G_{1}(\nu,Q^2)+2 E G_{2}(\nu,Q^2)\bigg)
\end{split}
\end{equation}
Notably, these structure functions are dimensionful quantities. So, there is a convention to redefine several dimensionless structure functions as a function of the Bjorken-x defined above~\cite{Thomas}:
\begin{equation}
\label{eqn:f1}
F_1(x,Q^2) = M_p W_1(\nu,Q^2)
 \end{equation}
 \begin{equation}
\label{eqn:f2}
F_2(x,Q^2) = \nu W_2(\nu,Q^2)
 \end{equation}
  \begin{equation}
\label{eqn:g1}
g_1(x,Q^2) = \nu M_p^2 G_1(\nu,Q^2)
 \end{equation}
 \begin{equation}
\label{eqn:g2}
g_2(x,Q^2) = \nu^2 M_p G_2(\nu,Q^2)
 \end{equation}
We can define the angle difference $\phi - \rho$ as the out-of-plane polarization angle, the angle between the polarization and scattering planes, $\theta_{OoP}$. This gives us a final definition of our polarized cross section differences:
 \begin{equation}
\begin{split}
\label{eqn:xs_poldiff_long2}
\Delta\sigma_\parallel = \frac{d^2\sigma^{\uparrow\Uparrow}}{d E' d \Omega} - \frac{d^2\sigma^{\downarrow\Uparrow}}{d E' d \Omega} =  \bigg(\frac{d\sigma}{d \Omega}\bigg)_{Mott}\frac{4 E^2}{Q^2}\tan^2\frac{\theta}{2}\sin^2\frac{\theta}{2}\times \\ \frac{1}{M_p}\bigg(\frac{1}{\nu} g_{1}(x,Q^2)\big[E+E'\cos\theta\big]-\frac{Q^2}{\nu^2} g_{2}(x,Q^2)\bigg)
\end{split}
\end{equation}
\begin{equation}
\begin{split}
\label{eqn:xs_poldiff_trans2}
\Delta\sigma_\perp = \frac{d^2\sigma^{\uparrow\Rightarrow}}{d E' d \Omega} - \frac{d^2\sigma^{\downarrow\Rightarrow}}{d E' d \Omega} =  \bigg(\frac{d\sigma}{d \Omega}\bigg)_{Mott}\frac{4 E^2}{Q^2}\tan^2\frac{\theta}{2}\sin^2\frac{\theta}{2}\times \\ \frac{E'}{M_p}\sin\theta\cos\theta_{OoP}\bigg(\frac{1}{\nu} g_{1}(x,Q^2)+2 \frac{E}{\nu^2} g_{2}(x,Q^2)\bigg)
\end{split}
\end{equation}
The former equation is dominated by the $g_1$ term, while the latter is dominated by the $g_2$ term. For this reason, we refer to $g_1$ as a longitudinal structure function, which can be measured with a longitudinally polarized target, and $g_2$ as a transverse structure function, which can be measured with a transverse polarized target. At this point we have two equations and two unknowns, so a measurement of both polarized cross section differences would let us solve for our structure functions $g_1$ and $g_2$. In practice, these are rarely both measured at the same kinematics, so we will need to solve for them slightly differently, a discussion which we will pick up in the Analysis chapter.

Let us now consider the meaning of these structure functions. It is clear from the equations above that $g_1$ and $g_2$ characterize the spin-dependent part of the scattering event which was introduced by our asymmetric tensors, and that they each act as a type of scaling on the basic Mott cross section. So, it is simplest to interpret that for a fixed momentum transfer $Q^2$, these structure functions represent the deviation of the proton's internal spin from point-like behavior, with measurements over the Bjorken-x at varying $Q^2$ allowing us to map out this spin structure over a range of distance scales. 
\section{Moments and Sum Rules}
The natural question now is how we can relate these structure functions back to theoretical predictions of QCD. The structure functions themselves are difficult to calculate directly from the known theory of the standard model because they encompass all possible unknown states X for the final state of the hadronic system. Instead, we can investigate a class of kinematically weighted integrals of the spin structure functions known as moments or sum rules~\cite{Zielinski:2017gwp}.

To handle the unknown states X in a way that we can calculate we can employ the optical theorem~\cite{jacksonT}. The optical theorem states that the total scattering cross section is proportional to the imaginary part of the forward scattering amplitude:
\begin{equation}
\label{eqn:opticaltheorem}
\sigma_{tot} = \frac{4 \pi}{K} \operatorname{Im}\big(f(0)\big)
 \end{equation}
 Here K is the virtual photon flux, the definition of which is solely based in convention. Commonly, and in this analysis, we will employ the Hand convention~\cite{KarlThesis}:
 \begin{equation}
\label{eqn:HandConvention}
K = \nu - \frac{Q^2}{2M_p}
 \end{equation}
 In Compton scattering, a high frequency photon scatters off of a stationary target. Specifically, we are interested in virtual-virtual Compton scattering (VVCS), which concerns the scattering of a virtual photon as both the initial and final state. By using completeness and writing out the amplitude of Figure~\ref{fig:compscat} in terms of the probability currents, much as we did for the general form of the hadron tensor, we can write the Compton tensor as~\cite{Drechsel2}:
 \begin{equation}
\label{eqn:forward_compton}
T^{\mu \nu} = i \mathcal{T}\int e^{i q x}<P,S|J^\mu(x)J^\nu(0)|P,S>d^4x
 \end{equation}
 We note that this is nearly identical to the generic form of our hadron tensor in (\ref{eqn:hadron_tensor_simpler}), with the addition of a time-ordered product $\mathcal{T}$ which demands the chronological ordering of the probability current operators. Consequently, we can relate the hadron and Compton tensors with~\cite{KarlThesis}:
 \begin{equation}
\label{eqn:opticaltheorem_hadron}
W_{\mu \nu} = \frac{1}{2 \pi M_p} \operatorname{Im}\big( T_{\mu \nu}\big)
 \end{equation}
 This means instead of the unknown states of Figure~\ref{fig:epscattering}, we can investigate the amplitude of the simpler diagram in Figure~\ref{fig:compscat} to derive the measurable quantities of interest.
 \vspace{0.2in}
 
\begin{figure}[htb]
\centering
 \begin{fmffile}{feyngraph2}
  \begin{fmfgraph*}(280,160)
    \fmfleft{ip,il}
    \fmfright{op,x1,x2,K,x3,x4,o2,o3,o4,ol}
    \fmfset{arrow_len}{10}
    
    \fmflabel{\large $\gamma^*$}{il}
    \fmflabel{\large $\gamma^*$}{ol}
    \fmf{photon}{il,vp,ol}
    \marrow{pa}{ up }{ top}{$q^{\mu}$}{il,vp}
    \marrow{ea}{ up }{top}{$q^\mu$}{vp,ol}
    \fmf{phantom}{ip,vp,op}
    \fmffreeze
    \fmf{phantom}{ip,vp}
    \fmfi{fermion}{vpath (__ip,__vp) scaled 1.01}
    \fmfi{fermion}{vpath (__ip,__vp) scaled 1.01 shifted (-1.7, 6)}
    \fmfi{fermion}{vpath (__ip,__vp) scaled 1.01 shifted ( 1.7,-6)}
    \marrowz{ed}{ up }{top}{$p^\mu$}{ip,vp}
    \marrowzz{ex}{ up }{top}{$p^\mu$}{vp,op}
    \fmf{phantom}{vp,op}
    \fmfi{fermion}{vpath (__vp,__op) scaled 1.05 shifted (-5.1, 6)}
    \fmfi{fermion}{vpath (__vp,__op) scaled 1.05 shifted (-7.0, 0)}
    \fmfi{fermion}{vpath (__vp,__op) scaled 1.05 shifted (-8.9,-6)}
    \fmffreeze
    \fmflabel{\large $p$}{ip}
    \fmflabel{\large $p$}{op}
    \fmfblob{70}{vp}
  \end{fmfgraph*}
\end{fmffile}

\caption{Virtual-Virtual Compton Scattering Interaction}
\label{fig:compscat}
\end{figure}
To begin, we can write down the decomposition of the VVCS forward scattering amplitude, which takes the form~\cite{Drechsel}:
 \begin{equation}
\label{eqn:VVCS}
T = \vec{\epsilon}' \cdot \vec{\epsilon} f_T + f_L + i \vec{S} \cdot (\vec{\epsilon}' \times \vec{\epsilon}) f_{TT} + - i \vec{S} \cdot \big[(\vec{\epsilon}' - \vec{\epsilon}) \times \hat{q} \big] f_{LT}
 \end{equation}
 Here the cross section is expanded in terms of the four amplitudes $f_{L,T,TT,LT}$ which describe the scattering amplitude of the photon's longitudinal polarization states and transverse polarization states respectively, as well as the transverse-transverse and longitudinal-transverse interference amplitudes. These interference amplitudes are produced by the interaction with the proton's spin, as without the additional spin vector the only nonvanishing terms would be $f_T$ and $f_L$. $\epsilon'$ and $\epsilon$ represent the transverse polarization vectors of the virtual photon, and $\hat{q}$ represents the longitudinal polarization vector of the virtual photon. These amplitudes are related by the optical theorem back to four partial virtual photon cross sections~\cite{Drechsel}: 
 \begin{equation}
\label{eqn:opticaltheorem2}
\operatorname{Im}\{f_{L},f_{T},f_{TT},f_{LT}\} = \frac{K}{4 \pi}\{\sigma_{L},\sigma_{T},\sigma_{TT},\sigma_{LT}\}
 \end{equation}
 These virtual photon cross sections are related back to the structure functions of the previous section by~\cite{Drechsel}:
 \begin{equation}
\label{eqn:vphot_struc}
\{\sigma_{L},\sigma_{T},\sigma_{TT},\sigma_{LT}\} = \frac{4 \pi^2 \alpha^2}{M_p K} \{F_1,\frac{M_p(1+\frac{Q^2}{\nu^2})}{\frac{Q^2}{\nu}}F_2-F_1,g_1 - \frac{Q^2}{\nu^2}g_2,\frac{Q}{\nu}(g_1+g_2)\}
 \end{equation}
 Because understanding the imaginary part of the amplitude will be difficult, it will be better to relate this back to the real part of the amplitude. We can do this by writing a dispersion relation with the Cauchy integral formula~\cite{Zielinski:2017gwp}:
 \begin{equation}
\label{eqn:cauchy}
f_i (\nu,Q^2) = \frac{2}{\pi} \int_{\nu_{pp}}^\infty \frac{\nu' \operatorname{Im}\big[f_i(\nu,Q^2)\big]}{\nu'^2 - \nu^2}d\nu'
 \end{equation}
 Here, $\nu_{pp}$ is the energy transfer of the pion production threshold discussed earlier. $F_1$ and $F_2$ are interesting for the insight they provide into the quark-gluon distribution of the nucleon, but here we are interested in the polarized structure functions, so we will focus on the interference amplitudes $f_{LT}$ and $f_{TT}$. By plugging (\ref{eqn:opticaltheorem2}) into (\ref{eqn:cauchy}) we obtain:
 \begin{equation}
\label{eqn:disp_TT}
f_{TT} (\nu,Q^2) = \frac{1}{2 \pi^2} \int_{\nu_{pp}}^\infty \frac{K \nu \sigma_{TT}(\nu,Q^2)}{\nu'^2 - \nu^2}d\nu'
 \end{equation}
 \begin{equation}
\label{eqn:disp_LT}
f_{LT} (\nu,Q^2) = \frac{1}{2 \pi^2} \int_{\nu_{pp}}^\infty \frac{K \nu' \sigma_{LT}(\nu,Q^2)}{\nu'^2 - \nu^2}d\nu'
 \end{equation}
 To study this integral, we perform a low-energy power series expansion on the real part of the amplitude. For $\nu < \nu_{pp}$ we obtain~\cite{Drechsel}:
 \begin{equation}
\label{eqn:disp_TT2}
\operatorname{Re}f_{TT} (\nu,Q^2) = \frac{2 \alpha}{M_p^2}I_{A}(Q^2)\nu + \gamma_0(Q^2)\nu^3 + \mathcal{O}(\nu^5)...
 \end{equation}
 \begin{equation}
\label{eqn:disp_LT2}
\operatorname{Re}f_{LT} (\nu,Q^2) = \frac{2 \alpha}{M_p^2}Q I_{LT}(Q^2) + Q \delta_{LT}(Q^2)\nu^2 + \mathcal{O}(\nu^4)...
 \end{equation}
 The terms here are described as moments of the spin structure functions, comparison to the integral and the use of~(\ref{eqn:vphot_struc}) yields a series of sum rules defining them. The first moment in the $f_{TT}$ low-energy expansion (LEX) is:
 
 \begin{equation}
\label{eqn:gdh}
I_A(Q^2) = \frac{2 M_p^2}{Q^2}\int_{0}^{x_{pp}} \bigg(g_1(x,Q^2) - \frac{4 M^2 x^2}{Q^2} g_2(x,Q^2)\bigg)dx
 \end{equation}
 This is known as the GDH Sum Rule. At the real photon point, $Q^2$ = 0, the GDH Sum Rule can be identified with the magnetic moment $\kappa$ of the proton, such that:
 \begin{equation}
\label{eqn:gdh2}
I_A(0) = -\frac{2 \pi^2 \alpha \kappa^2}{M_p^2}
 \end{equation}
 The second term in $f_{TT}$'s expansion yields the generalized forward spin polarizability:
 \begin{equation}
\label{eqn:gamma0}
\gamma_0(Q^2) = \frac{16 \alpha M_p^2}{Q^6}\int_{0}^{x_{pp}} x^2\bigg(g_1(x,Q^2) - \frac{4 M^2 x^2}{Q^2} g_2(x,Q^2)\bigg)dx
 \end{equation}
 As will be discussed later, this and other polarizabilities describe the nucleon's response to an external field, with spin polarizabilities correlating to a spin-dependent response. In both $I_A$ and $\gamma_0$, the integral is dominated by the $g_1$ term due to the strong kinematic weighting on the $g_2$ term, and thus is measured primarily with longitudinally polarized nucleon data.
 
 For the $f_{LT}$ expansion, the first term yields the moment:
 \begin{equation}
\label{eqn:Ilt}
I_{LT}(Q^2) = \frac{2 M_p^2}{Q^2}\int_{0}^{x_{pp}} \bigg(g_1(x,Q^2) + g_2(x,Q^2)\bigg)dx
 \end{equation}
 While the second term gives us the longitudinal-transverse spin polarizability:
 \begin{equation}
\label{eqn:dlt}
\delta_{LT}(Q^2) = \frac{16 \alpha M_p^2}{Q^6}\int_{0}^{x_{pp}} x^2\bigg(g_1(x,Q^2) + g_2(x,Q^2)\bigg)dx
 \end{equation}
 Unlike our first two moments, $g_1$ and $g_2$ contribute on a similar scale to these integrals, so a transverse measurement is vital to compute the moment. Further, it is worth noting that in the second moments $\gamma_0$ and $\delta_{LT}$, the strong $x^2$ weighting means that the part of the integral closest to the pion production threshold will dominate, and a measurement that goes out to x = 0 or W = $\infty$ is unnecessary. For the first moments it is necessary to calculate this low-x, DIS contribution, but for the higher moments the integral is completely dominated by the resonance region shown in Figure~\ref{fig:xs_elastic}.
 
 There are several other moments of interest for which this is not the case. These moments are obtained by attempting to construct the VVCS amplitudes which correspond directly to the structure functions, which is done by casting (\ref{eqn:VVCS}) into a covariant form~\cite{Drechsel2}:
 \begin{equation}
 \begin{split}
\label{eqn:VVCS}
T = \epsilon_\mu' \epsilon_\nu \Bigg\{\bigg(-g^{\mu \nu}+\frac{q^\mu q^\nu}{Q^2}\bigg)T_1(\nu, Q^2) + \frac{1}{\textbf{p} \cdot \textbf{q}}\bigg(p^\mu - \frac{\textbf{p}\cdot\textbf{q}}{Q^2}q^\mu\bigg)\bigg(p^\nu - \frac{\textbf{p}\cdot\textbf{q}}{Q^2}q^\nu\bigg)T_2(\nu, Q^2) \\
+ \frac{i}{M_p}\epsilon^{\mu \nu \alpha \beta}q_\alpha S_\beta S_1(\nu, Q^2) + \frac{i}{M_p^3}\epsilon^{\mu \nu \alpha \beta}q_\alpha\bigg(\textbf{p}\cdot\textbf{q}S_\beta - \textbf{S}\cdot\textbf{q}p_\beta\bigg)S_2(\nu, Q^2)
\end{split}
 \end{equation}
 This form introduces four covariant VVCS amplitudes, which are related to the previously introduced VVCS amplitudes by:
 \begin{equation}
\label{eqn:T1}
T_1 (\nu,Q^2) = f_T (\nu,Q^2)
 \end{equation}
 \begin{equation}
\label{eqn:T2}
T_2 (\nu,Q^2) = \frac{\nu Q^2}{M_p(\nu^2 + Q^2)}\big(f_T (\nu,Q^2)+f_L (\nu,Q^2)\big)
 \end{equation}
 \begin{equation}
\label{eqn:S1}
S_1 (\nu,Q^2) = \frac{\nu M_p}{\nu^2 + Q^2}\big(f_{TT} (\nu,Q^2)+\frac{Q}{\nu}f_{LT} (\nu,Q^2)\big)
 \end{equation}
 \begin{equation}
\label{eqn:S2}
S_2 (\nu,Q^2) = -\frac{M_p^2}{\nu^2 + Q^2}\big(f_{TT} (\nu,Q^2)-\frac{\nu}{Q}f_{LT} (\nu,Q^2)\big)
 \end{equation}
 If we employ our previous application of the optical theorem and take the imaginary part of these definitions, we can relate these amplitudes back to the structure functions:
 \begin{equation}
\label{eqn:T1_optical}
\operatorname{Im}T_1 (\nu,Q^2) = \frac{2 \pi}{M_p} F_1(x,Q^2)
 \end{equation}
 \begin{equation}
\label{eqn:T2_optical}
\operatorname{Im}T_2 (\nu,Q^2) = \frac{2 \pi}{\nu} F_2(x,Q^2)
 \end{equation}
 \begin{equation}
\label{eqn:S1_optical}
\operatorname{Im}S_1 (\nu,Q^2) = \frac{2 \pi}{\nu M_p^2} g_1(x,Q^2)
 \end{equation}
 \begin{equation}
\label{eqn:S2_optical}
\operatorname{Im}S_2 (\nu,Q^2) = \frac{2 \pi}{\nu^2 M_p} g_2(x,Q^2)
 \end{equation}
 Again, we will focus on the latter two amplitudes, as we can see they are directly related to our spin-dependent structure functions. Writing a dispersion relation for each with the Cauchy theorem:
 \begin{equation}
\label{eqn:disp_S1}
S_1 (\nu,Q^2) = \frac{2}{\pi} \int_{\nu_{pp}}^\infty \frac{\nu' \operatorname{Im}S_1}{\nu'^2 - \nu^2}d\nu'
 \end{equation}
 \begin{equation}
\label{eqn:disp_S2}
S_2 (\nu,Q^2) = \frac{2}{\pi} \int_{\nu_{pp}}^\infty \frac{\nu \operatorname{Im}S_2}{\nu'^2 - \nu^2}d\nu'
 \end{equation}
 It will also be useful to write a dispersion relation for the amplitude $\nu S_2$:
 \begin{equation}
\label{eqn:disp_nuS2}
\nu S_2 (\nu,Q^2) = \frac{2}{\pi} \int_{\nu_{pp}}^\infty \frac{\nu' \operatorname{Im}S_2}{\nu'^2 - \nu^2}d\nu'
 \end{equation}
 For $S_1$ we can perform a low-energy expansion just like we did for the other amplitudes:
 \begin{equation}
\label{eqn:S1_LEX}
\operatorname{Re}S_1 (\nu,Q^2) = \frac{2 \alpha}{M_p}\Gamma_1(Q^2)+\bigg[\frac{2\alpha}{M_p Q^2}\big(I_A(Q^2) - \Gamma_1(Q^2)\big)+M_p\delta_{LT}(Q^2)\bigg]\nu^2+\mathcal{O}(\nu^4)...
 \end{equation}
 Here we have several of the moments defined previously, as well as one new one. Using our definitions of $S_1$ in terms of the structure functions:
 \begin{equation}
\label{eqn:Gamma1}
\Gamma_1 (Q^2) = \int_0^1 g_1(x,Q^2)dx
 \end{equation}
 It is worth noting that this moment contains the full integral from x of 0 to 1, and thus requires a calculation of the elastic contribution. We can obtain a similar moment, and with it an important sum rule, by subtracting (\ref{eqn:disp_nuS2}) from (\ref{eqn:disp_S2}) to obtain~\cite{Drechsel2}:
 \begin{equation}
\label{eqn:Gamma2}
\Gamma_2 (Q^2) = \int_0^1 g_2(x,Q^2)dx = 0
 \end{equation}
 This ``Super-Convergence relation'' is known as the Burkhardt-Cottingham sum rule, an identity which has power in equating the elastic and inelastic parts of the integral, stating that they ought to cancel each other for any value of $Q^2$. Due to the lack of any kinematic weighting, $\Gamma_1$ and $\Gamma_2$ require every part of the integral to be accounted for, including whatever unmeasured low-x part exists.
 
 One final set of moments is of interest to us, and can be produced by considering the full, general form of the forward Compton scattering amplitude in (\ref{eqn:forward_compton}). One way to obtain a useful expansion of this integral is with the Operator Product Expansion (OPE)~\cite{Wilson}. At small distances, the OPE allows the following general expansion of two operators~\cite{Zielinski:2017gwp}:
 \begin{equation}
\label{eqn:OPE}
\lim_{x\rightarrow0}\mathcal{O}_{a}(x)\mathcal{O}_{b}(0) = \sum_{k}C_{abk}(x)\mathcal{O}_{k}(0)
 \end{equation}
 Taking the Fourier transform of this equation yields:
 \begin{equation}
\label{eqn:OPE_fourier}
\lim_{x\rightarrow0}\int e^{iqx}\mathcal{O}_{a}(x)\mathcal{O}_{b}(0)d^4x = \lim_{Q^2\rightarrow\infty}\sum_{k}C_{abk}(q)\mathcal{O}_{k}(0)
 \end{equation}
 This allows us to apply the OPE directly to our forward Compton amplitude, but as we can see, the method becomes valid only in the high $Q^2$ regime. This yields a Compton amplitude in the high $Q^2$ limit of:
 \begin{equation}
\label{eqn:compton_OPE}
T^{\mu \nu} = \sum_{k}C_{\mu \nu k}(Q^2)J_{k}(0)
 \end{equation}
 The exercise of calculating the Wilson coefficients $C_{abk}(q)$ is carried out in~\cite{Thomas}, and yields the general form:
 \begin{equation}
\label{eqn:wilson_coeff}
C_{\mu \nu k} \approx \bigg(\frac{1}{x}\bigg)^{n}\bigg(\frac{M_p}{\sqrt{Q^2}}\bigg)^{\tau - 2}
 \end{equation}
 Here n is defined as the spin of the relevant operator, and with D as dimension of the operator, we additionally define the \textit{twist} $\tau$ as $\tau = D - n$. Twist is a spin-dependent quantity which helps to quantify the order in $Q^2$ that a particular effect contributes for a given interaction~\cite{Jaffe:1996zw}. At high $Q^2$, the leading twist is twist-2, but at lower $Q^2$, higher twist terms can contribute. If we expand up to twist-3, we can write our forward Compton amplitude expansion as a function of the matrix elements $a_{n}$ and $d_{n}$, the twist-2 and twist-3 local operators, such that:
 \begin{equation}
 \begin{split}
\label{eqn:localops}
T^{\mu \nu} = C^a_{n}(Q^0)a_n(Q^2) + C^d_{n}(Q^1)d_n(Q^2) + \mathcal{O}(Q^2) +...\\
\approx \bigg(\frac{1}{x}\bigg)^n a_n(Q^2) + \bigg(\frac{1}{x}\bigg)^n \frac{M_p}{\sqrt{Q^2}} d_n(Q^2)
 \end{split}
 \end{equation}
 We can extract a relation between these matrix elements and our structure functions by writing a dispersion relation, an exercise performed in ~\cite{Jaffe:1996zw}. The result gives us the general form:
 \begin{equation}
 \begin{split}
\label{eqn:an_disp}
\int_0^1 x^n g_1(x,Q^2) = \frac{1}{4}a_{n}(Q^2) ; n = 0,2,4
 \end{split}
 \end{equation}
 \begin{equation}
 \begin{split}
\label{eqn:dn_disp}
\int_0^1 x^n g_2(x,Q^2) = \frac{1}{4}\frac{n}{n+1}\big(d_n - a_n\big) ; n = 2,4
 \end{split}
 \end{equation}
 The n=0 case of the second relation is ignored because the OPE assumes implicitly that the B.C. sum rule previously introduced holds. Of interest to us is the matrix element $d_2$. For the n=2 case, we can solve the previous two equations to obtain:
 \begin{equation}
\label{eqn:d2}
d_2(Q^2) = \int_0^1 x^2\bigg(2g_1(x,Q^2) + 3g_2(x,Q^2)\bigg)dx
 \end{equation}
 At high $Q^2$, this matrix element is identified as a color polarizability, which quantifies a color-dependent response to an external field. At lower $Q^2$, we can still employ this sum rule to look at the inelastic part:
 \begin{equation}
\label{eqn:d2bar}
\overline{d}_2(Q^2) = \int_0^{x_{pp}} x^2\bigg(2g_1(x,Q^2) + 3g_2(x,Q^2)\bigg)dx
 \end{equation}
 As the elastic part does not contribute to this moment, at low $Q^2$ we can still view this as a ``pure polarizability''~\cite{Alarcon}. Although the color polarizability definition fails and we can no longer identify $\overline{d}_2$ with $d_2$, the low $Q^2$ regime of $\overline{d}_2$ contains information about the quark-gluon correlations of the nucleon.

\section{Chiral Perturbation Theory}

\begin{figure}[htb]
\centering
\includegraphics[width=0.5\textwidth]{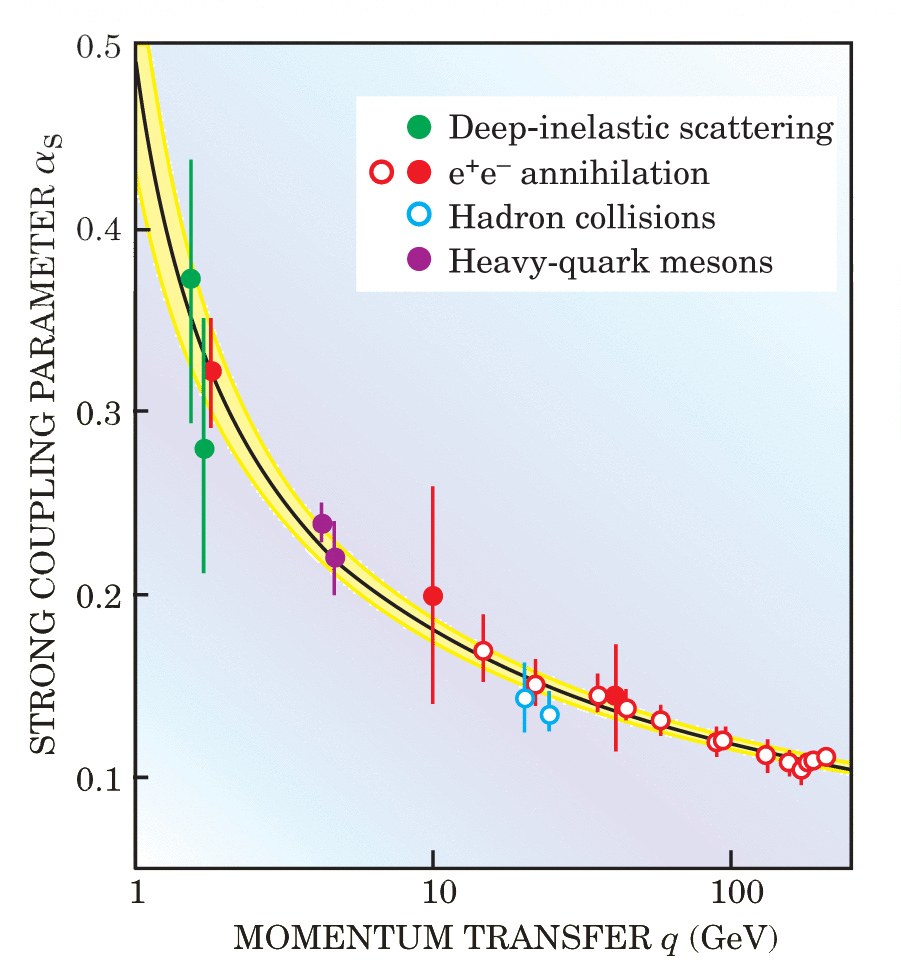}
\caption{Asymptotic freedom and confinement. Adapted from a figure in \cite{asymptotic_freedom}}
\label{fig:strongforce}
\end{figure}

The moments introduced in the last section show various ways that we can calculate meaningful quantities from our structure functions, but the question remains how we can compare these quantities to modern theories of Quantum Chromodynamics (QCD). QCD is the theory of the strong nuclear force, and unlike the comparatively simple coupling of the electromagnetic force, it has been observed to couple quarks and gluons more strongly together in response to attempts to extract them from the nucleon, a phenomenon known as confinement. In contrast, at small length scales they are bound very weakly in what is known as `asymptotic freedom'~\cite{Scherer}. We know that our $Q^2$ controls what length scale we are probing, which means that at small $Q^2$, the strong force becomes very powerful, as shown in Figure ~\ref{fig:strongforce}.

At this length scale, rather than characterizing the dynamics of individual quarks and gluons, QCD handles how they are bound together into a larger hadron. In the realm of asymptotic freedom, where the strong force is weaker, QCD is treated perturbatively, but at low $Q^2$ such treatments become impossible. As it is currently impossible to find an analytical solution at this energy scale, it is most useful to examine low $Q^2$ QCD with an effective field theory. In quantum field theory, an effective field theory is a subset of a larger theory which details its behavior in a specific energy regime. Low energy effective theories allow for a perturbative approach in terms of the momentum, instead of the coupling constant as is used for the full theory of QCD at high energy. There are a number of effective theories which present predictions of strong force behavior at low $Q^2$, such as Lattice QCD, but so far the most successful at very low energies is known as Chiral Perturbation Theory, or $\chi$PT. 

To begin, let us consider QCD as a gauge theory, meaning that it is invariant under local transformations at a given energy scale. QCD is based on the color gauge group SU(3). In other words, the special unitary group of rank 3 parameterizes many of the symmetries of the strong force. From the gauge attributes of SU(3) it is possible to obtain the general QCD Lagrangian~\cite{Scherer}:
\begin{equation}
\label{eqn:lagrangian_QCD}
\mathcal{L}_{QCD} = \sum_{f=u,d,s,c,t,b}\overline{q}_f (i\gamma^\mu D_\mu - m_f)q_f - \frac{1}{4}\mathcal{G}_{\mu \nu} \mathcal{G}^{\mu \nu}
 \end{equation}
 Here we sum over the six known quark flavors, $q_f$ and $\overline{q}_f$ are the quark fields, $\mathcal{G}_{\mu \nu}$ is the gluon tensor, $m_f$ is the mass of the relevant quark flavor, and $D_\mu$ is the covariant derivative. The heavier quarks (charm, top, and bottom) occur much less frequently so their contributions may be largely disregarded. For the remaining quarks (up, down, strange), their mass $m_f \ll M_p$, so it may be a natural assumption to treat them as massless.
 
\begin{figure}[htb]
\centering
\includegraphics[width=0.5\textwidth]{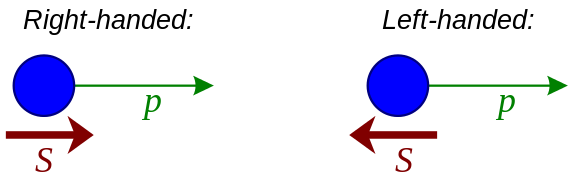}
\caption{Definitions of handedness in helicity. Adapted from a figure in \cite{helicity_wiki}}
\label{fig:helc}
\end{figure}

 $q_f$ are color triplets based on the three possible color charges, but we must also consider left and right handed chirality of the quarks. For massive particles, chirality refers to the handedness of the group representation, but for massless particles we can easily understand it as being the same as the helicity. The helicity of a particle refers to whether its spin is aligned or antialigned with its momentum, as shown in figure~\ref{fig:helc}.

We define the chiral quark fields in terms of a left or right-handed projection operator~\cite{Scherer}:
\begin{equation}
\label{eqn:chiral_P}
P^{L,R} = \frac{1}{2}(1\mp\gamma_5)
 \end{equation}
 \begin{equation}
\label{eqn:chiral_q}
q^{L,R}_f = P^{L,R}q_f = P^{L,R}\begin{pmatrix}
\textcolor{red}{q_f^{red}} \\
\textcolor{green}{q_f^{green}} \\ 
\textcolor{blue}{q_f^{blue}}\end{pmatrix}
 \end{equation}
 \begin{equation}
\label{eqn:chiral_qbar}
\overline{q}^{L,R}_f = \overline{q}_f  P^{L,R} = \begin{pmatrix}
\textcolor{cyan}{\overline{q}_f^{antired}} \\
\textcolor{magenta}{\overline{q}_f^{antigreen}} \\ 
\textcolor{yellow}{\overline{q}_f^{antiblue}}\end{pmatrix}P^{L,R}
 \end{equation}
 In the chiral limit of $m_f = 0$, the left and right handed quark interaction term vanishes. Disregarding the mass term and plugging these into our general QCD Lagrangian, we obtain:
 \begin{equation}
 \begin{split}
\label{eqn:lagrangian_QCD_0}
\mathcal{L}_{QCD}^0 = \sum_{f=u,d,s}\overline{q}_f^L (i\gamma^\mu D_\mu)q_f^L + \overline{q}_f^R (i\gamma^\mu D_\mu)q_f^R - \frac{1}{4}\mathcal{G}_{\mu \nu} \mathcal{G}^{\mu \nu}
\\ =\sum_{f=u,d,s}\overline{q}_f i (\gamma^\mu D_\mu + \gamma_5 \gamma^\mu D_\mu \gamma_5)q_f - \frac{1}{4}\mathcal{G}_{\mu \nu} \mathcal{G}^{\mu \nu}
\end{split}
 \end{equation}
 This chiral Lagrangian is said to have a global $U(3)_L \times U(3)_R$ symmetry~\cite{Scherer}. The progression of an effective field theory is to take a symmetric or analytically solvable term and add terms that are functions of a small parameter to reach the full Lagrangian through perturbative effects. Here, that is done by considering the missing part of the Lagrangian, namely, the mass term:
 
\begin{equation}
\label{eqn:lagrangian_QCD_I}
\mathcal{L}_{QCD}^I = -\sum_{f=u,d,s}\overline{q}_f^L m_f q_f^R + \overline{q}_f^R m_f q_f^L
 \end{equation}
 We can see that this part of the Lagrangian contains an explicit breaking of the chiral symmetry in $\mathcal{L}_{QCD}^0$. Because the quarks are weakly interacting at low energy, in a low energy effective theory it is possible to treat this term perturbatively with a low energy expansion. This allows us to consider the total effective $\chi$PT Lagrangian as~\cite{BernardChiral}:
 \begin{equation}
\label{eqn:lagrangian_chiPT}
\mathcal{L}_{\chi PT} = \mathcal{L}_{QCD}^0 + \mathcal{L}_{QCD}^I = \mathcal{L}_{QCD}^0 + \mathcal{L}_2 + \mathcal{L}_4 + ...
 \end{equation}
 Where the index on $\mathcal{L}_2$, etc refers to the low energy dimension in terms of derivatives or powers of the quark mass. Each of these terms can be written explicitly by considering all possible diagrams of a given dimension, they are constructed by ordering diagrams in powers of momentum. In this $\chi$PT Lagrangian, corrections are expanded in terms of pion loops, whereas in the full QCD Lagrangian, quark and gluon loops are used explicitly. At dimension 2, the tree level diagrams are of order $Q^2$ while the one loop corrections are of order $Q^4$, at dimension 4 ($\mathcal{L}_4$) the tree level diagrams are order $Q^4$ and so on. This means that to consider a full Lagrangian up to order $Q^4$ we must include tree level and one-loop corrections of $\mathcal{L}_2$, and tree level diagrams of $\mathcal{L}_4$.
 
 To expand these terms in orders of pion loops, the order in $Q^2$ also corresponds to the included order of the pion mass, $m_\pi$. The effective Lagrangian, and all quantities derived from it, is constructed by ordering the result in powers of the pion mass. The specific way in which this `power-counting' is done is an important point, and one which may have an effect on the final calculation, as assumptions must be made to handle the inclusion of the $\Delta$-resonance. One such method which is important here is the $\epsilon$-counting scheme, where it is assumed that the pion mass $m_\pi$ is of similar scale to the difference between the mass of the $\Delta$-resonance, $M_\Delta - M_p = \varDelta \approx m_\pi$. This allows $\varDelta$ and $m_\pi$ to be expanded to the same order~\cite{Bernard_moments}. The other method which will be discussed in the following chapters is the $\delta$-counting method, in which a different assumption is made; here it is assumed that the ratio $\frac{m_\pi}{\varDelta}$ is of similar scale to the ratio $\frac{\varDelta}{M_p}$, and consequently used to simplify and expand the symmetry breaking terms to like powers~\cite{Alarcon}.
 
 Once an effective Lagrangian is obtained, it can be used to obtain the equations of motion, and consequently, calculate the VVCS amplitudes of the previous section by varying the action such that $
\delta \mathcal{Z} = 0$:
\begin{equation}
\label{eqn:action}
\mathcal{Z} = \int d^4 x \mathcal{L}_{\chi PT}
 \end{equation}
 
 The calculation to obtain the Lagrangian expansion and compute the VVCS amplitudes is an elaborate one far beyond the scope of this thesis, but it is core to understand that differences such as the choice of power counting cause the cutting-edge chiral perturbation theory calculations to differ starkly in their predictions for the proton's spin structure function moments. These differences shall be discussed further in the analysis chapter, and the full details of these calculations can be found in \cite{Alarcon,Bernard_moments}. But the fact that such differences exist further illustrates the importance of experimental measurements in the region where $\chi$PT is applicable.
 
\chapter{Existing Data}
Before we begin to discuss the particular measurement which is the topic of this thesis, let us first consider existing relevant measurements of the structure functions and their moments.  Both structure functions for the neutron have been measured with high precision, as discussed in chapter 1, relevant for their revealing of the $\delta_{LT}$ puzzle in Figure~\ref{fig:dltpuzzle}. Jefferson Lab E94010, published in 2004, ran down to a minimum $Q^2$ of 0.1 GeV$^2$, and showed a disagreement with the predictions of $\chi$PT at the time~\cite{E94010}. The cutting edge $\chi$PT calculations\cite{Bernard_moments,Alarcon} have since remedied this discrepancy, but are still in tension with the recently published Small-Angle GDH neutron data from Jefferson Lab~\cite{saGDH}, which revealed a new disagreement at a lower $Q^2$ of 0.05 GeV$^2$. At higher $Q^2$, there is even more neutron structure function data of less relevancy to this thesis.

For the proton, the earliest structure function measurements in the late 1980s and early 1990s focused on the DIS region and were collected at the Stanford Linear Accelerator (SLAC) and the European Organization for Nuclear Research (CERN). These early measurements covered a broad range from medium to high $Q^2$ but were largely focused at higher $Q^2$. More recently, a number of experiments at Jefferson Lab (JLab) have delved to lower $Q^2$ and higher $x$, investigating the resonance region. This list of experiments was originally compiled by Ryan Zielinski in his thesis~\cite{Zielinski:2017gwp} and has been updated for the state of the field in 2022. For reference, $\chi$PT is valid below a $Q^2$ of around 0.3. The complete list of proton structure function data is summarized and organized by average $Q^2$ in Table~\ref{table:sfdata}. Though many of these experiments have used various methods to compute both $g_1$ and $g_2$, for the purpose of this table, an experiment which measured a longitudinally polarized target is marked having measured $g_1$, and one which measured a transverse-polarized target is marked as having measured $g_2$.

\begin{table}[h!]
\centering
\begin{tabular}{|c c c c c c|} 
 \hline
 Experiment & Facility & Avg. $Q^2$ (GeV$^2$) & $x_{bj}$ Range & $g_1$ & $g_2$ \\ [0.5ex] 
 \hline
 
 COMPASS\textsuperscript{\cite{COMPASS}} & CERN &  15.7 &  0.004-0.7 & \checkmark &  \\ 
 EMC\textsuperscript{\cite{Ashman2}} & CERN &  10.7 &  0.01-0.7& \checkmark &  \\ 
  E155\textsuperscript{\cite{E155}} & SLAC & 10.0 & 0.02-0.8 & \checkmark  &  \\ 
   E143\textsuperscript{\cite{E143}} & SLAC & 5.0 & 0.03-0.8 & \checkmark & \checkmark \\ 
   SANE\textsuperscript{\cite{warmstrong}} & JLab & 4.5 & 0.3-0.8 &  & \checkmark \\
 SMC\textsuperscript{\cite{SMC2}} & CERN & 4.0 & 0.003-0.7 & \checkmark & \checkmark \\ 
  E155x\textsuperscript{\cite{E155x}} & SLAC & 4.0 &  0.02-0.8 & & \checkmark\\
 HERMES\textsuperscript{\cite{Hermes}} & HERA & 3.0 & 0.004-0.9 & \checkmark & \checkmark \\
  EG1b\textsuperscript{\cite{EG1b2}} & JLab &  2.1 &  0.02-0.8& \checkmark &  \\ 
  RSS\textsuperscript{\cite{RSS}} & JLab & 1.3 & 0.03-0.8 & \checkmark & \checkmark \\
  EG4\textsuperscript{\cite{EG4_final}} & JLab &  0.4&  0.01-0.9& \checkmark &  \\ 
  \textcolor{blue}{g2p\textsuperscript{\cite{g2p_nature}}} & \textcolor{blue}{JLab} & \textcolor{blue}{0.07} & \textcolor{blue}{0.01-0.8}& \textcolor{blue}{\checkmark} & \textcolor{blue}{\checkmark}\\
 [1ex] 
 \hline
\end{tabular}
\caption{Existing proton structure function data as of August 2022}
\label{table:sfdata}
\end{table}

The first spin structure function measurement for the proton was an observation of $g_1$ by the European Muon Collaboration (EMC) in 1988~\cite{Ashman2}. This measurement disagreed with the prediction of the Ellis-Jaffe sum rule, and showed that the proton's valence quarks carry only a small fraction of its spin~\cite{JaffeSum}. Rather than the electron scattering described so far, this measurement was performed with polarized muon-proton scattering. Additional $g_1$ and $g_2$ measurements were obtained at CERN with polarized muon scattering in the SMC~\cite{SMC2} and COMPASS~\cite{COMPASS} experiments. Due to the high energy of the CERN beam, all of these experiments operated at a relatively high $Q^2$.

Also at high to medium $Q^2$ are several experiments from SLAC in the 1990s which helped to provide clarity on the structure functions through electron-proton scattering. SLAC E143~\cite{E143} measured both structure functions and found at high $Q^2$ that the $g_2$ results had good consistency with the leading twist predictions of the Wandzura-Wilczek relations. SLAC E155~\cite{E155} and E155x~\cite{E155x} built on these results by measuring the structure functions with high precision, with E155x performing one of the highest precision $g_2$ measurements to date. These experiments found good agreement with a number of sum rules as well as the $g_2^{WW}$ predictions.

Starting in the early 2000s, Jefferson Lab began to produce a number of modern structure functions results at much lower average $Q^2$, making use of the high intensity CEBAF electron beam. The RSS~\cite{RSS,KarlThesis} experiment performed an extraction of both structure functions in the resonance region for both the proton and neutron from a scattering measurement of Helium-3, a nucleus with 2 protons and 1 neutron. The proton measurement showed promising agreement with predictions for the Burkhardt-Cottingham and Gerasimov-Drell-Hearn sum rules, while the neutron result revealed the surprising result discussed in chapter one, which showed a deviation from the predictions of chiral perturbation theory. Another more recent measurement of $g_2$ comes from the HERMES experiment with HERA at DESY. HERMES integrated both polarized electron and positron scattering to measure $g_2$, focused in the DIS regime~\cite{Hermes}.

At higher $Q^2$, the SANE experiment~\cite{warmstrong}, which recently finished analysis, also measured several kinematic points of $g_2$, finding a potentially surprising result for one of its moments, the $d_2$ twist-3 operator. In Hall B, the EG1b experiment likewise measured the $g_1$ structure function in the resonance region at medium $Q^2$ and extracted the moments dominated by it. This experiment is notable as it is largely the basis for the CLAS model discussed in later chapters~\cite{EG1b2}.

Most recently, a program at Jefferson Lab encompassing several experiments has worked to study the low-$Q^2$ behavior of the nucleons. The Small-Angle GDH experiment, discussed in the first chapter, performed a low $Q^2$ measurement of the neutron structure functions and moments which revealed the original $\delta_{LT}$ puzzle is alive and well with a new discrepancy from $\chi$PT~\cite{saGDH}. The reason for the neutron's disagreement is still unknown, as the newest calculations agree with the earlier E94-010 neutron data but still disagree with the small-angle GDH results at even lower $Q^2$.

The EG4 experiment~\cite{EG4_final} performed a low $Q^2$ measurement of the $g_1$ structure function of the proton with longitudinal scattering data, in which they directly extracted the polarized cross section differences instead of relying on the diluted asymmetry method of many prior experiments. The results show good agreement with calculations of $\chi$PT for several moments, but also reveal the current lack of precision in the cutting-edge calculations and the need to improve the uncertainty associated with each.

\begin{figure}[htb]
\centering
\includegraphics[width=0.7\textwidth]{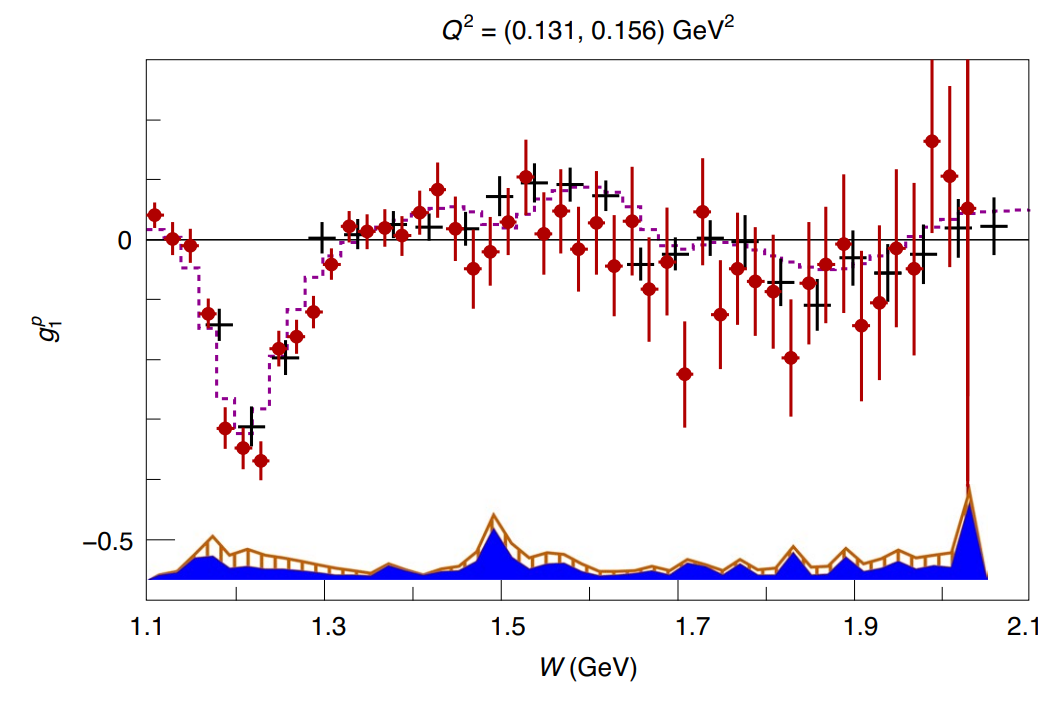}
\caption{The $g_1$ structure function of the proton measured by the EG4 experiment. The data (red circles) are compared to the CLAS Hall B phenomenological model~\cite{EG1b2}. The solid and hatched bands refer to the systematic uncertainty of the data and CLAS model parametrization respectively. The black crosses indicate earlier data from the EG1b~\cite{EG1b2} experiment. Reproduced from \cite{EG4_final}.}
\label{fig:eg4}
\end{figure}

Finally, the experiment which is the subject of this thesis, the E08-027 (or g2p) experiment ran in the spring of 2012 in Hall A at Jefferson Lab. The aim of this experiment was to fill out the experimental program discussed above with a very low $Q^2$ measurement of the $g_2$ structure function of the proton. This required transverse polarized target scattering data, which introduced a number of experimental challenges. We have recently finished analysis on this experiment and its first results were released in the fall of 2022~\cite{g2p_nature}.
\chapter{Experimental Setup}

\begin{figure}[htb]
\centering
\includegraphics[width=1.0\textwidth]{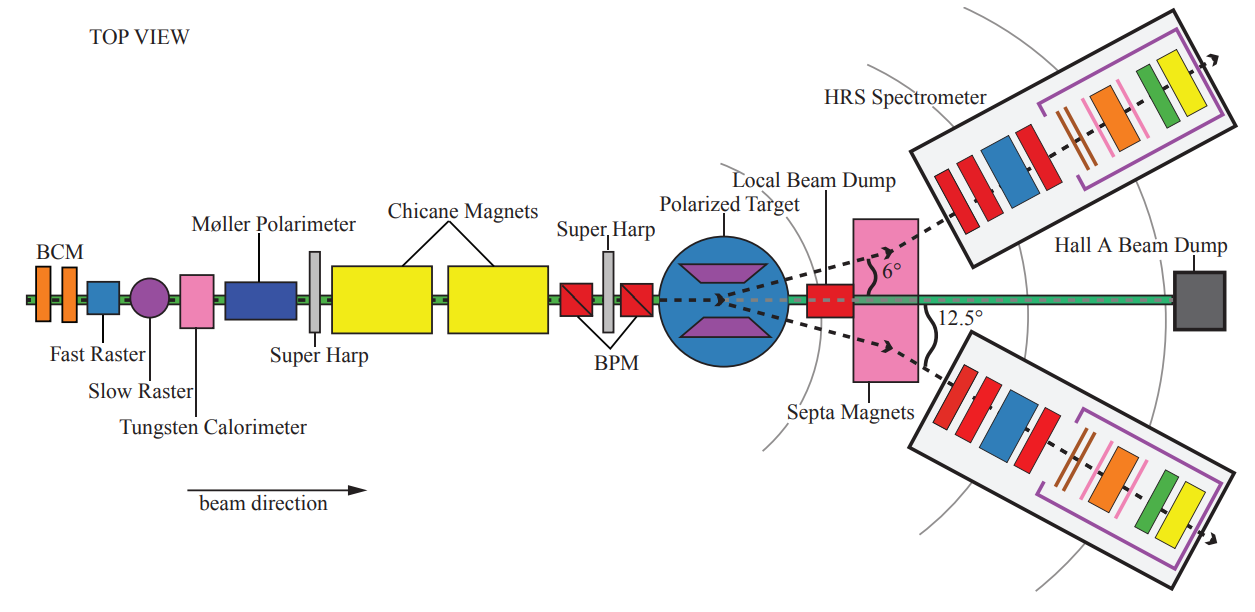}
\caption{The experimental setup of the g2p Experiment. Reproduced from Ryan Zielinski's thesis \cite{Zielinski:2017gwp}}
\label{fig:halla}
\end{figure}

The g2p experiment was an unprecedented effort in Hall A  at Jefferson Lab to measure the $g_2$ structure function of the proton at a very low $Q^2$ sufficient to test chiral perturbation theory predictions. The experiment used the inclusive electron scattering described in Chapter 2, with polarized electrons scattering off a spin-polarized target before being measured in the Hall A high resolution spectrometers (HRS). The experimental target was a cell of frozen ammonia (NH$_3$) beads polarized with the Dynamic Nuclear Polarization (DNP) process.

Data were taken over two target magnetic fields of 2.5T and 5T, and four beam energies ranging from 1.1 - 3.3 GeV. The target magnetic field was transverse to the electron beam for most settings, but longitudinal for one. This gave the experiment a total of six distinct kinematic settings, five transverse and one longitudinal. The lowest setting, at a field of 2.5T and a beam energy of 1.1 GeV, was taken primarily for the purpose of radiative corrections and is not otherwise used in the final results.

I was unfortunately not personally involved in the running of the experiment, but rather became involved in the intensive analysis process starting in 2015. However, I was given the opportunity to perform experimental work on a polarized target setup very similar to that discussed here, which is described in Chapter 5. Much of the information in this chapter comes originally from my collaborators who worked on the experiment directly in 2012, Ryan Zielinski~\cite{Zielinski:2017gwp}, Karl Slifer~\cite{g2pProposal}, Jian-Ping Chen~\cite{g2pProposal}, Toby Badman~\cite{TobyThesis}, and Chao Gu~\cite{Chao3}. The overall experimental setup is shown in Figure~\ref{fig:halla}, with each component discussed in the following sections.

\section{Electron Beam}

\begin{figure}[htb]
\centering
\includegraphics[width=0.8\textwidth]{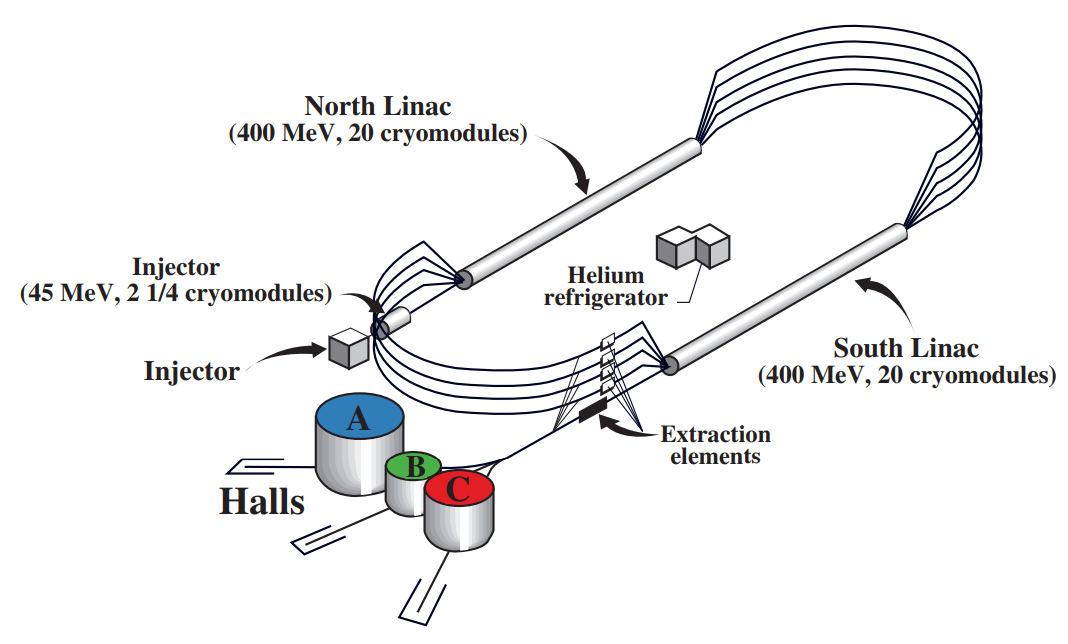}
\caption{The Jefferson Lab Electron Accelerator. Reproduced from \cite{CEBAF}}
\label{fig:jlab}
\end{figure}

At Jefferson Lab, experiments are performed with a high intensity electron beam linear accelerator, known as the Continuous Electron Beam Accelerator Facility (CEBAF). Currently the beam energy has a maximum of 12 GeV, but at the time that g2p ran, the maximum energy was 6 GeV. The beam current can be up to 200 $\mu$A, but was limited to 50 nA for this experiment to prevent depolarization of the target. 

\begin{figure}[htb]

\centering
    \begin{subfigure}[t]{0.45\textwidth}
        \centering
        \includegraphics[width=\linewidth]{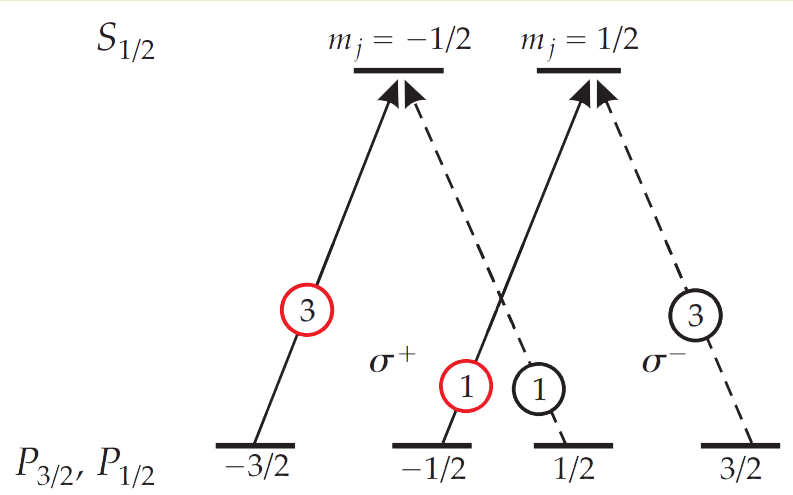} 
        \caption{Regular GaAs Cathode} \label{fig:esource1}
    \end{subfigure}
    \hspace{1.1em}
    \begin{subfigure}[t]{0.45\textwidth}
        \centering
        \includegraphics[width=\linewidth]{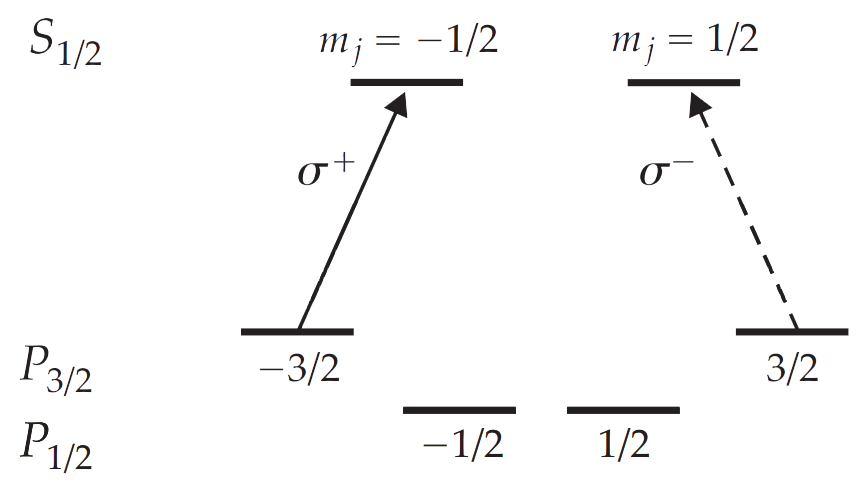} 
        \caption{Strained GaAsP Cathode} \label{fig:esource2}
    \end{subfigure}
\caption{The polarized electron generation technique. Reproduced from \cite{doped}.}
\end{figure}

Polarized electrons are initially generated by photoemission from a phosphorous-doped Gallium Arsenide cathode (GaAsP), which is stimulated with a circularly polarized laser at a frequency of 860 nm, the bandgap energy of the GaAs crystal~\cite{doped}. In pure, undoped Gallium Arsenide, this photoelectric stimulation introduces a degeneracy in the valence band states, exciting electrons out of the states $P_{1/2}$ and $P_{3/2}$. Three electrons are promoted to one spin direction from the $P_{3/2}$ state for every 1 promoted from $P_{1/2}$ to the other spin state, yielding a net 50\% polarization. The direction of this polarization is based on whether the polarized laser has left-handed helicity ($\sigma^-$) or right-handed helicity ($\sigma^+$). This effect is shown in Figure~\ref{fig:esource1}.

The introduction of the phosphorous doping causes a mechanical strain on the cathode, lifting the degeneracy and causing the bandgap from the $P_{1/2}$ state to be larger than that from the $P_{3/2}$ state. This means that light tuned to the bandgap of the latter state can theoretically promote 100\% of electrons to the desired spin state, again based on the helicity of the stimulating laser, as shown in Figure~\ref{fig:esource2}. In the Jefferson Lab source, a superlattice structure containing alternating layers of GaAs and GaAsP is used to create electrons with both a high polarization and high intensity. This routinely creates a polarization of around 85\%~\cite{doped}.

The direction of this polarization can be flipped from positive to negative then by changing the helicity of the stimulating laser. This is done with a voltage-controlled wave plate, when the sign of the voltage applied to this plate is flipped the helicity of a photon passing through the plate can also be flipped. This voltage is flipped in a pseudo-random fashion at a rate of 960.02 Hz, limiting time-dependent systematic uncertainties by frequently changing the direction of polarization. The beam helicity can also be flipped with an insertable half-wave plate, which was performed every few hours during the experiment to study and minimize helicity-dependent systematics~\cite{Zielinski:2017gwp}. 

The electrons generated by the cathode are pulled through an injector by an applied bias voltage, which gives the electrons an initial energy of 45 MeV~\cite{CEBAF}. These injected electrons then enter the paired linear accelerators shown in Figure ~\ref{fig:jlab}. Each linear accelerator consists of 20 cryomodules, units which consist of 8 superconducting niobium cavities each. These cavities are kept cooled to a temperature of 2K and powered to create an internal RF field, which accelerates the electrons as they pass through, with each cavity achieving an average gradient of 7 MeV/m. After their trip through the first linear accelerator, electrons achieve an energy of around 400 MeV, before entering a recirculating arc and being transferred to a second, identical linear accelerator. From here, the electrons can either be distributed to the experimental halls A, B, and C, or sent back to the first linear accelerator to achieve higher energies. By controlling the number of loops, it is possible to control the beam energy which reaches the hall~\cite{CEBAF}.

The beam energy is measured by observing the bend angle through the recirculation arc, in what is known as the ``Arc Method''~\cite{Chao3}. In principle, the momentum of the electrons traveling through a magnetic field is directly related to the bend angle of the electrons and the strength of the magnetic field:
\begin{equation}
\label{eqn:beamenergy}
E = k_c \frac{\int \vec{B} \times d\vec{l}}{\theta}
\end{equation}
Where $k_c$ = 0.299792 GeV$\cdot$rad$\cdot$T$\cdot$m corresponds to the speed of light, $\theta$ is the measured deflection as the electrons travel through the recirculation arc, and the integral runs over the measured magnetic fields of the dipoles in the arc~\cite{Chao3}. For the g2p experiment, a total of four different beam energies were used: 1.1 GeV, 1.7 GeV, 2.2 GeV, and 3.3 GeV. These differing beam energies combined with the direction and magnitude of the target field generate the different kinematic settings of the experiment.

\section{Beam Current Monitors (BCMs)}

Once the beam enters the hall, the first component in the experimental setup is the Beam Current Monitors (BCMs) used to measure the incident flux of electrons. The BCMs are located 23 m upstream of the target, and consist of a pair of stainless steel, cylindrical TM$_{010}$ mode waveguides tuned to the frequency of the beam, 1497 MHz~\cite{Zielinski:2017gwp,Chao3}. A receiver antenna inside each cavity is excited by the passing electrons, emitting a signal proportional to the beam current, or total flux of electrons flowing through the waveguide.

Traditionally this signal is processed with a RMS-to-DC converter, but the previously used receiver did not work with the low beam current necessary for the g2p experiment. Instead, a new receiver was designed by the Jefferson Lab instrumentation group specifically for the experiment~\cite{Chao3}. The new receiver employed an analog part, which used several analog amplifiers and filters to boost the strength of the signal, converting it from a 1497 MHz radiofrequency (RF) signal to a 45 MHz intermediate frequency (IF) signal. The system then uses an Analog-Digital-Converter (ADC) to digitize the signal and further boost the Signal-to-Noise ratio by using a digital filter tuned to cut off at 10.4 kHz. Once the signal has been fully processed, it is converted back to an analog 0-10V DC signal and sent to the Hall A Data Acquisition (DAQ) system~\cite{Zielinski:2017gwp}.

To calibrate the BCMs, a tungsten calorimeter was employed to measure the effective current based on heating effects of the beam~\cite{Bevins}. A tungsten slug is placed in the path of the beam, absorbing the full beam current and heating as a result. The number of electrons $N_e$ to hit the slug is obtained with:
\begin{equation}
\label{eqn:beamcurrent}
N_e = \frac{K_w \Delta T}{E}
\end{equation}
Where $K_w$ = 8555.5 J/K is the specific heat capacity of the tungsten, $\Delta T$ is the temperature change of the slug, and E is the beam energy measured with the arc method before entering the hall. The scaling factor on the signal conversion of the BCMs was fixed to normalize the BCM measured current to that obtained with the tungsten calorimeter~\cite{Zielinski:2017gwp,Chao3}.

By performing the current measurement with two independent cavities, it is possible to estimate the uncertainty of the BCM measurement by comparing the difference between the measurement of the upstream and downstream BCM. For a vast majority of the runs, the difference between the two BCMs was less than 0.7\%~\cite{Chao3}.

\section{Fast and Slow Raster}

After passing through the BCMs, the beam encounters a series of beam rasters. As was mentioned earlier, it was necessary to limit the beam current to 50 nA to prevent depolarization of the target, as the high intensity of of the Jefferson Lab beam would quickly destroy the spin polarization of the target and cause radiation damage to the material. On top of this current limiting, it is necessary to sweep the beam around the target so as to avoid concentrating it on a single spot for too long, ensuring the target can maintain its polarization and avoiding radiation effects.

\begin{figure}[htb]

\centering
    \begin{subfigure}[t]{0.45\textwidth}
        \centering
        \includegraphics[width=\linewidth]{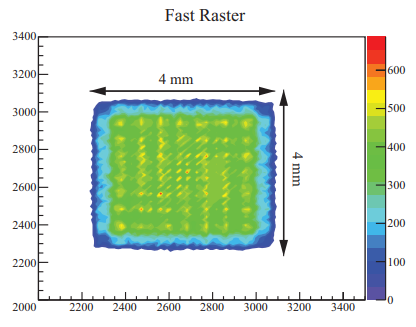} 
        \caption{Fast Raster} \label{fig:fastraster}
    \end{subfigure}
    \hspace{1.1em}
    \begin{subfigure}[t]{0.45\textwidth}
        \centering
        \includegraphics[width=\linewidth]{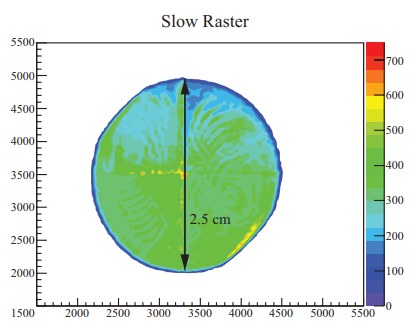} 
        \caption{Slow Raster} \label{fig:slowraster}
    \end{subfigure}
\caption{Raster swept beam profiles. Plots are a function of current in arbitrary units. Reproduced from \cite{Zielinski:2017gwp}}
\end{figure}

The fast raster increases the initial 200 $\mu$m beam to a 4mm $\times$ 4mm square profile. This is a standard piece of Hall A equipment, and employs two perpendicular dipole magnets to move the beam. Both dipoles are driven by the same 25 kHz triangle wave current, sweeping out the square area shown in Figure~\ref{fig:fastraster}.

The target cell for the g2p experiment however was a much larger cylinder with a diameter of 25 mm, requiring an additional raster to sweep out the full area of the target. A slow raster was added specially for the experiment, consisting of a second pair of dipoles and sweeping out the area shown in Figure~\ref{fig:slowraster}. However, to sweep out the circular profile of the target cell, a simple triangle wave could not be used as the driving current of the dipole magnets. Instead, the $\hat{x}$ and $\hat{y}$ dipoles were driven by a dual channel function generator, with respective forms of~\cite{Chao3}:
\begin{equation}
\label{eqn:xraster}
I_x(t) = A_x \sqrt{t} \sin (\omega t)
\end{equation}
\begin{equation}
\label{eqn:yraster}
I_y(t) = A_y \sqrt{t+t_0} \cos (\omega t)
\end{equation}
Here $A_x$ and $A_y$ are the amplitudes of the respective signals and t represents the time-based amplitude modulation. The frequency $\omega$ of the sin wave is equal to 99.412 Hz. This pattern allows the beam to sweep out a uniform 2.5 cm circular raster pattern to match the target cell.

\section{M\o{}ller Polarimeter}

The polarization of the electron beam was measured using a M\o{}ller Polarimeter, which employs the M\o{}ller process of electron-electron ($e^- + e^- \rightarrow e^- + e^-$) scattering to measure the beam polarization. The incident electrons scatter off polarized atomic electrons in a magnetized iron and copper foil, and are focused through three quadrupole magnets used to set the acceptance and one dipole magnet used to select for the electron momentum~\cite{Zielinski:2017gwp}. Scattered electrons are detected by a lead-glass calorimeter.

\begin{figure}[htb]
\centering
\includegraphics[width=0.8\textwidth]{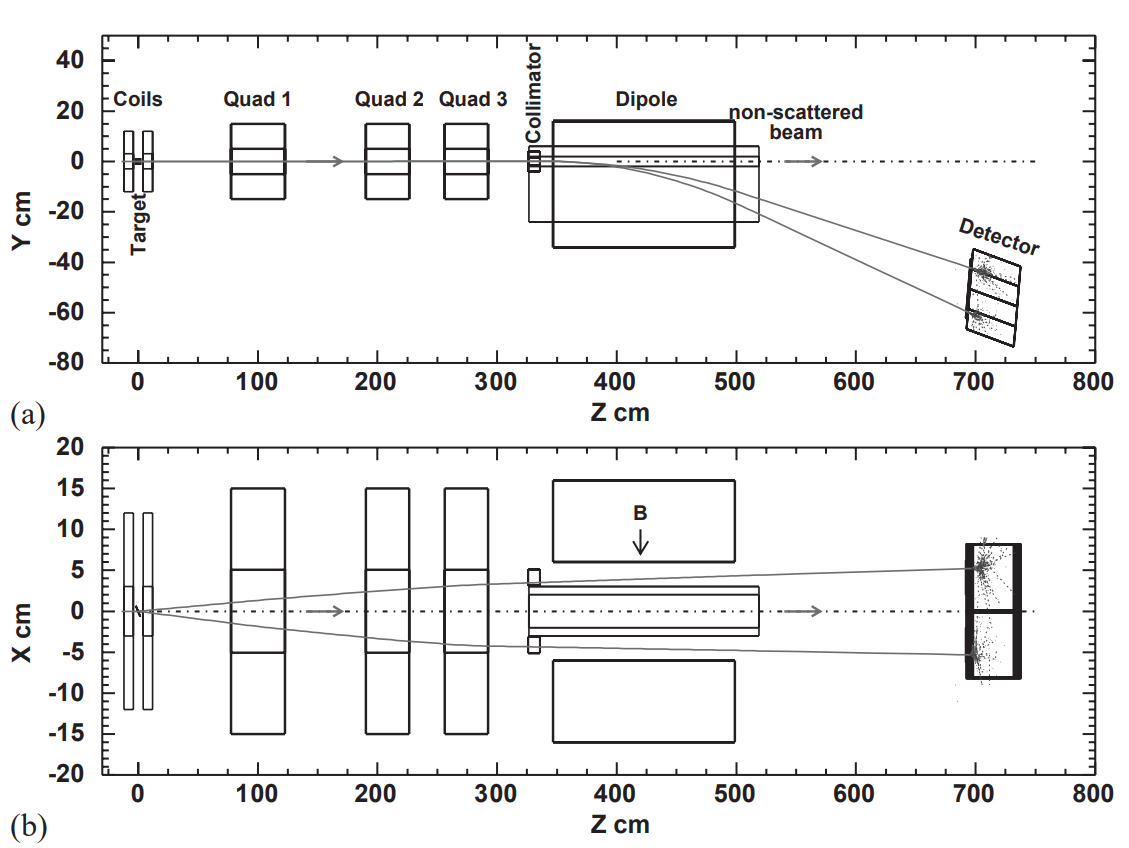}
\caption{M\o{}ller Polarimeter Schematic. (a) Side View. (b) Top View. Reproduced from \cite{CEBAF}}
\label{fig:mollerpol}
\end{figure}

We can write the M\o{}ller cross section as a function of the beam polarization $P_b$ and the target polarization $P_t$~\cite{CEBAF}:
\begin{equation}
\label{eqn:moller}
\sigma_{M\o{}ller} = \sigma_0 \bigg[\sum_{i,j = x,y,z}P^b_i A_{ij} P^t_j\bigg]
\end{equation}
Where $A_{ij}$ represents the asymmetry components and $\sigma_0$ is the unpolarized cross section. With the z-axis along the direction of the electron beam and the y-axis perpendicular to the scattering plane, we can write the longitudinal and transverse asymmetries as~\cite{CEBAF}:
\begin{equation}
\label{eqn:mollerazz}
A_{zz} = \frac{\sin^2\theta_{CM}\big(7 + \cos^2\theta_{CM}\big)}{\big(3+\cos^2\theta_{CM}\big)^2}
\end{equation}
\begin{equation}
\label{eqn:molleraxx}
A_{xx} = -A_{yy} = - \frac{\sin^4\theta_{CM}}{\big(3+\cos^2\theta_{CM}\big)^2}
\end{equation}
Here $\theta_{CM}$ is the center of mass scattering angle for the M\o{}ller interaction. $A_zz$ is our component of interest, as it tracks the longitudinal polarization. There is some transverse polarization, but it is small and can be corrected for by averaging one transverse direction with the other~\cite{Chao3}.

A schematic of the polarimeter is available in Figure~\ref{fig:mollerpol}. To measure the polarization, an asymmetry is measured at two center-of-mass angles of +20$^{\circ}$ and -20$^{\circ}$, the asymmetry between these two angles is averaged to cancel the transverse polarization effects. With the asymmetry A quantifying the difference between left ($N^+$) and right ($N^-$) handed helicity electrons counts, we can write the beam polarization as:
\begin{equation}
\label{eqn:beampol}
P_b = \frac{A}{P_t \cos\theta_t <A_{zz}>} = \frac{1}{P_t \cos\theta_t <A_{zz}>}\frac{N^+ - N^-}{N^+ + N^-}
\end{equation}
The average $A_{zz}$ is calculated with a Monte-Carlo simulation, and the target polarization $P_t$ and target scattering angle $\theta_t$ are extracted from the polarized target setup as described in the following sections. Nine measurements of the beam polarization were taken throughout the experiment, as is shown in Table~\ref{table:beampol}~\cite{Zielinski:2017gwp}:

\begin{table}[h!]
\centering
\begin{tabular}{|c|c|c|c|} 
 \hline
Date & $P_b$ (\%) & Stat. Error (\%) & Syst. Error (\%) \\ [0.5ex] 
 \hline
 
 03/03/2012 & 79.91 &  0.2 &  1.7 \\ 
 \hline
 03/30/2012 & 80.43 &  0.46 &  1.7 \\ 
  \hline
 03/30/2012 & 79.89 &  0.58 &  1.7 \\ 
  \hline
 04/10/2012 & 88.52 &  0.3 &  1.7 \\
  \hline
 04/23/2012 & 89.72 &  0.29 &  1.7 \\ 
  \hline
 05/04/2012 & 83.47 &  0.57 &  1.7 \\
  \hline
 05/04/2012 & 81.82 &  0.59 &  1.7 \\
  \hline
 05/04/2012 & 80.40 &  0.45 &  1.7 \\
  \hline
 05/15/2012 & 83.59 &  0.31 &  1.7 \\ 
 \hline
\end{tabular}
\caption{Electron Beam Polarization for the E08-027 Experiment}
\label{table:beampol}
\end{table}

\section{Chicane Magnets}
\begin{figure}[htb]
\centering
\includegraphics[width=0.8\textwidth]{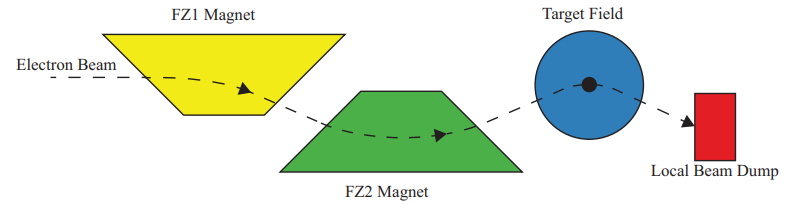}
\caption{Chicane magnet configuration for the g2p experiment. Reproduced from \cite{Zielinski:2017gwp}.}
\label{fig:chicanery}
\end{figure}

One of the most major experimental difficulties in the experiment was the beam deflection caused by the polarized target. To measure $g_2$, a transverse polarization is needed, and consequently, a transverse magnetic field which can cause a severe bending, especially at low beam energies. To compensate for this, a pair of chicane magnets were used ahead of the beam reaching the target, to pre-bend the beam and compensate for the deflection of the target field such that the beam strikes the center of the target as intended. The FZ1 dipole magnet, located 5.92 m before the target, bends the beam downwards out of its initial straight path~\cite{Chao3}. The FZ2 magnet then bends the beam back up 2.66 m before the target field~\cite{Zielinski:2017gwp}. The FZ2 magnet was mounted on a hydraulic support which could be used to move it up and down for optimization. The operation of the chicane magnets is shown in Figure~\ref{fig:chicanery}.

This configuration is only necessary for the transverse settings, for the longitudinally polarized setting the field is along the direction of the beam and subsequently does not deflect it significantly. The chicanes are operated in a 'straight-through' configuration for this setting~\cite{Zielinski:2017gwp}.

The chicane configurations for each transverse momentum setting are shown in Table~\ref{table:chicane}. $\theta_{in}$ and $\theta_{out}$ represent the angle of the beam when entering and leaving the target field, while $\theta_{target}$ is the angle of the beam when it strikes the target. For the 2.5T settings, the chicanes were calibrated to strike the target at an angle and leave along the normal beam path. For the 5T settings, the bending was significant enough that it was impossible to reach the main hall beam dump, so instead the configurations cause the beam to strike the target straight-on and then deflect into a local beam dump.

\begin{table}[h!]
\centering
\begin{tabular}{|c|c|c|c|c|} 
 \hline
E (GeV) & $B_t$ (T) & $\theta_{in}$ & $\theta_{target}$ & $\theta_{out}$ \\ [0.5ex] 
 \hline
 
 1.16 & 2.5 &  11.96$^\circ$ &  5.98$^\circ$ & 0.0$^\circ$ \\ 
 1.72 & 2.5 &  8.06$^\circ$ &  4.03$^\circ$ & 0.0$^\circ$ \\ 
 2.25 & 2.5 &  6.09$^\circ$ &  3.05$^\circ$ & 0.0$^\circ$ \\ 
 2.25 & 5.0 &  6.09$^\circ$ &  0.0$^\circ$ & -6.09$^\circ$ \\ 
 3.35 & 5.0 &  4.09$^\circ$ &  0.0$^\circ$ & -4.09$^\circ$ \\ 
 \hline
\end{tabular}
\caption{Chicane configurations for each transverse kinematic setting. Values reproduced from~\cite{Zielinski:2017gwp}.}
\label{table:chicane}
\end{table}

\section{Beam Position Monitors (BPMs)}

\begin{figure}[htb]
\centering
\includegraphics[width=0.75\textwidth]{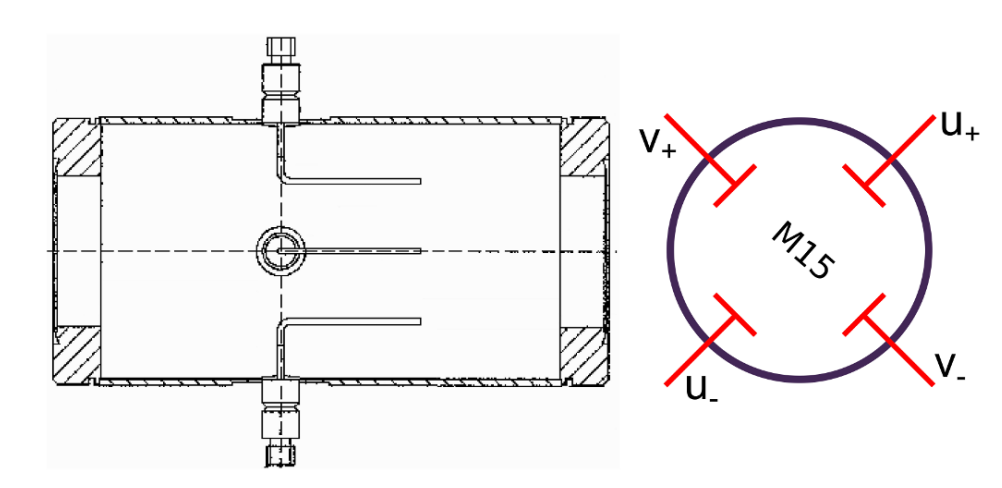}
\caption{Beam position monitor cavity from the side (left) and front (right). Reproduced from \cite{Chao3}.}
\label{fig:bpm}
\end{figure}

The Beam Position Monitors (BPM) are used to properly determine the beam's position and angle after the deflection from the chicane magnets. The beam passes through a cavity which contains four antennas similar to the one used in the BCM. These antennas are open-ended and lie coaxial to the beam, and are excited by the passage of the beam, stimulating a signal proportional to the distance of the beam from that particular antenna. By using these four antennas at right angles around the cavity, the beam angle and position can be measured, as shown in Figure~\ref{fig:bpm}.

\begin{figure}[htb]
\centering
\includegraphics[width=0.5\textwidth]{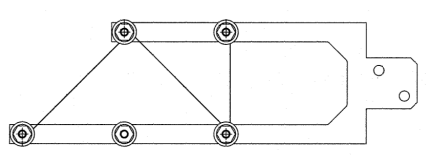}
\caption{Superharp wire scanner used to calibrate the BPMs. Reproduced from \cite{Chao3}.}
\label{fig:superharp}
\end{figure}

As with the BCM receiver, the low current necessary for E08-027 necessitated a new BPM cavity designed by the Jefferson Lab instrumentation group~\cite{Chao3}. These BPM cavities were calibrated with the use of a superharp, shown in Figure~\ref{fig:superharp}. One of these superharps was inserted between the two BPM cavities, while the other was placed directly before the chicane magnets. These superharps contain three 50 $\mu$m wires angled at -45$^\circ$, 0$^\circ$, and 45$^\circ$ respectively. When these wires are passed through the beam by a motor control on the harp, they produce a shower of radiation measured in nearby photomultiplier tubes (PMTs). This radiation can be used to calculate an absolute measurement of the beam position and angle, which is used to calibrate the BPM cavities.

\section{Polarized Target}

\begin{figure}[htb]
\centering
\includegraphics[width=0.6\textwidth]{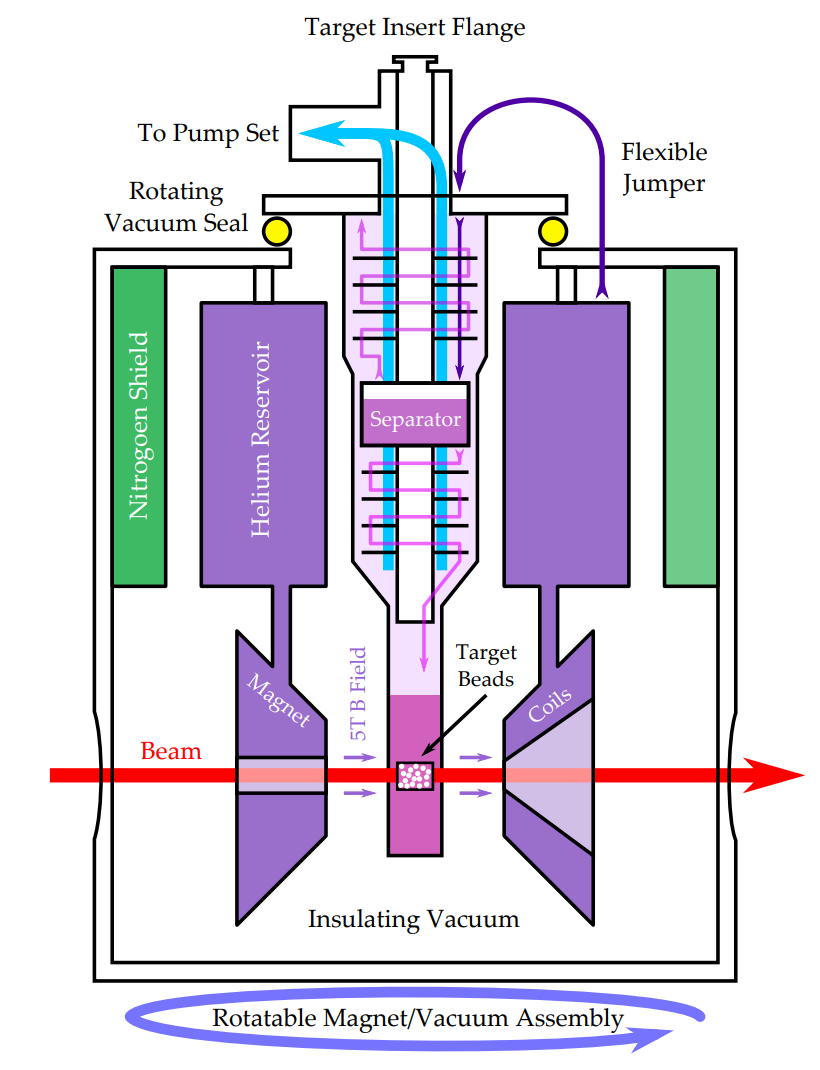}
\caption{Polarized target for the g2p experiment. Reproduced from  \cite{TargetPol}.}
\label{fig:poltarget}
\end{figure}

The polarized target is the central feature of the g2p experiment, being necessary to produce the structure functions introduced in Chapter 2. The target was a spin polarized solid ammonia target, doped with irradiation and polarization enhanced with the Dynamic Nuclear Polarization (DNP) process. The overall experimental setup is shown in Figure~\ref{fig:poltarget}.

The target itself consists of solid ammonia (NH$_3$) beads, contained within a cup made out of Kel-F plastic, which is used primarily for its absence of free protons so as not to contribute any background signal. Ammonia is an ideal target material in this case because it has a short time-constant, allowing maximum polarization in a relatively short amount of time, and because it is relatively durable against the depolarizing effect of the electron beam. The target ammonia is frozen initially using liquid nitrogen, (LN$_2$) and then pre-irradiated at a linear accelerator at the National Institute of Standards and Technology (NIST)~\cite{TargetPol}. This irradiation produces an excess of paramagnetic centers in the ammonia, which is necessary for the polarization enhancement process. The target is sealed off with thin aluminum end-caps to hold the material in place. 

Initial polarization is created by cooling the target to 4 K by submerging it in a bath of Liquid Helium (LHe), and applying a strong magnetic field with a superconducting magnet. The superconducting magnet is a split-pair Helmholtz configuration coil, to allow the beam to pass unimpeded through it, and is composed of niobium-titanium wire, which becomes superconducting below 10 K and is also cooled to 4 K with LHe. For this experiment, the magnet was run at strengths of 2.5 and 5.0 T, and was able to be swapped between transverse and longitudinal polarization by way of a rotating vacuum seal beneath the coils. 

Due to the strong magnetic field and cold temperature, magnetically interacting particles in the target align themselves with the direction of the magnetic field. The Zeeman Effect introduces a spin-dependent energy level splitting for these target particles, dividing the spin-$\frac{1}{2}$ protons and electrons in the target into spin up ($\uparrow$) and spin down ($\downarrow$) states, designating particles which are aligned or anti-aligned with the B-field. The vector polarization of these target particles is defined as the fraction of nuclei which are spin-polarized in a particular direction~\cite{CrabbMeyer}:
\begin{equation}
\label{eqn:vectorpol}
P_{\frac{1}{2}} = \frac{N^\uparrow - N^\downarrow}{N^\uparrow + N^\downarrow}
\end{equation}
Where $N^\uparrow$ and $N^\downarrow$ are the number of nuclei aligned or anti-aligned with the magnetic field. The polarization induced by the magnetic field alone can be determined at thermal equilibrium through the use of Boltzmann statistics to be:
\begin{equation}
\label{eqn:tepol}
P^{TE}_{\frac{1}{2}} = \tanh{\bigg(\frac{\mu B}{k_B T}\bigg)}
\end{equation}
Where $\mu$ is the magnetic moment of the relevant particle, $k_B$ is the Boltzmann constant, B is the magnetic field, and T is the temperature. The electron has a small mass and consequently, a large magnetic moment of 9.3e-24 J/T, so at 5 T and 4 K, we get a very large electron spin polarization exceeding 99\%. The proton, with its much smaller magnetic moment of 1.4e-26 J/T, does not polarize nearly so easily, so we only get a proton polarization of around 0.5\% in this fashion.

\begin{figure}[htb]
\centering
\includegraphics[width=0.8\textwidth]{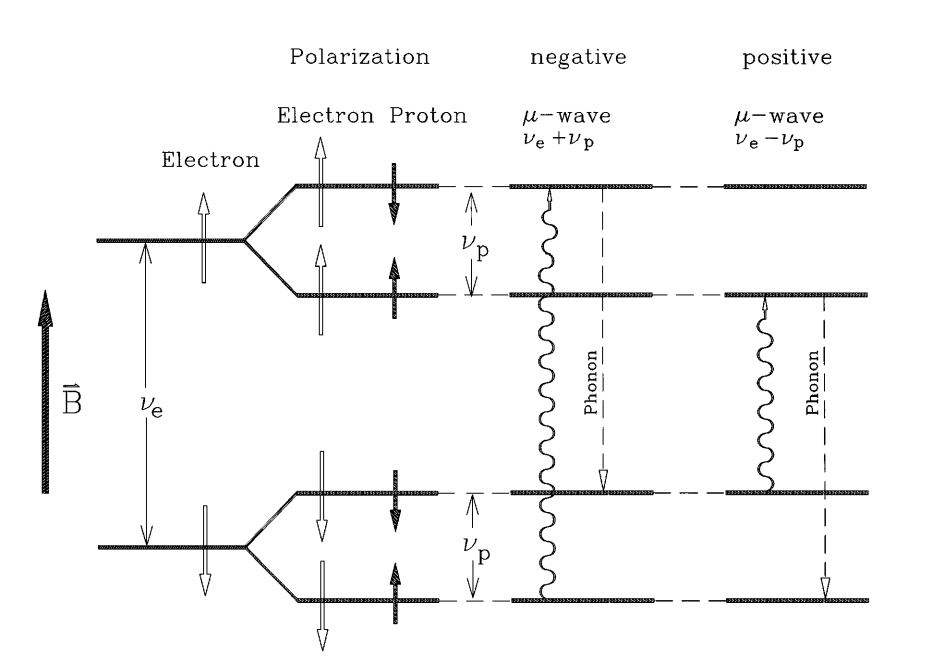}
\caption{Energy levels of the dynamic nuclear polarization process. Reproduced from  \cite{CrabbMeyer}.}
\label{fig:dnplevels}
\end{figure}

The DNP process takes advantage of the high electron polarization induced in this process. Internal to the target, unpaired electrons and protons interact, creating a hyperfine splitting into four distinct energy levels comprising all possible combinations of proton and electron electron spins, as shown in Figure~\ref{fig:dnplevels}. The irradiation doping process described above ensures there is a large number of unpaired electrons to interact in this fashion in the target.

To enhance the proton polarization, microwaves are applied tuned to the energy level gaps shown in the figure. This causes the electrons and protons to couple, such that a polarizing electron will also cause a coupled proton to polarize. Since the electron has a much smaller relaxation time, a single electron can repeatedly relax and couple with new protons before the relaxation time of a single proton elapses. This allows the proton polarization to be enhanced to very high values, upwards of 70\% or more at 5 T.

The polarization is also enhanced from thermal equilibrium with the use of a LHe evaporation refrigerator, shown in Figure~\ref{fig:poltarget}. By pumping on the bath of liquid helium immersing the target with high flow roots pumps, the temperature can be lowered from the base of 4 K, while fresh liquid helium is fed to the target by way of a separator, which allows liquid helium to drip through a fine mesh while hot gaseous helium is pumped out~\cite{CrabbMeyer,TargetPol}.  The helium which drips through then must pass through a heat exchanger before reaching the target, which serves to create a temperature gradient such that the coldest helium submerges the target cell without allowing warm 4 K helium to reach the bottom of the refrigerator. This process serves to cool the target to a temperature of 1 K, enhancing the polarization greatly.

The polarization is measured by way of a Nuclear Magnetic Resonance (NMR) system. In this system, polarizing protons either emit or absorb a photon at the Larmor frequency of the proton, depending on the direction of polarization. This can be observed with a Q-Meter, a high precision RLC (Resistance-Inductance-Capacitance) circuit, which has its inductance coupled to the photons emitted by the target through a coil inductor placed inside the cup surrounding the target material. This coupling can be seen as a signal in the real part of the circuit's impedance, which is centered at the 2.5/5T proton Larmor frequency, and the area of which is directly proportional to the proton polarization of the target~\cite{CrabbMeyer}.

\begin{figure}[htb]

\centering
    \begin{subfigure}[t]{0.45\textwidth}
        \centering
        \includegraphics[width=\linewidth]{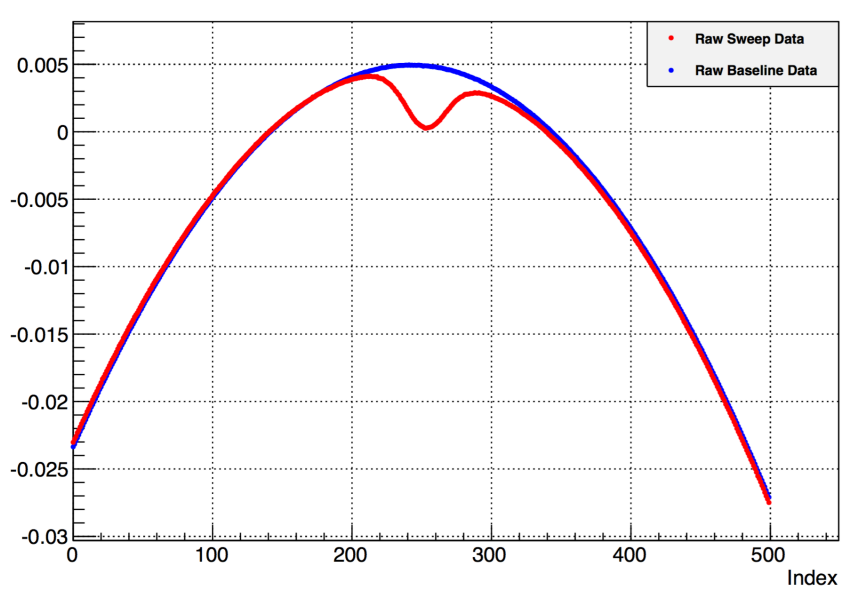} 
        \caption{Raw NMR Signal on Real Q-Meter Impedance} \label{fig:nmr1}
    \end{subfigure}
    \hspace{1.1em}
    \begin{subfigure}[t]{0.45\textwidth}
        \centering
        \includegraphics[width=\linewidth]{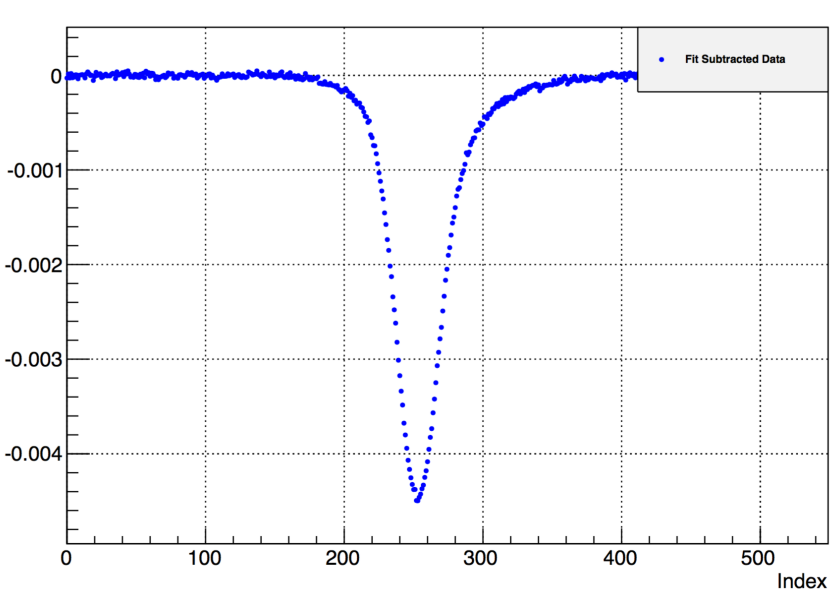} 
        \caption{Background-subtracted NMR Signal} \label{fig:nmr2}
    \end{subfigure}
\caption{NMR Signal examples from the g2p experiment. Reproduced from \cite{Chao3}.}
\end{figure}

To extract the signal, an example of which is shown in Figure~\ref{fig:nmr1}, a baseline of the circuit background near the correct frequency is taken at a lower magnetic field and subtracted from the final result. Due to temperature shifts in the target and other effects, this baseline is rarely perfect, so it is necessary to also fit a cubic polynomial to the wings of the signal, and subtract it to obtain a raw signal as is shown in Figure~\ref{fig:nmr2}. Since the polarization is only known explicitly at thermal equilibrium, we calibrate our signal with a long measurement of the thermal equilibrium polarization in equation~\ref{eqn:tepol}. We can then use the area of the thermal equilibrium signal to find the enhanced polarization based on the measured area A:
\begin{equation}
\label{eqn:tepol}
P = \frac{P_{TE}}{A_{TE}}A
\end{equation}
This method works for both positive and negative spin polarization. Both directions of polarization are employed in the experiment to try and remove systematic effects dependent on the direction. Further discussion of the underlying principles of NMR can be found in Chapter 5, where the UNH Polarized Target is discussed. 

\begin{figure}[htb]
\centering
\includegraphics[width=0.9\textwidth]{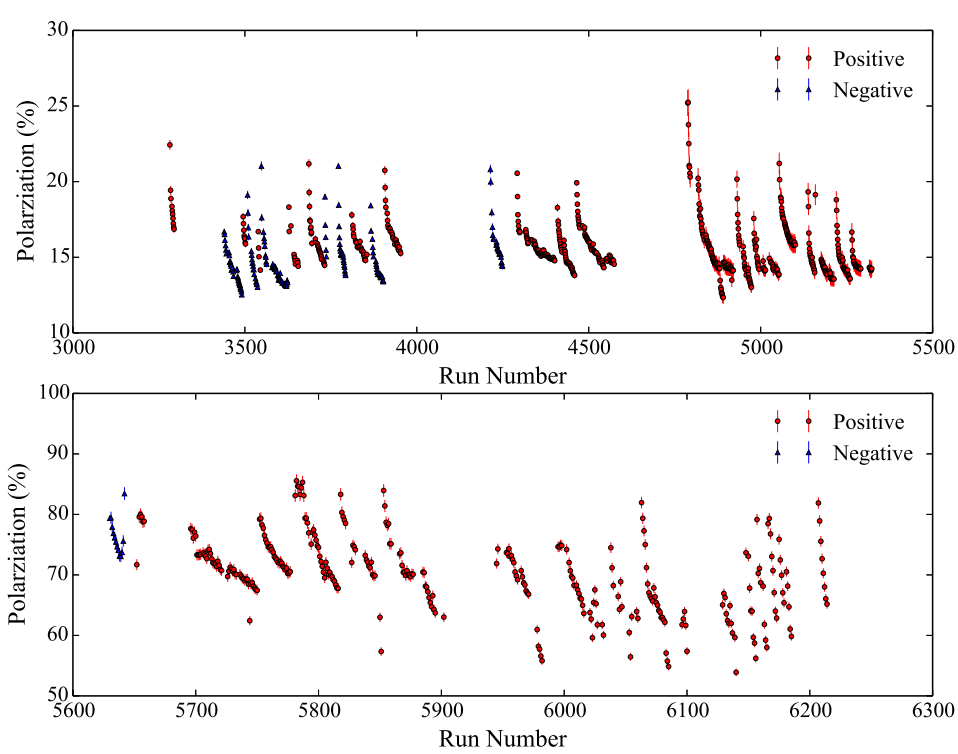}
\caption{Target polarizations for the 2.5 T (top) and 5 T (bottom) settings. Reproduced from  \cite{TECHNOTE:tpol}.}
\label{fig:targetpol}
\end{figure}

Measured polarizations for the experiment are shown in Figure~\ref{fig:targetpol}. Polarizations for the 5 T settings averaged between 60\% and 80\%, acceptable values but not always reaching the full potential of the material as measured by other groups~\cite{CrabbMeyer}. For the 2.5 T settings, the target polarization was between 40\% and 50\%, but a year after the experiment's run, it was discovered that the T.E. calibration constant used was still set to 5 T for all settings, meaning the 2.5 T settings were less by a factor of 2. This results in a larger statistical error on the results of the 2.5 T kinematic settings~\cite{TECHNOTE:tpol}.

\begin{figure}[htb]
\centering
\includegraphics[width=0.8\textwidth]{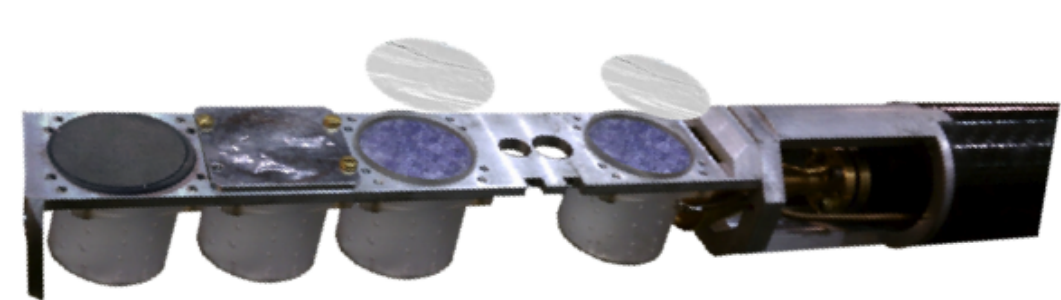}
\caption{Target insert for E08-027 showing Carbon (First), Dummy (Second), and Ammonia (Third and Fourth) targets. Reproduced from \cite{Zielinski:2017gwp}.}
\label{fig:targetstick}
\end{figure}

The target cup must be held in the region of the superconducting magnet with the highest field uniformity, the more uniform the magnetic field, the sharper the measured NMR signal will be. The target ladder which holds the Kel-F cup is made of aluminum and suspended from a carbon fiber target insert which can be removed to replace the target material or moved up and down to change to another target cup. For the purposes of the dilution analysis, several non-ammonia targets were featured on the target ladder as shown in Figure~\ref{fig:targetstick} ~\cite{Zielinski:2017gwp}. The right two targets are filled with frozen ammonia, the clear beads of which are turned a deep purple by the irradiation process at NIST. The second target from the left is the `Dummy' target, left empty to measure the yield which results from the aluminum end-caps and an NMR coil inside. The leftmost target is a carbon disk, which is used to estimate indirectly the nitrogen contribution to the scattering yields, as discussed in Chapter 6. The full target cup is a cylinder with diameter of 2.72 cm and length of 2.83 cm, though a shorter cup of length 1.3 cm was used for the 1.2 GeV beam energy runs to lessen radiation lengths~\cite{Zielinski:2017gwp}. An additional `Empty' target was employed for the dilution analysis of Chapter 6, for which a production cup was simply left empty and filled with LHe to measure the helium contribution.

\section{Beam Dumps}

\begin{figure}[htb]
\centering
\includegraphics[width=0.8\textwidth]{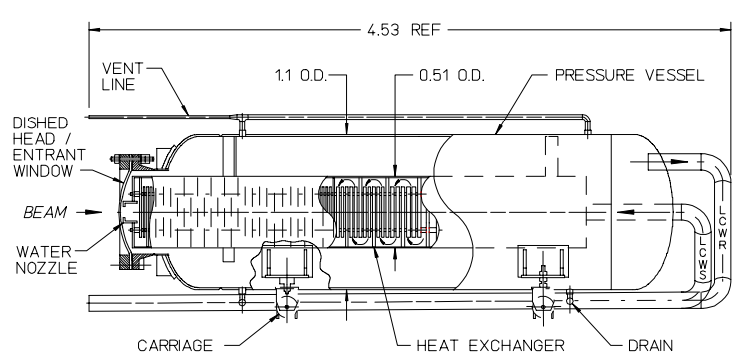}
\caption{Hall A main beam dump schematic. Reproduced from \cite{TECHNOTE:beamdump}.}
\label{fig:beamdump}
\end{figure}

Though many electrons in the beam will interact within the target and scatter into the detectors, a large portion of the beam intensity will pass through the target without significant interaction and must reach a final destination which can safely dissipate the beam energy. For standard experiments in Hall A, a hall beam dump is located at the end of the hall. This beam dump is composed of an aluminum plate heat exchanger inside an aluminum pressure vessel. The heat exchanger is cooled by a water recirculator, which feeds a bath of water that helps to dissipate beam energy and heat~\cite{TECHNOTE:beamdump}. A cross sectional schematic of the Hall A beam dump is shown in Figure~\ref{fig:beamdump}. The beam dump is optimized such that 70\% of the power is dissipated directly in the aluminum, minimizing the production of tritium and other elements which might serve to activate the cooling water.

\begin{figure}[htb]
\centering
\includegraphics[width=0.4\textwidth]{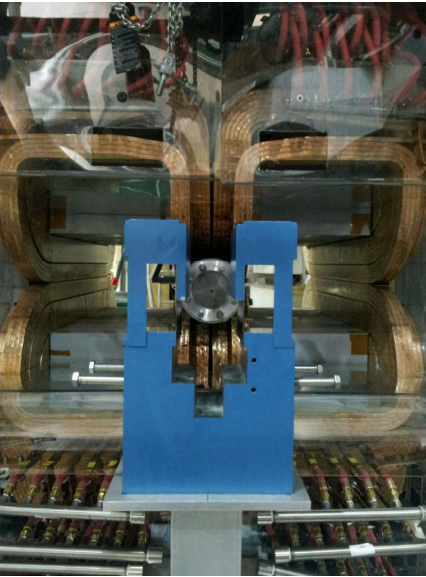}
\caption{Local beam dump for E08-027. The septa magnets are visible in the background. Holes on the left and right of the blue beam dump support structure allow scattered electrons to pass through to the detectors. In the center, copper and tungsten plates absorb the beam energy. In this photo, the insert is not inserted into the beam dump. Reproduced from \cite{Zielinski:2017gwp}.}
\label{fig:beamdump2}
\end{figure}

For the 2.5 T settings, the chicane magnets were optimized such that the beam exits the target field at a 0$^\circ$ angle, to reach the beam dump. At the 5 T settings, the high field combined with the low beam energy made it impossible to configure the chicanes to both hit the target in the center and reach the main hall beam dump, so a local beam dump had to be commissioned to dissipate the beam energy for the 5 T transverse settings. A photo of the local beam dump is shown in Figure~\ref{fig:beamdump2}. 

The local beam dump is composed of a series of copper and tungsten plates, sufficient to diffuse the beam energy for this experiment due to the low beam current. It contained a removable insert which could be used to control radiation levels in the hall, and was located 0.64 m downstream of the polarized target~\cite{Chao3}.

\section{Septa Magnets}

\begin{figure}[htb]
\centering
\includegraphics[width=0.4\textwidth]{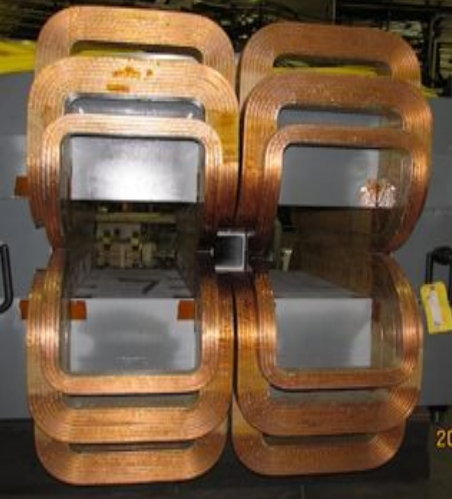}
\caption{Septa magnets for E08-027. Reproduced from \cite{Zielinski:2017gwp}.}
\label{fig:septa}
\end{figure}

To obtain the desired low $Q^2$ of the experiment, it is necessary to measure scattered electrons at an angle of around 6$^\circ$. However, due to space constraints in the hall, the Hall A spectrometers could not be placed closer than 12.5$^\circ$ from the center. It was therefore necessary to employ a pair of dipole septa magnets after the polarized target, to bend scattered electrons at 6$^\circ$ into the angular acceptance of the Hall A detectors. These magnets were composed of a pair of coils above and below the beamline designed to bend it the requisite amount, as shown in Figure~\ref{fig:septa}.

The top set of coils for the right spectrometer septa became damaged during the operation of the experiment several times, requiring multiple reconfigurations of the three top coils for the RHRS. This required changing the number of loops of wire in the coils. The ideal configuration is 48-48-16, but on 3/18/2012 and 4/11/2012, the top right coils became damaged and were instead changed to 40-32-16 and then 40-00-16 loops for the latter sections of the run~\cite{Zielinski:2017gwp}.

The septa magnets were employed for all settings with a transverse target field, but for the longitudinal setting, counts were instead measured at an acceptance of 11$^\circ$, for which the septa were not used.
\section{Spectrometers}

\begin{figure}[htb]
\centering
\includegraphics[width=0.6\textwidth]{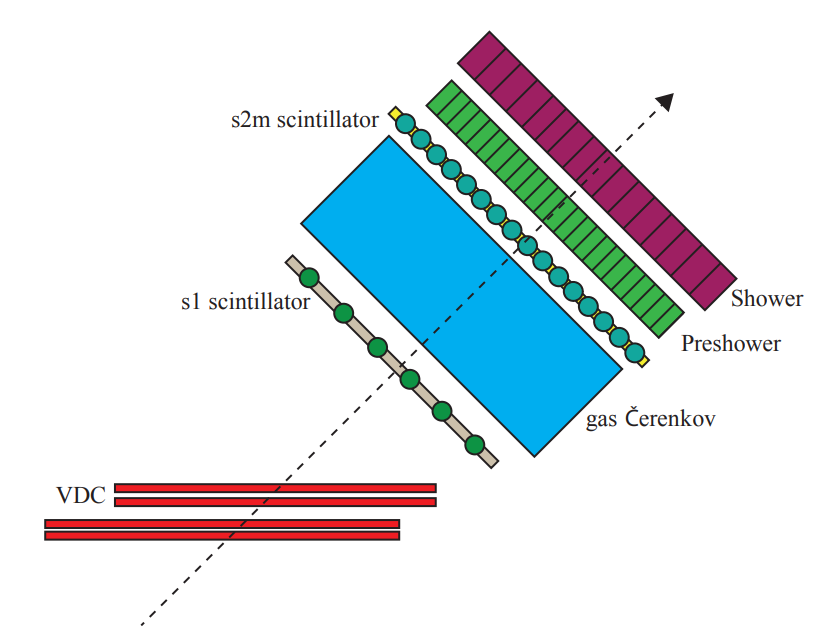}
\caption{Detector stack for the Hall A High-Resolution Spectrometers. Reproduced from \cite{Zielinski:2017gwp}.}
\label{fig:detstack}
\end{figure}

Electrons and other particles which scatter from the target at the appropriate angle are directed into the Hall A high resolution spectrometers, used to identify the measured particles and track their momentum and position. There are two identical detectors, referred to as the Left (LHRS) and Right (RHRS) arms. These spectrometers consist of three superconducting quadrupoles, which focus the scattered electrons, and one superconducting dipole, which is used to select the electron momentum which reaches the detector stack. After passing through the focusing magnets the scattered particles reach the detector stack shown in Figure~\ref{fig:detstack}.

Particles which enter the HRS first pass through two vertical-drift wire-chambers (VDCs). These chambers consist of cells filled with a gas mixture of argon and ethane, as well as an array of wires used to sense the approach of nearby electrons~\cite{Zielinski:2017gwp}. The cell is operated under an electric field produced by a gold-plated Mylar foil connected to a voltage differential of -4 kV. There are two VDCs in each spectrometer, each composed of two wire planes in the UV configuration. Each plane consists of 368 sense wires spaced 4.2 mm from each other~\cite{Chao3}.

When an electron passes through the VDC cell, it ionizes the gas around it, creating a shower of radiation. The argon is used as a quenching agent to suppress secondary emissions and limit the produced radiation to that which comes from scattered electrons. The produced radiation, stimulated by the electric field, couples with the nearest sense wire, creating a signal that can be measured by the Data-Acquisition system (DAQ). A single scattered electron triggers signals on multiple sense wires as it passes, with the shortest drift time associated showing what wire corresponds to the electron's closest approach. From the measured drift distances and the known position of the wires, the electron's trajectory can be reconstructed back to the target~\cite{Zielinski:2017gwp,Charpak}. This, in combination with the HRS dipole, allows the spectrometers to measure and select the particle's momentum and exact scattering angle.

\begin{figure}[htb]
\centering
\includegraphics[width=0.8\textwidth]{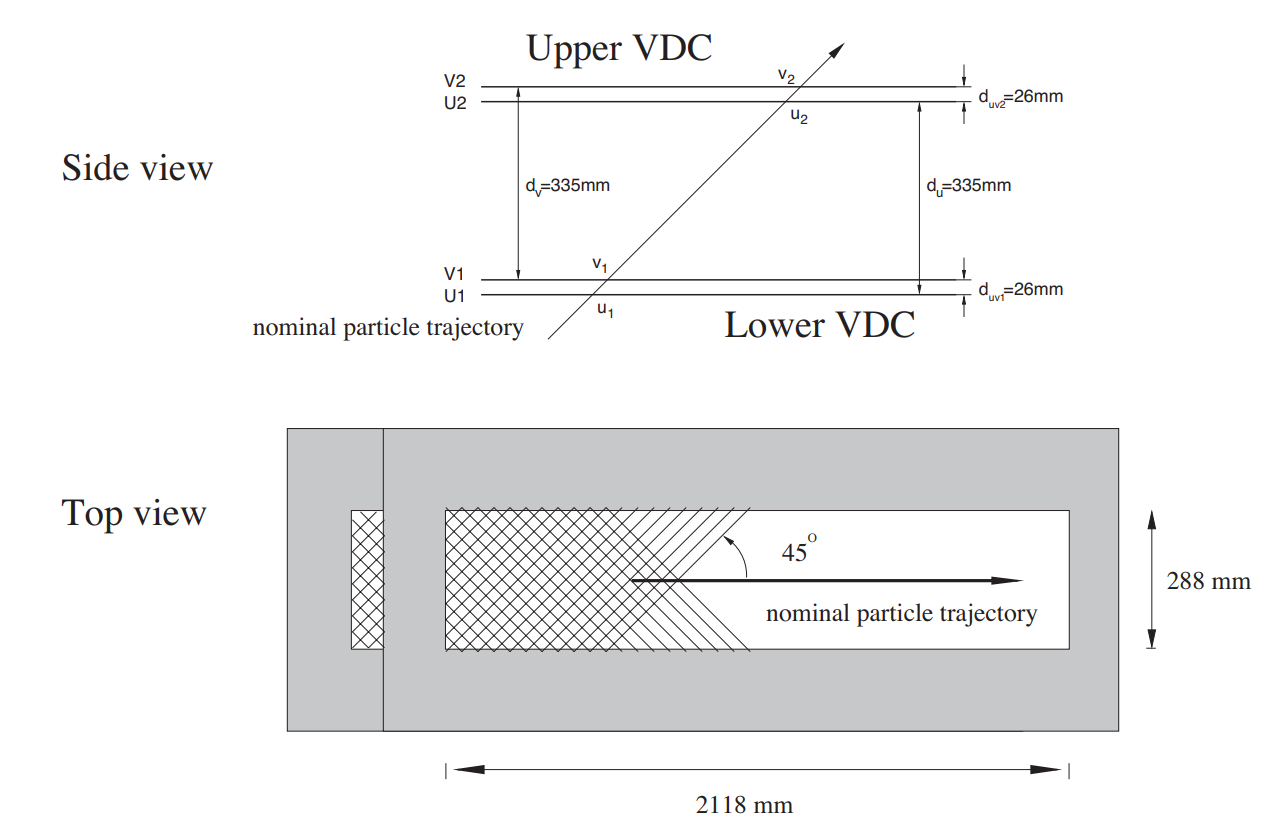}
\caption{VDC Diagram. Reproduced from \cite{CEBAF}.}
\label{fig:vdc}
\end{figure}

A diagram of the VDCs is shown in Figure~\ref{fig:vdc}. The upper and lower chambers for each VDC have the sensing wires oriented 90$^\circ$ from each other. The U and V planes are oriented at a 45$^\circ$ angle from the direction of the travelling electrons. 

After passing through the VDCs, scattered particles pass through the first of two plastic scintillators. These plastic scintillators function to trigger photomultiplier tubes (PMTs) on each side. These photomultiplier tubes are calibrated with cosmic rays and used to form the Data Acquisition trigger. When both the left and right PMTs are fired for each scintillator, the DAQ trigger registers a scattering event. 

Between the two plastic scintillators is a gas \v{C}erenkov detector, used for particle identification. Because we are only interested in the scattered electron, we will need a way to eliminate events which come from other negatively charged particles that may result from scattering, such as $\pi^-$. The detector is filled with CO$_2$ gas, and operates on the measurement of \v{C}erenkov radiation, which is produced when a particle travels faster than the phase velocity of light in a given medium~\cite{GASC}. \v{C}erenkov radiation can only be produced by a particle with a minimum momentum of~\cite{GASC}:
\begin{equation}
\label{eqn:gascerenkov}
p = \frac{m c}{\sqrt{n^2 - 1}}
\end{equation}

Where p is the particle's momentum, m is its rest mass, c is the speed of light, and n is the index of refraction for the medium, in this case n = 1.00041 for CO$_2$. Because electrons are very light, they have a low momentum threshold of 0.017 GeV. Pions are much heavier and have a momentum threshold of 4.8 GeV. Because of the momentum selected by the dipole magnet in the HRS, any pions or heavier particles which reach the \v{C}erenkov detector will have insufficient momentum to produce this radiation, so a signal will only be produced by a travelling electron.

\begin{figure}[htb]
\centering
\includegraphics[width=0.65\textwidth]{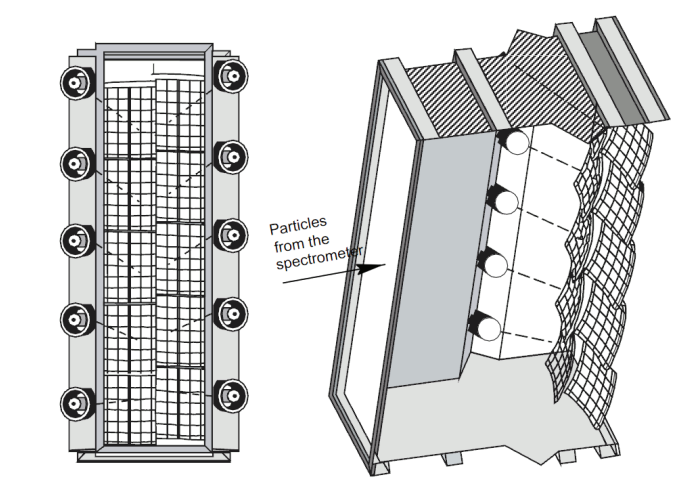}
\caption{Gas Cerenkov diagram. Reproduced from \cite{GASC,Zielinski:2017gwp}.}
\label{fig:gasc}
\end{figure}

The produced radiation is measured by way of ten overlapping mirrors at the end of the detector, which focus produced radiation into a corresponding series of photomultiplier tubes. The signal from each PMT is summed to get the total radiation measured in the detector. A diagram of the \v{C}erenkov detector is shown in Figure~\ref{fig:gasc}. Though heavier particles do not produce \v{C}erenkov radiation directly, they may produce some electron radiation due to ionization of the CO$_2$ gas, which may be mistaken for an electron scattering event.

To isolate and remove these events, a lead-glass calorimeter is used after the second plastic scintillator. This calorimeter is a series of lead glass blocks at the end of the detector stack which are stimulated by energetic particles to produce e$^-$-e$^+$ pairs and bremsstrahlung radiation photons. The radiation produced is measured in a series of photomultiplier tubes, and is directly proportional to the energy deposited by the particle, which can be used in combination with the \v{C}erenkov to definitively rule out heavier negatively charged particles~\cite{Chao3}.

\begin{figure}[htb]
\centering
\includegraphics[width=0.65\textwidth]{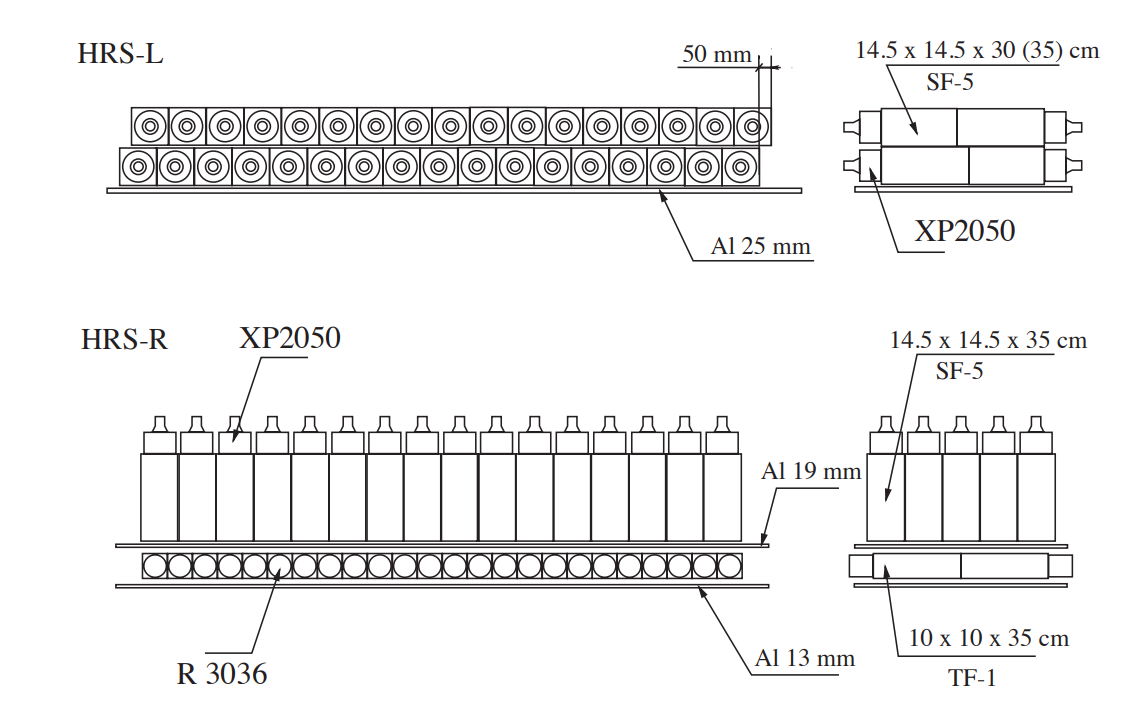}
\caption{Lead calorimeter for HRS detectors. Reproduced from \cite{CEBAF}.}
\label{fig:calorimeter}
\end{figure}

The number of lead-glass blocks differs slightly between the left and right HRS. For the left HRS, the calorimeter is composed of two layers each with 34 blocks perpendicular to the incoming particle trajectory, each with a dimension of 14.5$\times$14.5$\times$30 cm for the first layer and  14.5$\times$14.5$\times$35 cm for the second. The right HRS instead has on the first layer 48 10$\times$10$\times$35 cm perpendicular blocks, and on the second, 80 14.5$\times$14.5$\times$35 cm blocks oriented parallel to the incoming particle. The increased number of blocks on the right HRS means that it completely absorbs the energy of the incoming particle, while this is not true of the left HRS. The calorimeter setup is shown in Figure~\ref{fig:calorimeter}.

\section{Data Acquisition System}

\begin{figure}[htb]
\centering
\includegraphics[width=0.9\textwidth]{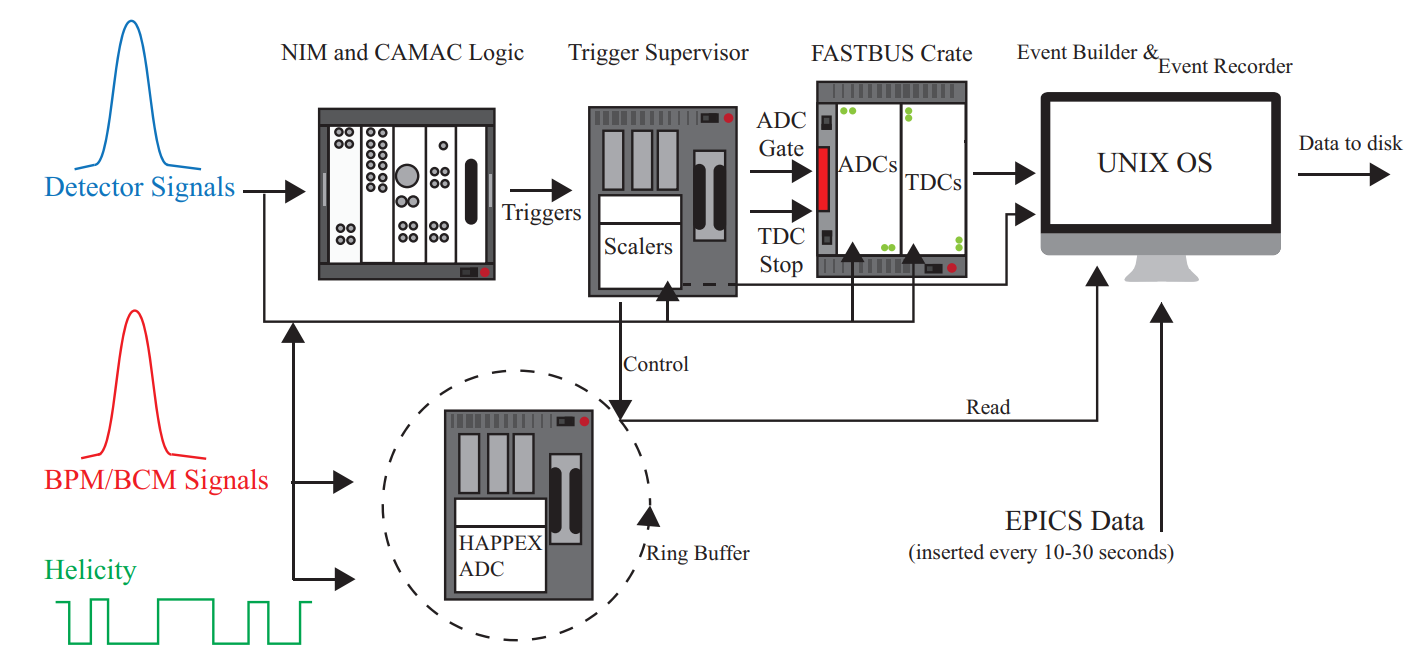}
\caption{Overall Data Acquisition setup for E08-027. Reproduced from \cite{Zielinski:2017gwp}.}
\label{fig:g2pdaq}
\end{figure}

Scattering events which activate the trigger by tripping a PMT in both s1 and s2m scintillators and which fall within the particle identification and momentum acceptance of the detector are recorded in the data acquisition system. The overall DAQ system is shown in Figure~\ref{fig:g2pdaq}. 

Signals from the detectors are directly received by trigger modules formed by several NIM and CAMAC electronics crates~\cite{Zielinski:2017gwp}. These determine the primary trigger, which is whether both scintillators were tripped as discussed above, as well as a secondary trigger used to measure the trigger efficiency, which trips when only one scintillator plane measures an event instead of both, but the event also registered in the \v{C}erenkov detector. These triggers are sent to the trigger supervisor, which determines whether or not to accept an event based on if the upstream data acquisition (DAQ) system is busy processing a previous event, and the prescale factor, chosen to reduce the missed events due to busy time as much as possible. The trigger supervisor maintains the maximum efficiency by keeping the DAQ busy for the maximum time possible.

Accepted events are passed to a FASTBUS crate containing analog-to-digital (ADC) and time-to-digital (TDC) converter modules. These modules serve to digitize the signals and pass them to the CEBAF Online Data Acquisition (CODA) software on a connected Unix computer.

In parallel, signals from the BCMs and BPMs, as well as the beam helicity measured at the source are recorded in ADC modules re-used from the HAPPEX experiment, before being paired with matching events based on time synchronization in the CODA system.

The CODA files are decoded by the Hall A Analyzer software, which is a set of C++ libraries which interface with CERN's ROOT particle physics libraries. These libraries help to decode the files into raw counts information. This allows the reconstruction of spectra for scattering counts versus scattering energy, momentum, and other kinematic variables, which can ultimately be used to form the physical quantities of interest.

\chapter{UNH Polarized Target}

Before delving into the details of the data analysis and results, let us take a brief diversion from discussion of the E08-027 experiment to focus on my own hands-on experimental experience. Though the timeline didn't work out such that I could participate in the experimental run described in the previous chapter, I was able to spend several years working in the UNH Polarized Target lab, featuring an experimental setup very similar to the polarized target described above.

Like the g2p target, the UNH polarized target lab focuses on the DNP process for polarizing protons, as well as deuterons. When I joined the lab, no polarization signal had successfully been created and observed for either target particle, but as of 2022, both have been measured and the lab is focusing on new ways to optimize the polarized target process. This means I was able to participate in and contribute to a number of major milestones over my time working in the UNH lab.

The UNH setup features a similar target insert, LHe evaporation fridge, and superconducting magnet to the g2p target. However, though the g2p target focused on vector polarization for spin-$\frac{1}{2}$ particles defined in (\ref{eqn:vectorpol}), one of the primary goals of the UNH polarized target lab is to measure and improve the polarization of spin-1 particles such as the deuteron, known as tensor polarization. This is valuable for studying a number of spin-dependent observables, but one of the major goals of the UNH lab is preparing for an upcoming experiment measuring a different structure function than E08-027. This experiment's primary goals include a measurement of the $b_1$ structure function of the deuteron, a structure function only present for spin-1 particles which may be instrumental in revealing hidden color effects~\cite{hiddencolor}.

\begin{figure}[htb]
\centering
\includegraphics[width=0.5\textwidth]{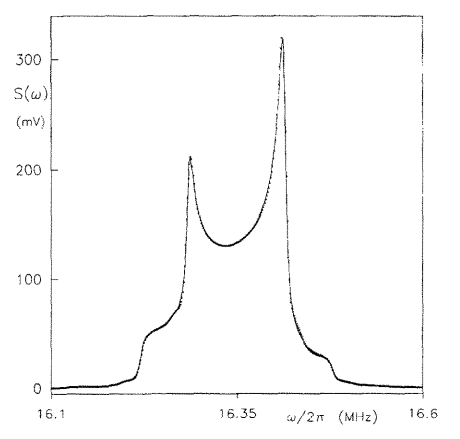}
\caption{Example of a deuteron signal and high-accuracy fit. Reproduced from \cite{dulya}.}
\label{fig:deuteron}
\end{figure}

Deuteron polarization measurements are complicated, because the deuteron signal is at least two orders of magnitude smaller than the proton, and has a more complex structure composed of two peaks in frequency space which become separated in materials where the deuteron forms a quadrupole. An example of this structure is shown in Figure~\ref{fig:deuteron}. To study these signals has required a number of changes and advancements from the experimental setup of E08-027.

\section{Superconducting Magnet}
\begin{figure}[htb]
\centering
\includegraphics[width=0.7\textwidth]{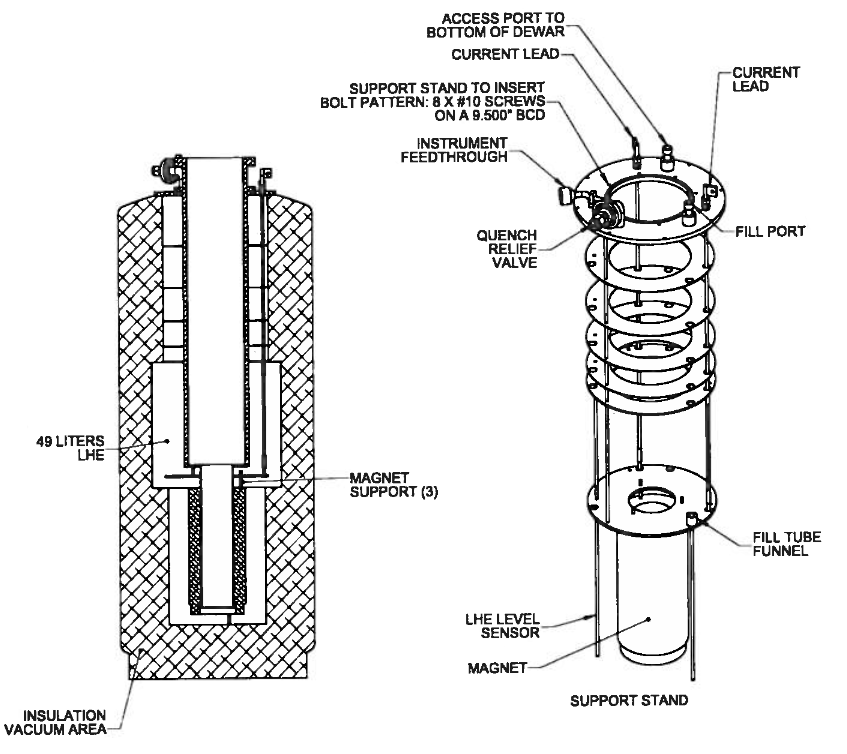}
\caption{UNH solenoid and insulating cryostat.}
\label{fig:unhmagnet}
\end{figure}

The UNH magnet is a superconducting solenoid custom designed to have a high magnetic field homogeneity. This is important for polarized target studies as the resolution of a polarization signal is proportional to the field homogeneity over the target, making the very small deuteron signals easier to resolve cleanly. The Helmholtz-configuration split coil magnet used in g2p is only necessary to allow the passage of the electron beam, meaning that the local setup at UNH is better served by a simple solenoid. The UNH magnet is shown in Figure~\ref{fig:unhmagnet}. The coils are cooled with liquid helium and become superconducting below 10~K.

\section{Vacuum and Cryogenics}
\begin{figure}[htb]
\centering
\includegraphics[width=0.85\textwidth]{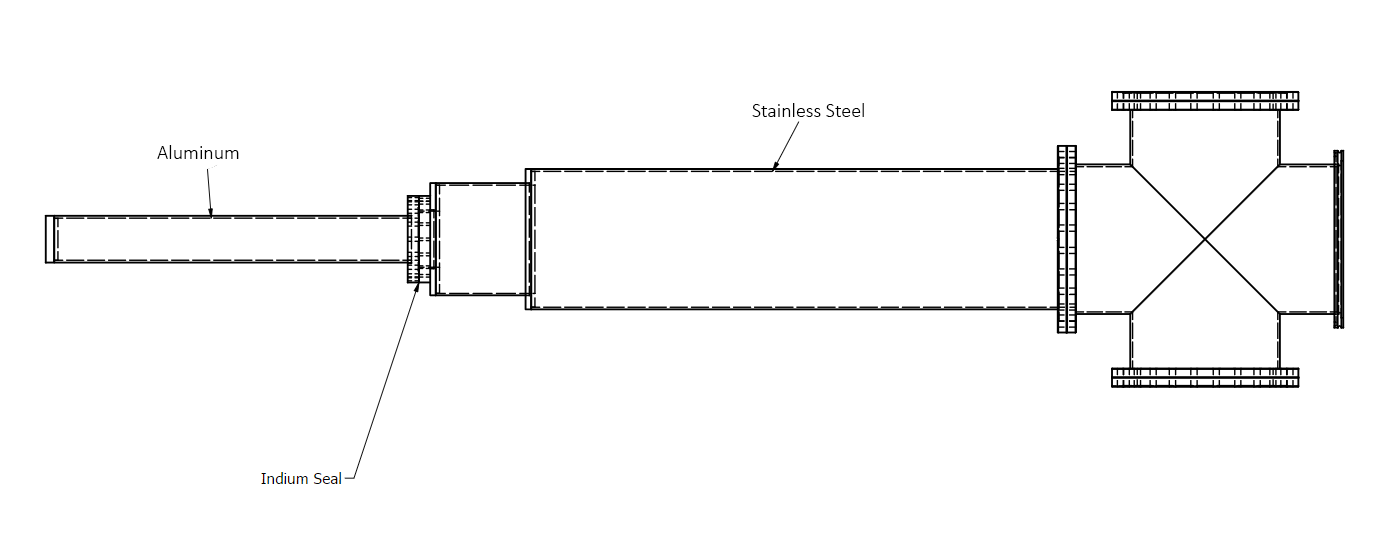}
\caption{Inner vacuum shell. Outer shell is similar but with larger diameter, designed to join with the bottom of the large T-piece via a Copper Flange (CF) seal.}
\label{fig:shells}
\end{figure}

Inside the magnet is a set of insulating vacuum shells, designed to thermally separate the magnet space from a vertical liquid helium evaporation refrigerator which rests in the center of the magnet. These vacuum shells are shown in Figure~\ref{fig:shells}, and have proven a crucial part of the system, since a loss of vacuum results in dangerous quench conditions for the superconducting magnet and an environment too warm for enhanced polarization in the helium refrigerator. This vacuum is also difficult to maintain due to the aggressive temperature gradient the shells are put under, and the need for them to contain superfluid helium inside the LHe evaporation refrigerator.

The top part of the shells is constructed of stainless steel, to prevent high thermal conduction, while the nose of each shell must be formed from aluminum, a less magnetic material which will not create stray currents when placed inside the solenoid. The forming of a vacuum seal which does not permit superfluid helium to pass between these two disparate materials is very challenging, and necessitates the use of an indium seal. I was given the opportunity to develop and improve the lab's procedures for creating these seals, as we had limited experience in creating them at the time that I started. The seals are formed from a length of indium wire, which, when compressed tightly between flanges on each side, flows like a liquid to form a vacuum seal. The challenge of the indium seal method is how to join the wire with itself in a way that will make a smooth circular seal, without creating an excess of material on one side that would prevent the flanges from applying even pressure.

\begin{figure}[htb]
\centering
\includegraphics[width=0.4\textwidth]{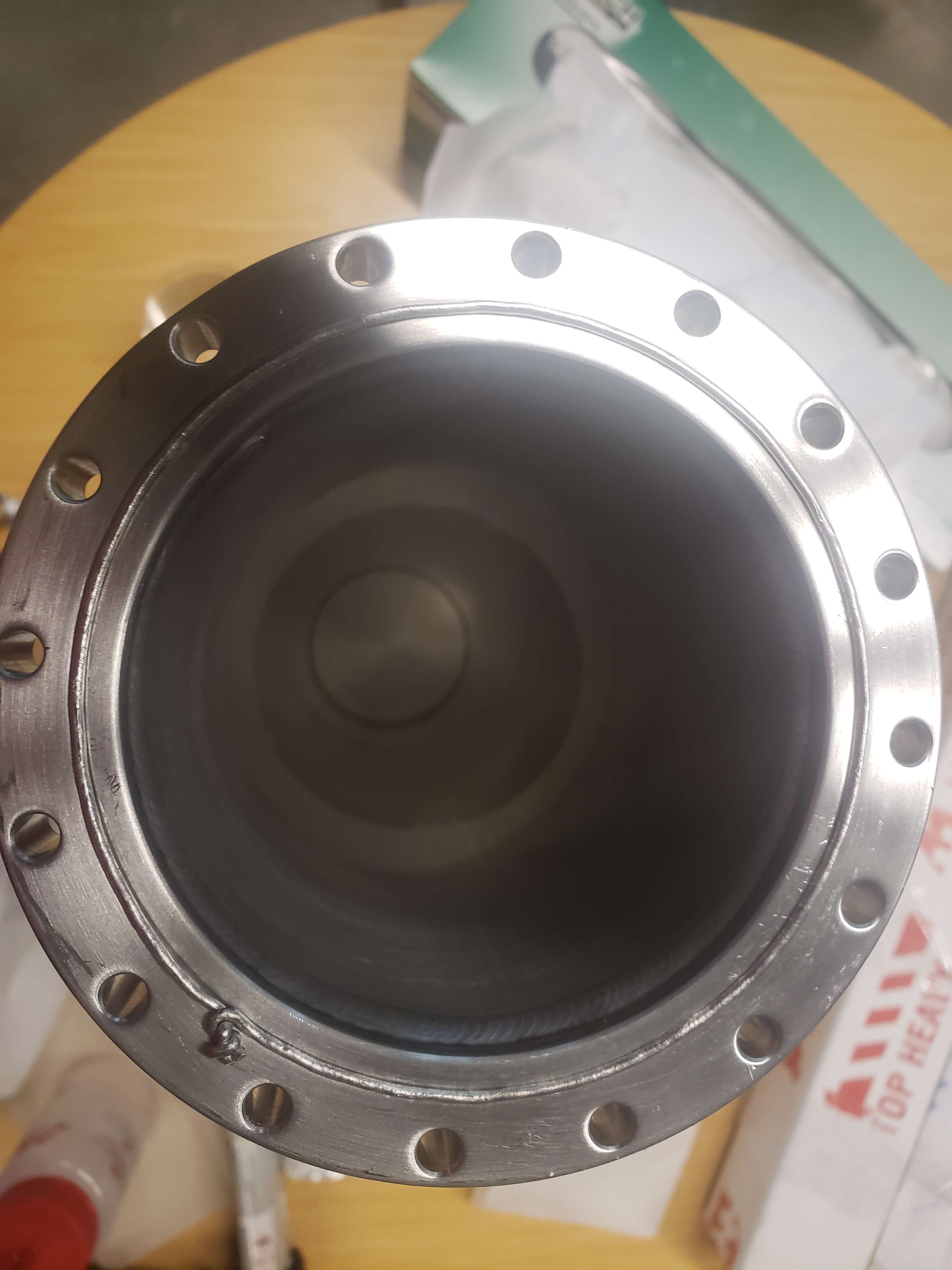}
\caption{Indium seal in progress, showing the folded knot structure}
\label{fig:indium}
\end{figure}

An indium seal in progress is shown in Figure~\ref{fig:indium}. Through repeated iterations to improve the procedure, the optimal method has been determined to require carefully cleaning both the indium wire and flanges with acetone before forming the seal. The indium is then formed into a loop with long tails, which are then twisted tightly to form a long `knot'. This knot is then cut around 4 mm away from the main loop to remove excess material, and then folded back to rest against the side of the primary loop. This seal is then compressed between both flanges, taking care that it stays vertical and that pressure is distributed evenly. The flanges are screwed together, with the screws being tightened evenly in the `star pattern', before allowing the indium to flow internally for around 20 minutes. The screws are then tightened again before allowing the indium to flow once more for 24 hours, and being tightened a third and final time. These indium seals are used both in the vacuum shells for the cryostat, and for smaller vacuum cylinders used for material production.

\begin{figure}[htb]
\centering
\includegraphics[width=0.5\textwidth]{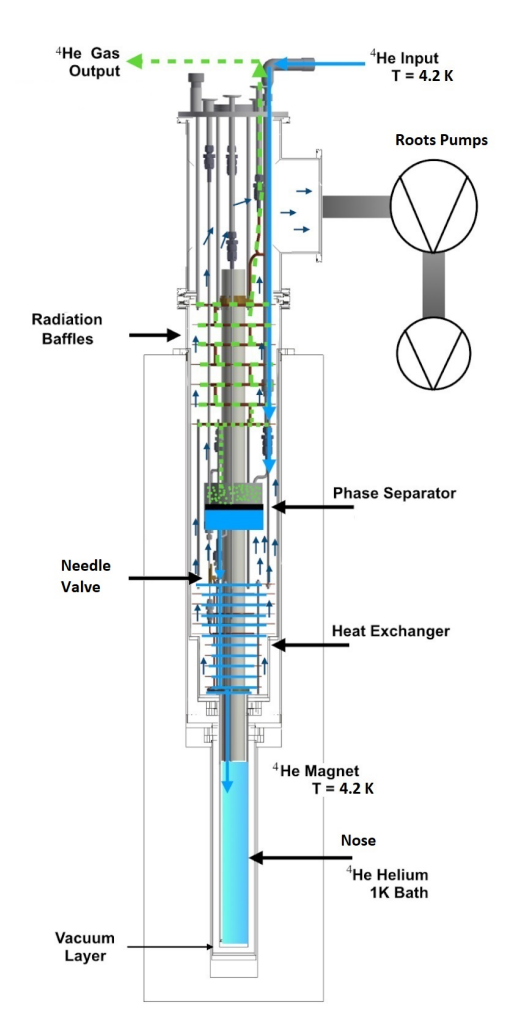}
\caption{LHe Evaporation Refrigerator for the UNH System. Reproduced from~\cite{ChileFridgeSim}.}
\label{fig:fridge}
\end{figure}

Inside the inner vacuum shell is a liquid helium evaporation refrigerator, designed to cool 4 K liquid helium to 1 K, very similar to what was used during the g2p experiment. The refrigerator is shown in Figure~\ref{fig:fridge}. This figure is taken from a detailed simulation of the UNH refrigerator published in Nuclear Instruments and Methods~\cite{ChileFridgeSim}. It is connected to the liquid helium reservoir in the magnet, and helium driven up into the fridge by way of a pump connected to the separator, which pulls helium out of the magnet and into the separator. The top baffles serve to create a temperature gradient from the 300 K world at the top plate to a 4 K liquid helium temperature at the separator. Helium then drips through a fine mesh in the separator, while warm gas helium is pumped out by the separator pump. Helium which collects in the nose, which is the part of the refrigerator surrounded by the superconducting magnet, is pumped on by a large set of high-flow Roots pumps to cool it to 1 K, with the heat exchanger serving to cool liquid helium down to the current temperature of the nose before it reaches the helium reservoir in the nose.

In addition to the other projects and tasks discussed here, I and many of the other members of the lab were responsible for operating, maintaining, and upgrading the refrigerator and connected vacuum and cryogenic equipment, as well as the superconducting magnet discussed in the previous section. Over my time working in the UNH lab, we have tremendously improved the leak-tightness of the system, resulting in better cryogenic operation and lower vacuum in the vacuum spaces.

\section{NMR}

One of the primary responsibilities I had in the UNH lab was to become the local expert on our main NMR system. This is a VME crate based system designed at Los Alamos National Laboratory (LANL)~\cite{MCGAUGHEY2021165045}. In 2018 I travelled to LANL to be trained on the system for proton NMR, and in 2020, I visited LANL again to pick up and be trained on the deuteron NMR system. These systems are designed to be an updated and modern version of the original Liverpool-designed ``Q-Meter'' NMR system, and are responsible for finding the UNH lab's first proton and first deuteron signal. 

\begin{figure}[htb]
\centering
\includegraphics[width=0.5\textwidth]{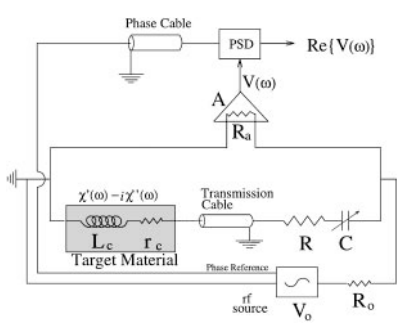}
\caption{Liverpool Q-Meter System. Reproduced from~\cite{CrabbMeyer}.}
\label{fig:qmeter}
\end{figure}

The innovation of the Liverpool system, a schematic of which is shown in Figure~\ref{fig:qmeter} is the integration of a `mixer'.  In passing through the RLC circuit part of the system, the supplied RF signal spreads out in phase, where the polarization signal of interest is only in the real part of the inductance. The complex part of the inductance instead contains dispersive effects driven by a mistune of the system. In the Liverpool system, a reference signal and the returning signal from the circuit are mixed together, selecting only the phases in common between the two. The reference signal passes through a length of cable which can be specifically tuned to ensure its central phase is 0$^\circ$, corresponding to the real part. The length of the cable which connects the inductor to the rest of the circuit is also crucial, as properly tuning the circuit requires the cable to be a multiple of the $\frac{\lambda}{2}$ wavelength of the target particle, on the order of 55 cm for the proton and several meters for the deuteron. Tuning this length is accomplished with the use of an internal diode.

The LANL system uses this same core concept of a mixer, but with many updated electronic components, and improves upon both the phase and diode tuning described above. Phase tuning is done by way of an internal phase shifter connected to the reference line, instead of by adjusting a cable length. And though the length of the signal cable is still important for the diode tune, an internal varactor diode allows for some shifting of the effective `length' of the cable by changing the capacitance of the circuit. This means that both tunes can mostly be accomplished by providing voltages, and controlled by a connected computer. The frequency $f$ the circuit is tuned to look at is a function of the coil inductance L and the capacitance of the circuit C:
\begin{equation}
\label{eqn:NMRf}
f = \frac{1}{2 \pi \sqrt{L C}}
\end{equation}

\begin{figure}[htb]

\centering
    \begin{subfigure}[t]{0.35\textwidth}
        \centering
        \includegraphics[width=\linewidth]{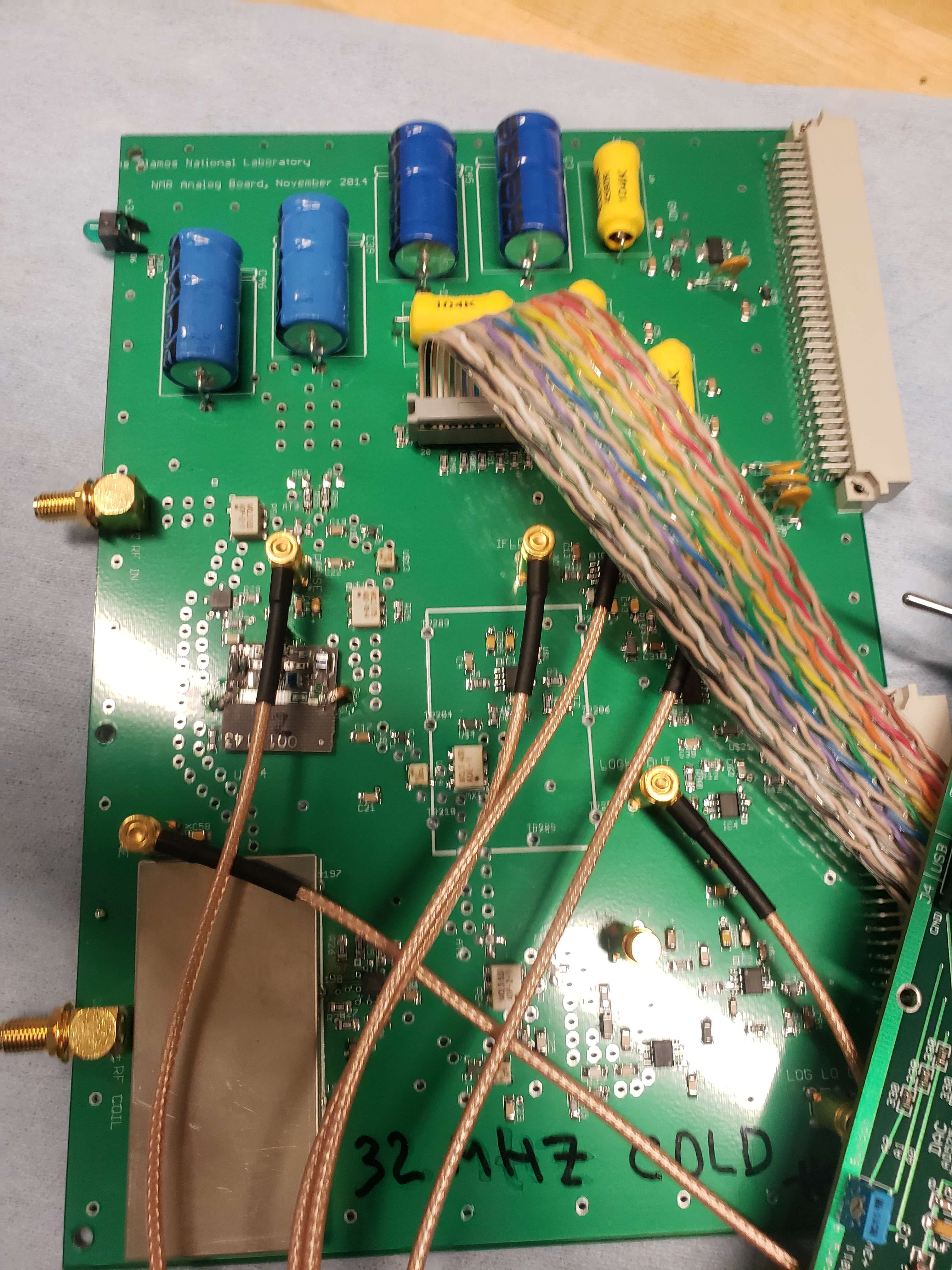} 
        \caption{LANL System Analog Board} \label{fig:vmeanalog}
    \end{subfigure}
    \hspace{1.1em}
    \begin{subfigure}[t]{0.35\textwidth}
        \centering
        \includegraphics[width=\linewidth]{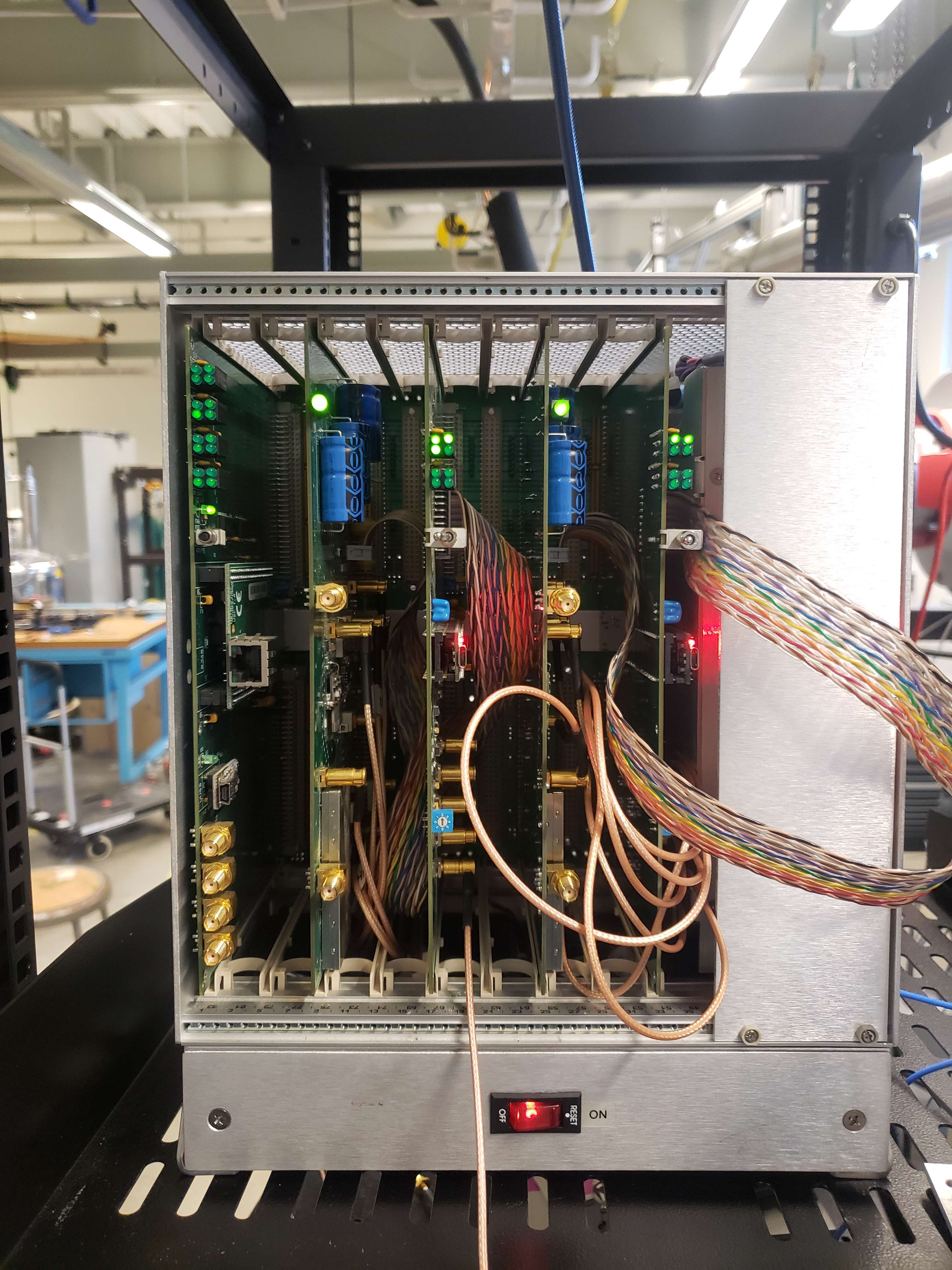} 
        \caption{LANL System in VME Crate} \label{fig:vmecrate}
    \end{subfigure}
\caption{The LANL VME-Based NMR System.}
\end{figure}

For the proton NMR system, the function is otherwise extremely similar to the NMR used in the g2p experimental setup, though it is designed with the intent of optimizing a better signal-to-noise ratio. Though I was trained on the system's operation at LANL, I was responsible for wiring, building, and commissioning it at UNH. The full LANL system is shown in Figure~\ref{fig:vmecrate}. Each `Q-Meter' equivalent is reproduced with two boards, an analog board that holds the actual RLC circuit and mixer, and a digital board that contains analog-digital converters (ADCs) which digitize the signal and digital-analog converters (DACs) which send digital signals as voltages to the various components. The crate also has one controller board, which coordinates the other two boards and the connection to the computer.

The controller board also generates a $\pm$1 V triangle wave at a frequency of 20 Hz, which is sent to a Rohde \& Schwarz SMT03 Signal Generator to program an RF sweep over the frequency range of interest near the Larmor frequency of the target particle. The resulting RF signal enters a splitter, with one side traveling to the reference line, and the other going to the tank circuit, which contains a connection to the coil inductor and most of the other components of the core RLC circuit.

Due to the much smaller size of the deuteron signal, the deuteron system requires the modification of a cold NMR chip. In principle, to do NMR, the system is tuned to a given frequency based on the inductance and capacitance of the circuit. But the ideal tune frequency can shift due to changes in temperature of the circuit, because capacitors are especially susceptable to variation based on thermal shifts. For a larger signal, these shifts are irrelevant, but for a very small signal, these shifts can completely obliterate it. The principle of cold NMR is that most of the actual tank circuit is moved into the main cryostat, to be immersed in liquid helium alongside the target material. Due to sharing the constant temperature of the target and being shielded from noise in the outside world, this can improve the signal-to-noise ratio of the deuteron signal tremendously.

\begin{figure}[htb]

\centering
    \begin{subfigure}[t]{0.25\textwidth}
        \centering
        \includegraphics[width=\linewidth]{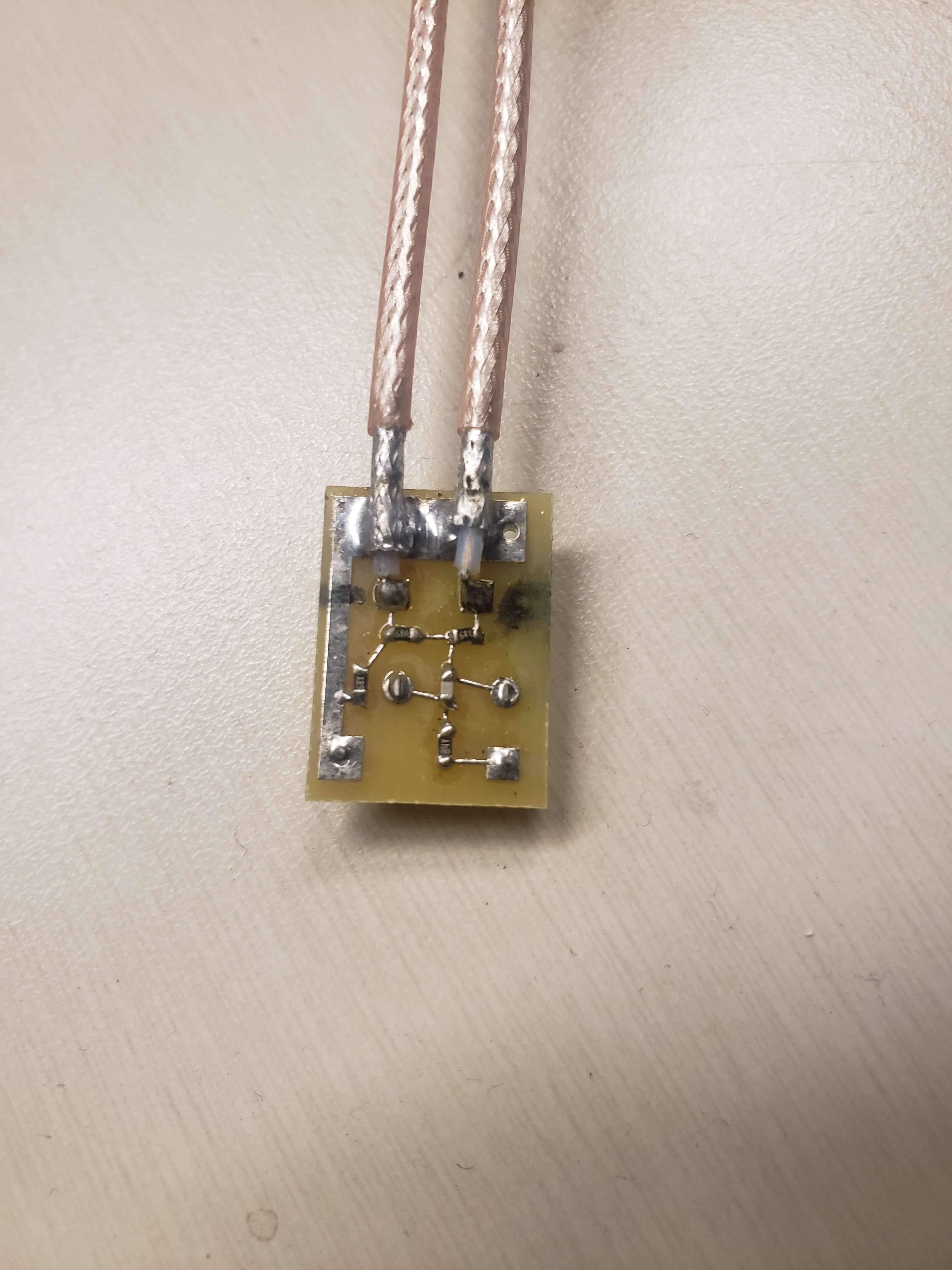} 
        \caption{Cold Board Photo} \label{fig:coldboard}
    \end{subfigure}
    \hspace{1.1em}
    \begin{subfigure}[t]{0.65\textwidth}
        \centering
        \includegraphics[width=\linewidth]{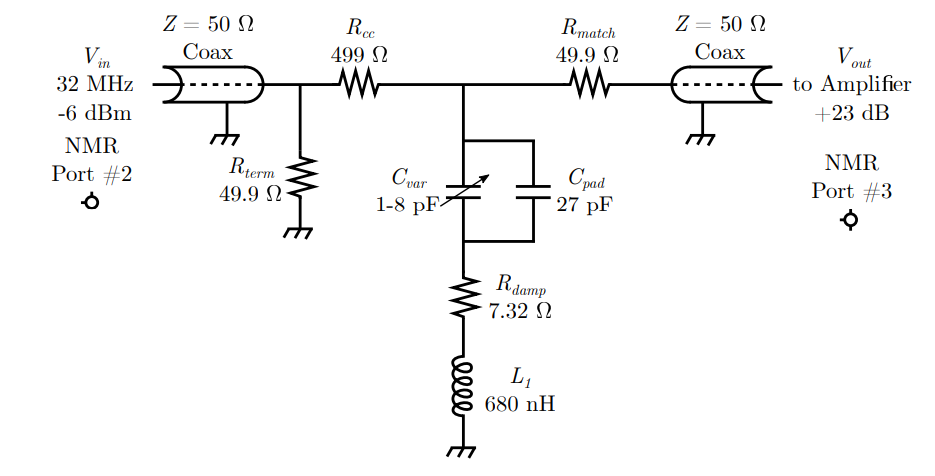} 
        \caption{Cold Board Circuit Diagram. Reproduced from~\cite{TECHNOTE:tstick}} \label{fig:coldboardschematic}
    \end{subfigure}
\caption{The Cold Board for the Deuteron NMR System}
\end{figure}

The cold NMR system for the deuteron requires the tank circuit to be incredibly small so that it can fit alongside the target. The LANL NMR design features a cold NMR board no larger than a fingernail, with surface mount SMD-0603 components. At LANL I learned how to solder these boards and made a functioning board there, and have since made multiple modifications to the UNH boards as well as created more when they become damaged. The cold NMR board is shown in Figure~\ref{fig:coldboard}. Because the circuit is placed alongside the coil, the functional length in $\frac{\lambda}{2}$ of the deuteron is zero. Rather than the varactor diode of the proton design, the capacitance of the circuit is controlled by an adjustable sapphire bearing 1-8 pF capacitor, which must be tuned prior to being immersed in cryogen. A padding capacitor initially designed at 27 pF controls the number and spacing of coils needed to generate the correct frequency within the tuneable range of the sapphire capacitor.

\section{DAQ}

Another major responsibility I have had in the UNH lab is developing the data acquisition (DAQ) system for the lab. The system is a National Instruments Labview program located on a central computer, and connected to more than 60 instruments throughout the lab through various serial standards. Though this has largely been a software project, it has required me to gain familiarity operating every instrument in the lab and consequently practice a number of hardware and electronic skills to commission each one. 

Devices which output or are controlled by a simple voltage are connected via coaxial BNC-standard cables to a National Instruments USB-6363 data acquisition device. Those with reliable RS232 serial standards, such as a number of pressure and flow gauges and a gaussmeter for measuring the fringe field near the cryostat, are connected through a DB-9 RS232 hub. Those devices which have firmware to establish a serial port through USB are connected to a large USB hub. Several more devices, most prominently the superconducting magnet power supply and the Lakeshore temperature monitors, are connected to a central GPIB hub. Finally, several crucial instruments such as the NMR system described in the previous section are connected through a TCP connection over the local network. 

\begin{figure}[htb]
\centering
\includegraphics[width=1.0\textwidth]{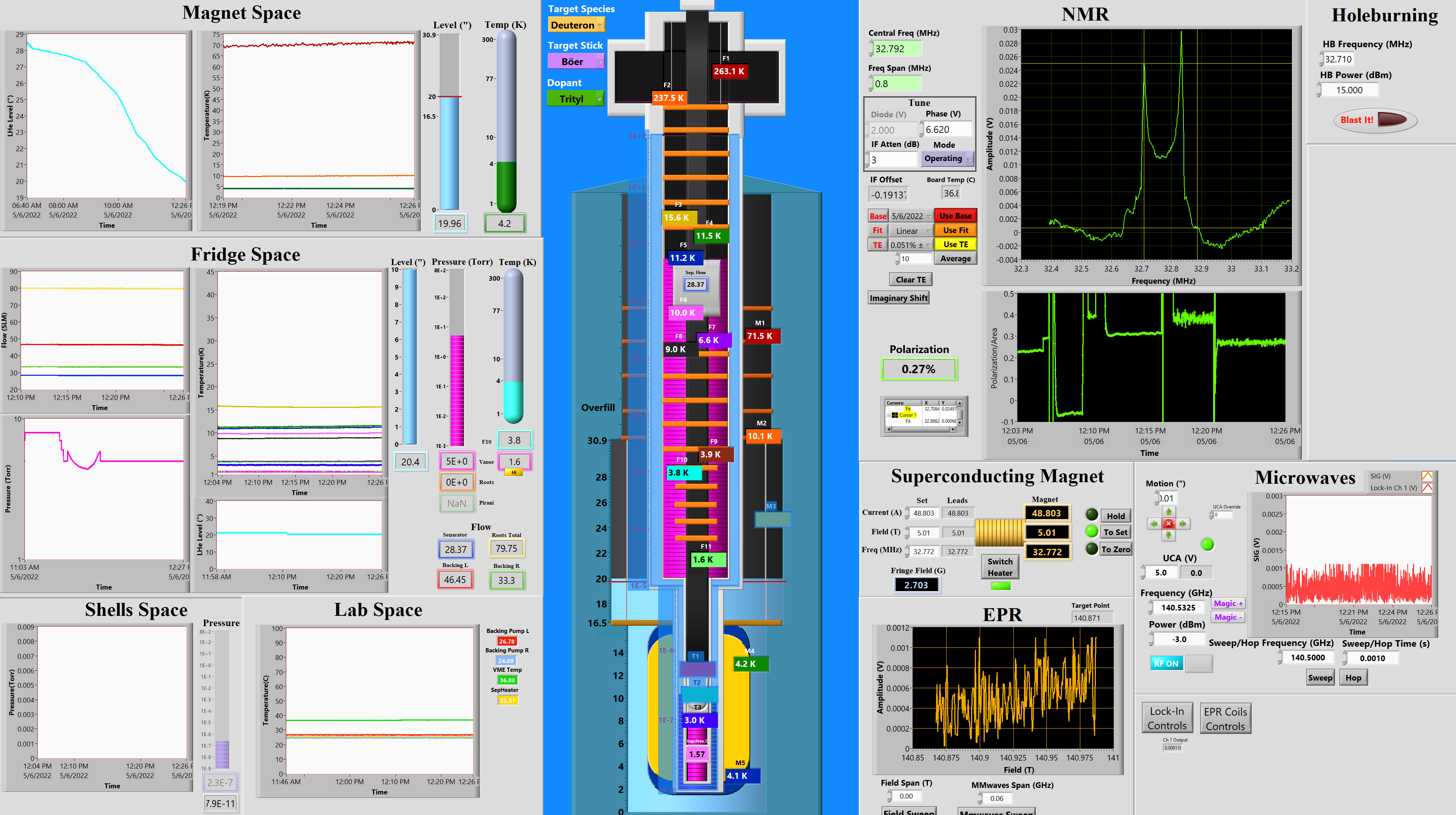}
\caption{DAQ System designed for the UNH Polarized Target Lab.}
\label{fig:daq}
\end{figure}

The main front panel of the DAQ software is shown in Figure~\ref{fig:daq}. This program is normally shown on several screens throughout the lab and controlled by a central computer. The center of the program features a `panopticon', showing a cross sectional reproduction of the full polarized target setup, with temperatures, pressures, and other relevant readouts shown in appropriate locations, to allow an at-a-glance view of the system's status. The left side features cryo-and-pressure readouts, organized into the individually separated spaces of the system, primarily the space inside the superconducting magnet and that inside the liquid helium evaporation refrigerator. The right side contains physics readouts and controls for the NMR and microwave system, as well as the superconducting magnet.

The physics controls section features a number of tools to make online analysis easier, allowing quick baseline subtraction and wing fitting for NMR signals, as well as Thermal Equilibrium calibration. It also contains a number of tools for optimizing the microwave system and finding ideal polarizing frequencies, as well as a number of safety measures designed to prevent an unsafe situation. The superconducting magnet cannot be ramped at an unsafe ramp rate, nor can it be ramped at all unless the level meter and temperatures confirm the coils are in superconducting state. If the liquid helium level in the magnet dips to a dangerous level, the DAQ system will attempt to take control and perform a safe ramp-down of the magnet to prevent a quench situation.

The DAQ averages the output of all instruments every 20 seconds, and records them to a data file for later analysis, including the NMR sweep and EPR sweep data. It is also integrated with a Slack group used by the group for communication, pushing updates on the status of instruments to a Slack channel every 15 minutes and alerting there if an alarm indicating a dangerous situation is triggered. It also harvests posts from a dedicated logbook Slack channel for posterity, and records them alongside logbook entries made on the lab computer. As the main data recording and control unit for all the instruments in the lab, it remains an essential tool for the UNH lab's work.

\section{Cooldowns \& Results}

\begin{figure}[htb]

\centering
    \begin{subfigure}[t]{0.4\textwidth}
        \centering
        \includegraphics[width=\linewidth]{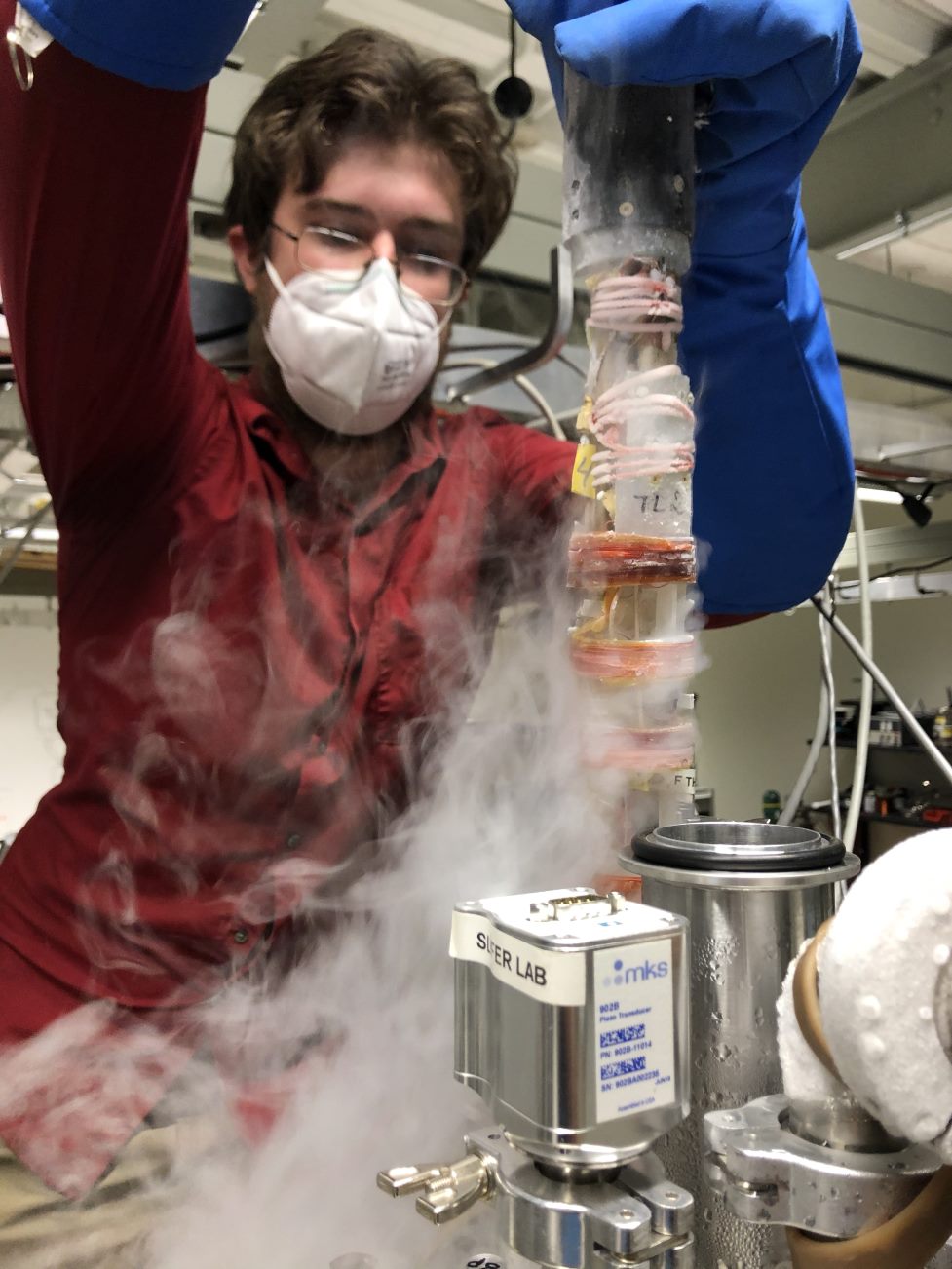} 
        \caption{Author loading a target insert into the cryostat}
    \end{subfigure}
    \hspace{1.0em}
    \begin{subfigure}[t]{0.4\textwidth}
        \centering
        \includegraphics[width=\linewidth]{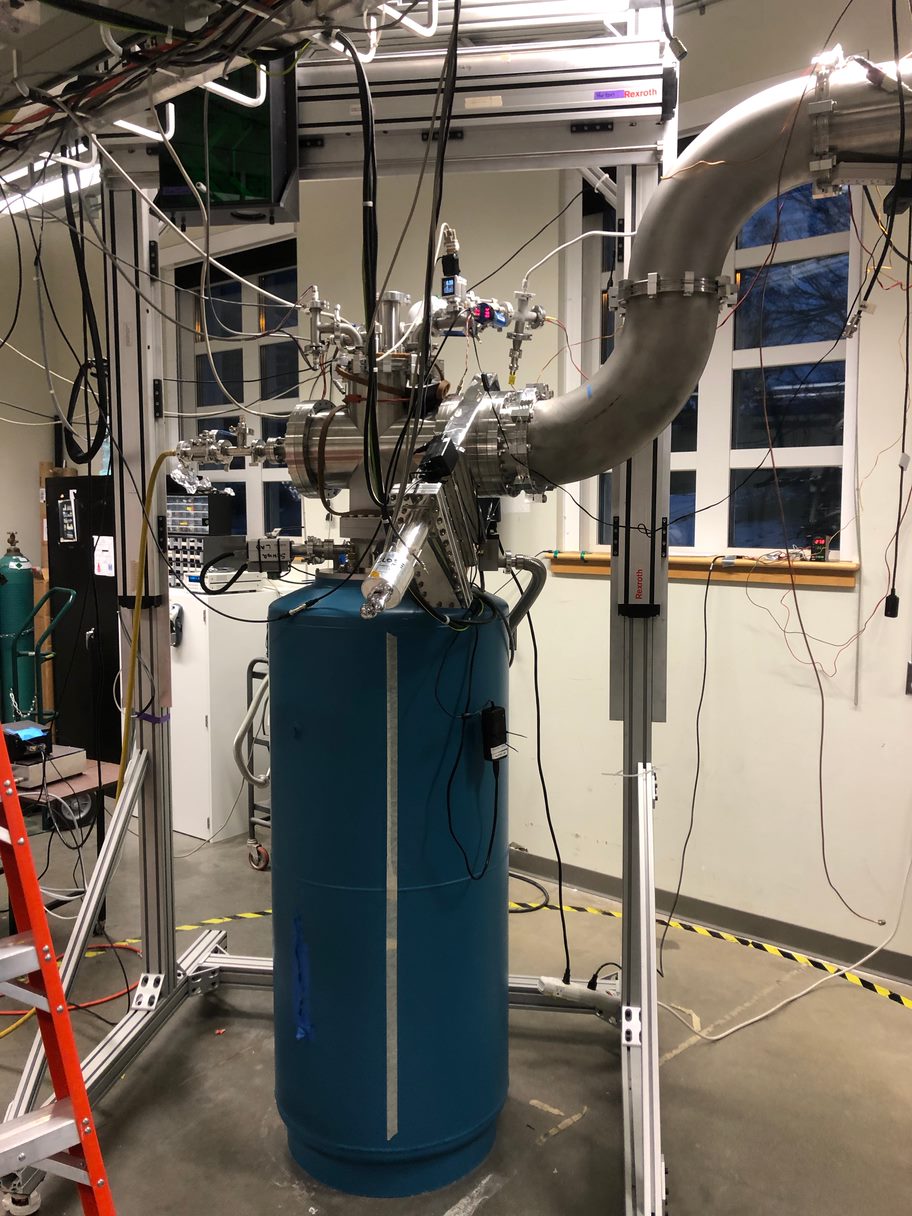} 
        \caption{Main UNH Cryostat containing superconducting magnet and LHe refrigerator}
    \end{subfigure}
    \hspace{1.0em}
    \begin{subfigure}[t]{0.7\textwidth}
        \centering
        \includegraphics[width=\linewidth]{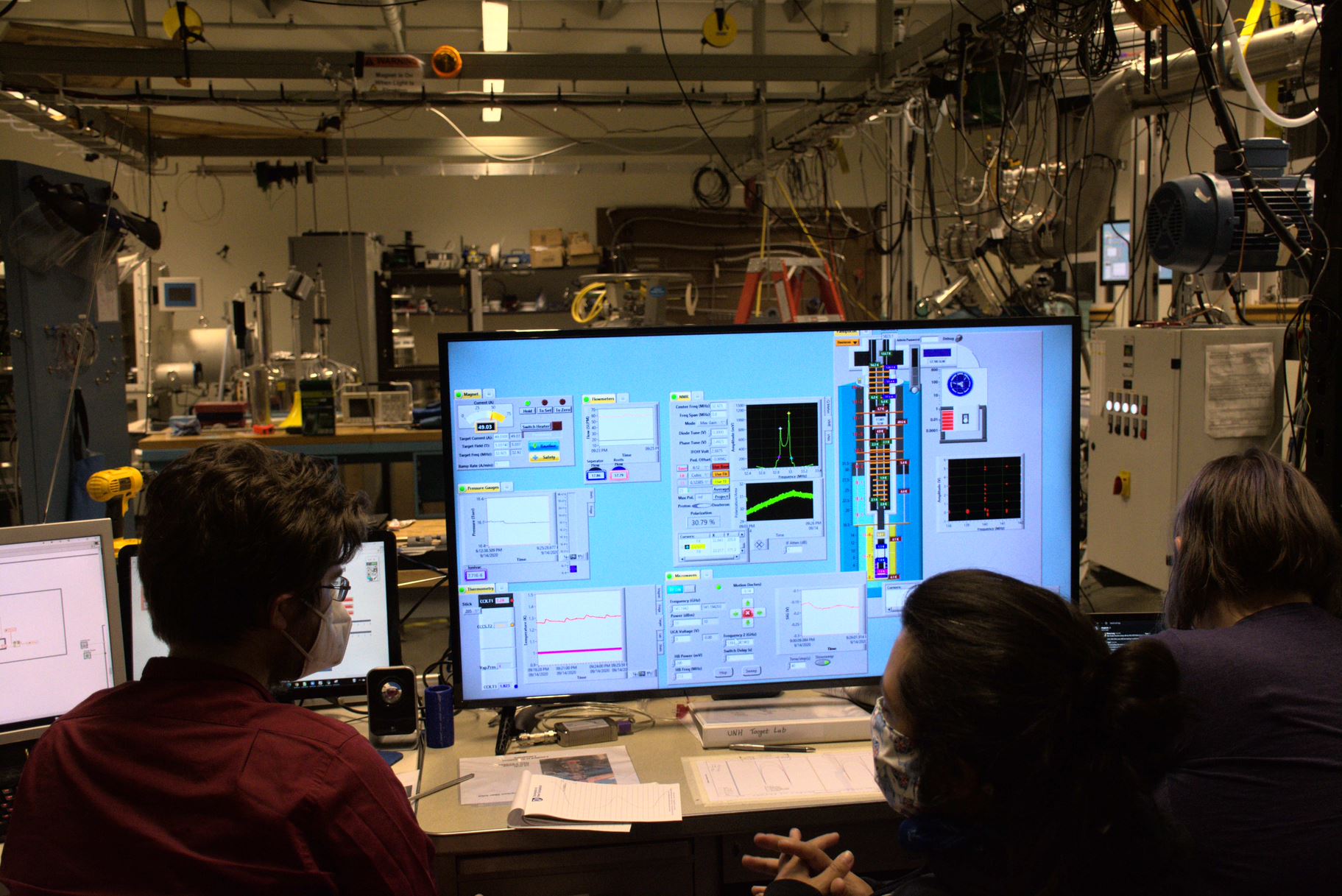} 
        \caption{UNH Group monitoring the DAQ system during a cooldown}
    \end{subfigure}
\caption{Photos from the UNH Polarized Target Cooldowns.}
\label{fig:cooldowns}
\end{figure}

Due to the expensive price associated with the liquid helium needed to run the system, physics results are collected several times a year during a `cooldown'. These cooldowns can last anywhere from a few days to over a week, and require constant work to operate all the subsystems in the lab at once. All group members participate in handling cryogen, loading and unloading target material, operating the pumps and the helium refrigerator, and using the NMR and EPR subsystems to collect physics results. Some photos from the UNH group's cooldowns are shown in Figure~\ref{fig:cooldowns}.

As of 2022, the UNH Group is focusing almost exclusively on deuteron polarization, and has observed polarized deuteron signals from the alcohols Butenol and Propendiol. Materials in the UNH lab are chemically doped, due to a current lack of access to the NIST facility used to irradiate the g2p target. Chemical doping is performed with a number of proven radicals, such as TEMPO, a relatively inexpensive radical which can obtain good polarization, and Trityl, a much more expensive radical supposed to produce very sharp ESR-lines for high deuteron polarization and easier access to the deuteron signal's complex structure. 

Currently, the UNH lab has obtained a deuteron polarization of around 27\% vector polarization, and 4-5\% tensor polarization. This vector polarization value is considered to be quite good for an evaporation refrigerator, as higher polarizations of around 50\% were observed for the same materials using a dilution refrigerator, which can obtain much colder temperatures than 1~K. The group's current goal is enhancing this tensor polarization further, to meet the target value of 30\% for the upcoming tensor experiments. This enhancement is being tested with several methods, including the ssRF `hole burning' method which the polarized target group at University of Virginia (UVA) has proven to be very effective, as well as several experimental techniques involving the UNH group's solid state microwave system.

\chapter{Analysis}
\begin{figure}[htb]
\centering
\includegraphics[width=0.9\textwidth]{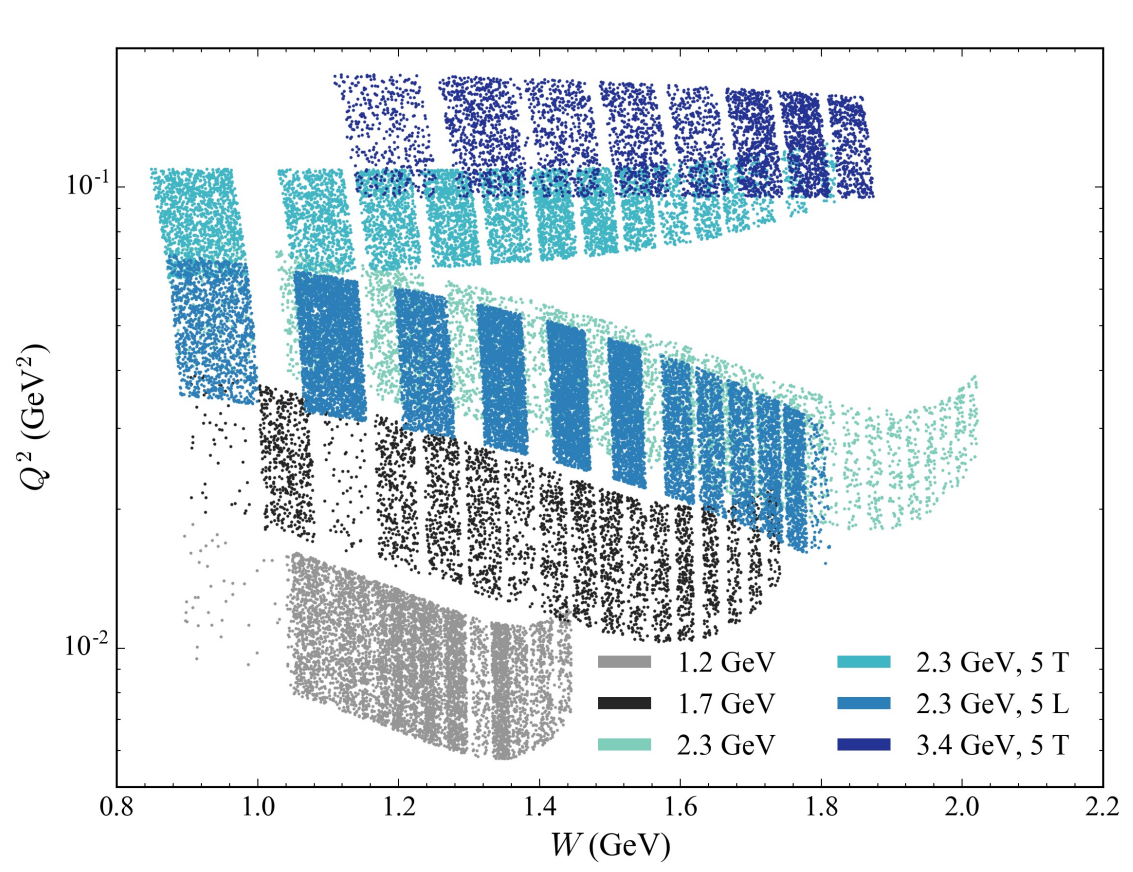}
\caption{Kinematic coverage of the E08-027 experiment, in terms of invariant mass and momentum transfer. Reproduced from~\cite{Zielinski:2017gwp}.}
\label{fig:kinematic}
\end{figure}

Let us now return to a discussion of the data collected in the E08-027 experiment. Raw data is encoded as pure counts in a series of ROOT files, comparing various kinematic variables to how many scattered electrons are associated with a given value. This allows a reconstruction of the electron spectra as a function of energy, scattering angle, and other kinematic variables. Data were taken at target fields of 2.5T and 5T, and beam energies of 1.2, 1.7, 2.2, and 3.3 GeV. In total, there are six different kinematic settings, of which five have a transversely polarized target and one has a longitudinally polarized target. The lowest energy setting, with beam energy of 1.2 GeV, was taken primarily for the purpose of doing radiative corrections, so the analysis focuses on the other five settings. The kinematic coverage of the experiment is shown in Figure~\ref{fig:kinematic}. Two energy settings, the 2.5T, 2.2 GeV Transverse and 5T, 2.2 GeV Longitudinal setting, fall on very similar kinematics and so can ultimately be used together in the final results, as will be discussed in the following sections.

To extract the structure functions, our quantity of interest is the polarized cross section differences of~(\ref{eqn:xs_poldiff_trans}). A convenient way to generate these is with an experimental asymmetry A and an unpolarized cross section $\sigma_0$:

\begin{equation}
\label{eqn:polxsdiff_perp}
\Delta \sigma_{\perp} = 2 A^{exp}_{\perp} \sigma_{0}
\end{equation}
\begin{equation}
\label{eqn:polxsdiff_par}
\Delta \sigma_{\parallel} = 2 A^{exp}_{\parallel} \sigma_{0}
\end{equation}

Here, the experimental asymmetry $A^{exp}$ represents a comparison between scattered electrons with forward helicity, and those with backward helicity. The construction of this quantity will be discussed further in section 6.3, but for now it is necessary to understand that the experimental asymmetry must be scaled from the raw, measured asymmetry by several quantities:
\begin{equation}
\label{eqn:asymmetry_meas}
A_{exp} = \frac{1}{f P_b P_t}A_{meas}
\end{equation}

$P_b$ and $P_t$ are the beam and target polarizations measured with the M\o{}ller polarimeter and NMR system respectively. $f$ represents a quantity called a dilution factor, which is used to eliminate contributions to the data from electron scattering off other materials in the target besides the proton of interest.

To extract our polarized cross section differences, then, we need to produce a dilution factor, a raw asymmetry, and an unpolarized proton cross section. The next sections will focus on the extraction of each of these quantities sequentially.

\section{Experimental Cross Section}

To form several of these quantities, let us discuss the formation of an experimental cross section, one of the primary measured quantities of interest in nuclear physics experiments. In equation~(\ref{eqn:xs}), we defined the cross section in terms of the number of of interactions, initial flux, and available states. Re-contextualizing this in experimental terms, we can understand each of these quantities to be related to the number of scattered electrons measured, initial flux of electrons, and number of available scattering centers in the target respectively. The initial flux of electrons can be determined from the beam current (measured by BCM) and the charge of an electron, and the number of scattering centers is related to the density of the target material. Taking these quantities into account and considering the efficiencies of relevant detectors, we write the experimental cross section as:
\begin{equation}
\label{eqn:exp_xs}
\frac{d^2\sigma^{exp}}{dE'd\Omega} = \frac{ps N}{\frac{Q}{e} \rho_{target} LT \epsilon_{det}} \frac{1}{\Delta\Omega \Delta E' \Delta Z}
\end{equation}
The terms of this equation are as follows:
\begin{itemize}
    \item N is the number of scattered electron counts measured in the detectors
    \item $\frac{Q}{e}$ represents the total number of incoming electrons, where Q is the total charge deposited by the beam and $e$ is the fundamental charge of an electron.
    \item $\rho_{target}$ is the density of protons in the target, taking into account the packing fraction of the target cell and the density of solid ammonia.
    \item $ps$ is the prescale factor. This determines how many scattering events are actually processed by the DAQ: for every $ps$ events that occur, 1 event is registered in the DAQ.
    \item $LT$ is the Livetime of the detectors. Due to the trigger pattern of the DAQ trigger, the detectors are not live to catch 100\% of events, this factor represents the fraction of time which they are.
    \item $\epsilon_{det}$ is the total sum of all detector efficiencies, the determination of which is discussed in detail in~\cite{Zielinski:2017gwp}.
    \item $\Delta Z$ is the length of the target cell
    \item $\Delta E'$ is the energy acceptance, interpreted as the width of the energy bin being used to calculate the cross section.
    \item $\Delta \Omega$ is the angular acceptance of the detector, corresponding to the overall area cut in $\theta$ and $\phi$.
\end{itemize}
These cross sections can be difficult to calculate due to the large number of inputs, and the difficulty in reconstructing a precise angular acceptance for the transverse settings, where a significant out of plane scattering angle contribution makes understanding the acceptance nontrivial. In many cases below, it will be more useful to look at a yield. When we are considering a ratio of two cross sections, or comparing two different linearly related cross sections, we can consider a version of the cross section where we cancel all terms which will be identical for all possible cross sections, known as a raw yield:
\begin{equation}
\label{eqn:yield}
Y = \frac{N}{Q\times LT}
\end{equation}
This is very useful to analyze because it cancels many of the factors which can have large systematics associated with them. 

\section{Packing Fractions}

Before discussing the dilution factor directly, it is first necessary to determine a packing fraction ($pf$). The packing fraction is defined simply as the fraction of the target cell volume which is full of frozen ammonia beads, where the remaining volume is filled with liquid helium. This packing fraction is a necessary input to the dilution factor and the experimental cross section.

The packing fraction was determined with a method designed by Oscar Rondon for the RSS experiment~\cite{TECHNOTE:rondonpf} and modified for use here. In principle, this method is based on the idea that the cross section and yield defined above should have a linear relationship with the packing fraction. If it is possible to reproduce the scattering count results with a simulation, then, we can simulate yields at several different packing fractions to determine a linear fit to the yield-dependence of the packing fraction. We can then use the production yields to determine the exact packing fraction for each material. In g2p, 2 different materials were used for each energy setting, with a `material' being used to refer to a separate iteration of the target cell, though all are filled with frozen ammonia.

For this method, a Monte-Carlo simulation called g2psim~\cite{Chao3} was used to reproduce the data. This simulation is able to reproduce the elastic yields with very good accuracy. Before performing the direct comparison to the production yields, it was necessary to calibrate the simulation and determine the proper input parameters through comparison to the Carbon dilution yields, for which the effective `packing fraction' is well understood, since the carbon target is a solid disk of known length. For the 3.3 GeV setting, no elastic carbon data were taken that could be used, meaning this method could not be used. However, the materials used during the 3.3 GeV runs were identical to those used during the 2.2GeV, 5T Transverse runs, meaning that the packing fraction should be identical. The same packing fraction is used with a larger systematic to account for possible losses due to the beam or motion in the target.

The carbon yields for the other settings were compared to carbon yields produced with the Monte-Carlo simulation. This was used to determine an appropriate radiative smearing factor for each energy setting by tweaking the input smearing parameter to make the width of both peaks match. They were also used to determine what normalization factor is necessary to scale the height of the simulated peak such that it matches the data as well as possible. Before comparison, an Empty cell yield for each energy setting was subtracted from the associated Carbon yield, scaled by the known fraction of the target cell filled with liquid helium, based on the length of the carbon disk. The comparison to resulting pure carbon yields and ratio with the average scale factor is shown in the following Figures.

\begin{figure}[htb]

\centering
    \begin{subfigure}[t]{0.7\textwidth}
        \centering
        \includegraphics[width=\linewidth]{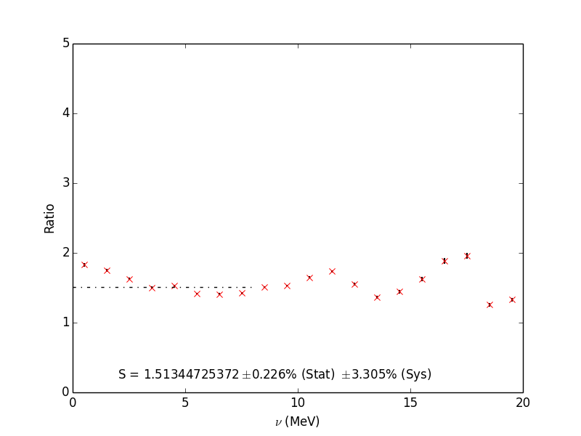} 
        \caption{Carbon Simulation/Data Ratio} \label{fig:carbonrat17}
    \end{subfigure}
    \hspace{1.1em}
    \begin{subfigure}[t]{0.7\textwidth}
        \centering
        \includegraphics[width=\linewidth]{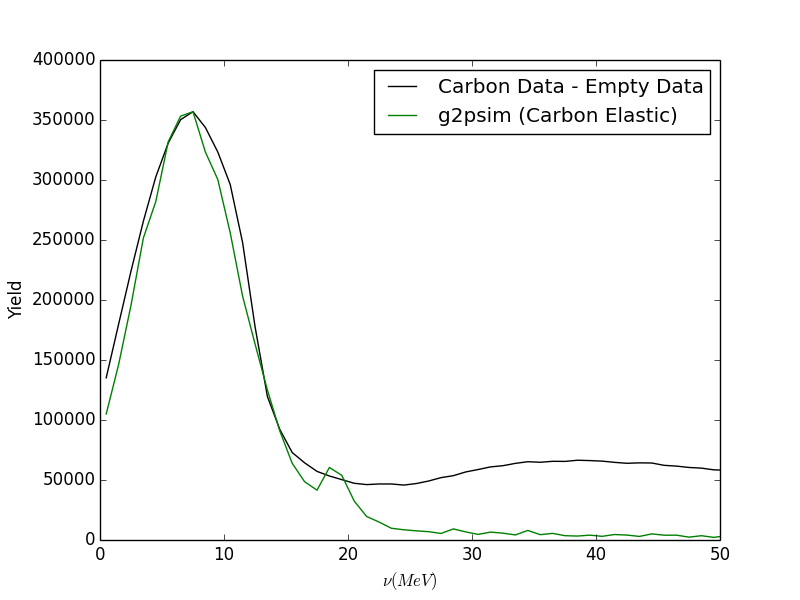} 
        \caption{Scaled Carbon Simulation \& Data Comparison} \label{fig:carboncomp17}
    \end{subfigure}
\caption{2.5T 1.7 GeV Transverse Carbon Comparison.}
\end{figure}

\begin{figure}[htb]

\centering
    \begin{subfigure}[t]{0.7\textwidth}
        \centering
        \includegraphics[width=\linewidth]{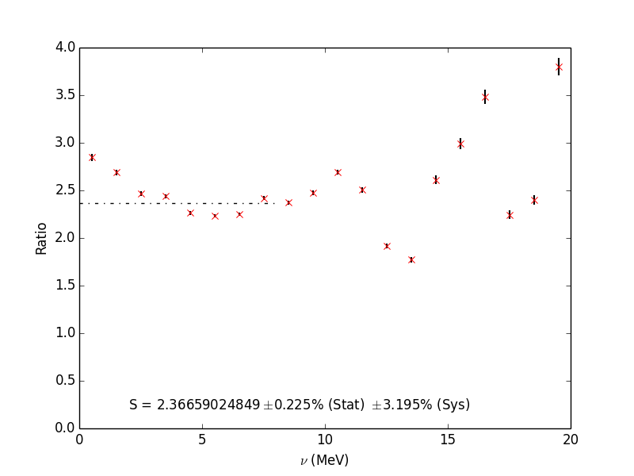} 
        \caption{Carbon Simulation/Data Ratio} \label{fig:carbonrat25T}
    \end{subfigure}
    \hspace{1.1em}
    \begin{subfigure}[t]{0.7\textwidth}
        \centering
        \includegraphics[width=\linewidth]{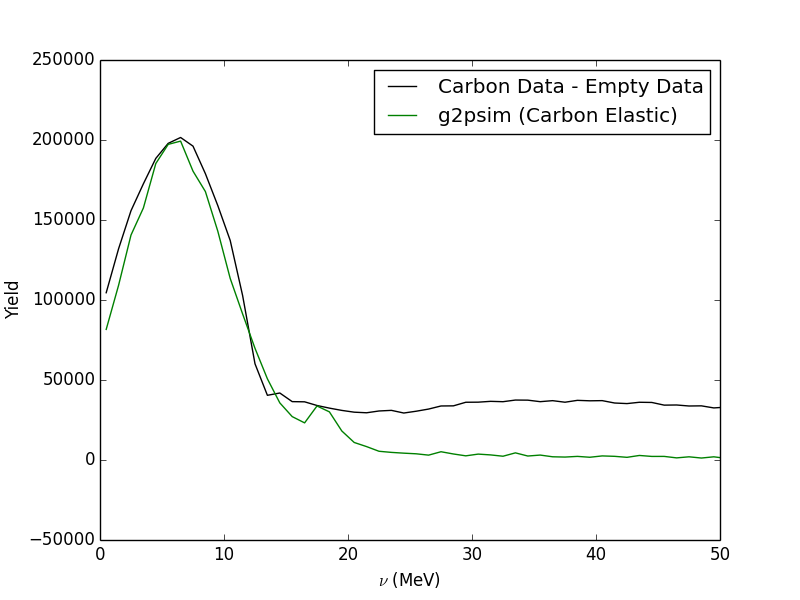} 
        \caption{Scaled Carbon Simulation \& Data Comparison} \label{fig:carboncomp25T}
    \end{subfigure}
\caption{2.5T 2.2 GeV Transverse Carbon Comparison.}
\end{figure}

\begin{figure}[htb]

\centering
    \begin{subfigure}[t]{0.7\textwidth}
        \centering
        \includegraphics[width=\linewidth]{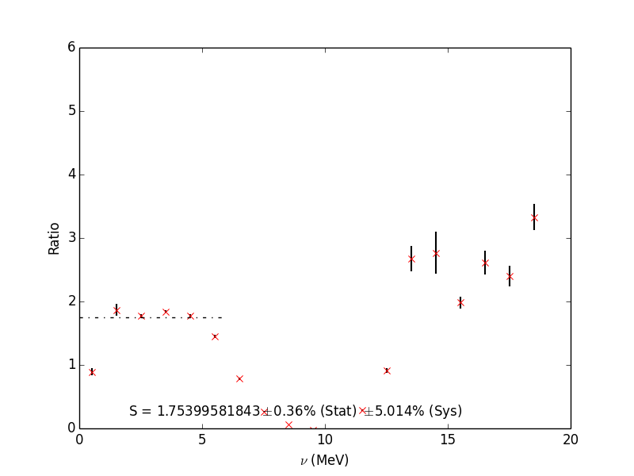} 
        \caption{Carbon Simulation/Data Ratio} \label{fig:carbonrat22}
    \end{subfigure}
    \hspace{1.1em}
    \begin{subfigure}[t]{0.7\textwidth}
        \centering
        \includegraphics[width=\linewidth]{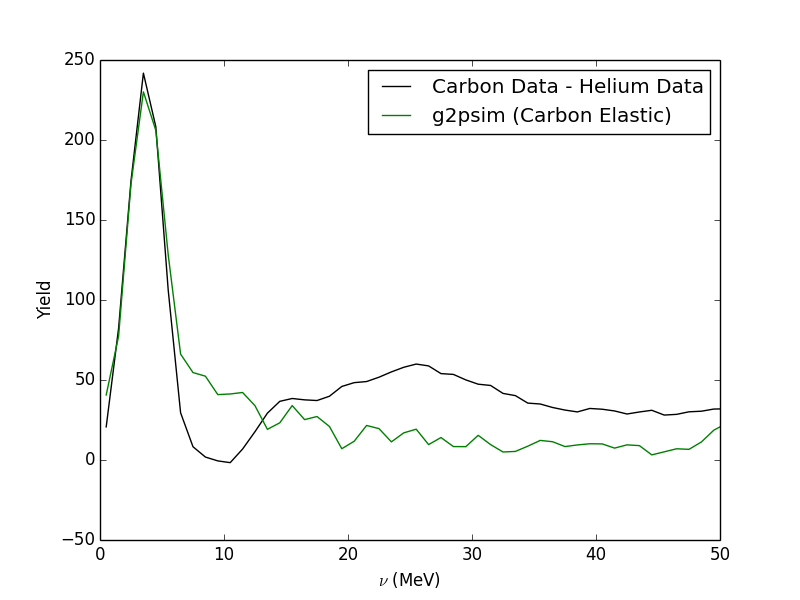} 
        \caption{Scaled Carbon Simulation \& Data Comparison} \label{fig:carboncomp22}
    \end{subfigure}
\caption{5T 2.2 GeV Transverse Carbon Comparison.}
\end{figure}

\begin{figure}[htb]

\centering
    \begin{subfigure}[t]{0.7\textwidth}
        \centering
        \includegraphics[width=\linewidth]{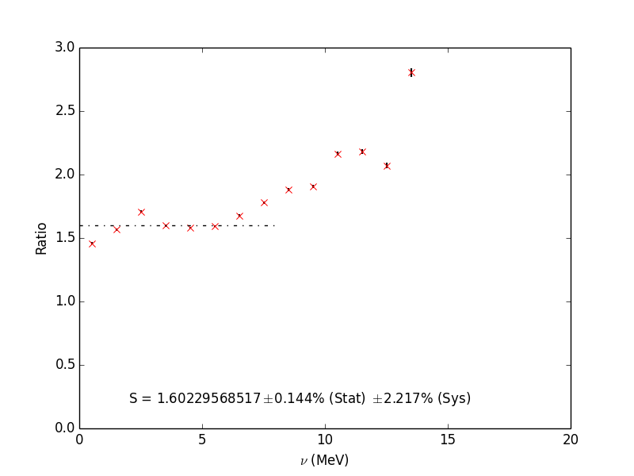} 
        \caption{Carbon Simulation/Data Ratio} \label{fig:carbonratL}
    \end{subfigure}
    \hspace{1.1em}
    \begin{subfigure}[t]{0.7\textwidth}
        \centering
        \includegraphics[width=\linewidth]{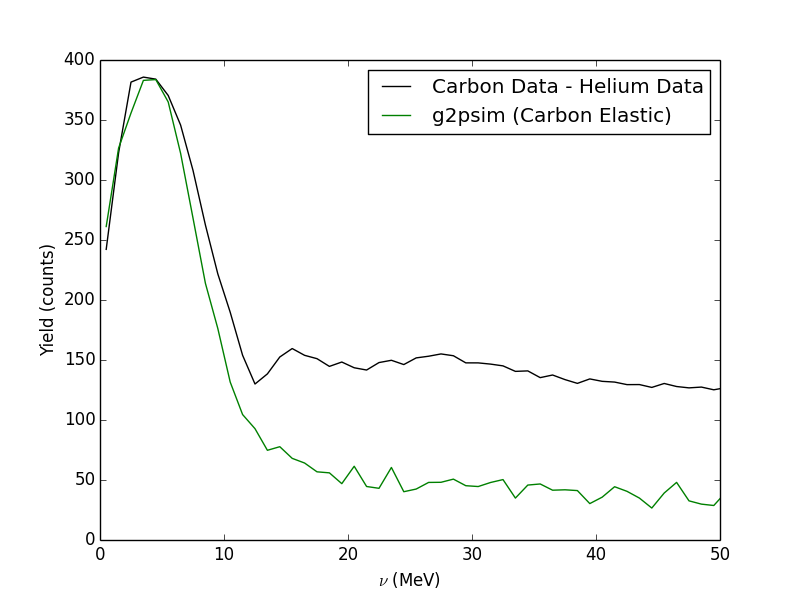} 
        \caption{Scaled Carbon Simulation \& Data Comparison} \label{fig:carboncompL}
    \end{subfigure}
\caption{5T 2.2 GeV Longitudinal Carbon Comparison.}
\end{figure}

\clearpage

Once the scaling factor and smearing input parameter are established for each setting, the simulation is used to generate nitrogen elastic yields at simulated packing fractions of 0.4 (40\%) and 0.6 (60\%). These can be compared to the data elastic yields, but first, it is necessary to try and isolate the nitrogen elastic peak in the production yields. A quasielastic response, which experiences a wide smearing due to Fermi-scale motion, and a proton elastic peak must be removed, as must any contribution from the helium in the target cell. This latter one is difficult because we don't know the packing fraction before this process, consequently, we don't know how much helium to remove from the production yield.

This issue is solved by employing an iterative procedure. An initial packing fraction of 50\%, or 0.5, is assumed, and an Empty run yield scaled by 0.5 is subtracted from the production data, leaving theoretically only nitrogen and proton data in the region of interest, if the packing fraction guess is correct. The following steps are performed to isolate the nitrogen peak, compare to the simulation, and get a final packing fraction result. This result will then be plugged back in as the new packing fraction guess, and the following steps repeated. This iterative procedure was performed up to 100 times, but the packing fraction generally converged to a constant value after around 4 iterations. 

To eliminate the quasielastic and proton elastic contribution, the elastic and quasielastic are each fit with a separate functional form. The helium elastic and nitrogen elastic peak are smeared enough that together they can be fit with a single functional form, which is a convolution of a Gaussian with a Landau tail on the high $\nu$ side. The quasielastic peak can be fit with a simple Gaussian. For the 5T 2.2GeV setting, the helium elastic peak is pronounced enough that it cannot be ignored, and separate Gaussian-Landau convolution is fit to it and subtracted. The initial fits are very low accuracy due to the strong mixing between the two peaks, so this procedure is iterated: a fit to the quasielastic peak is subtracted from the yield before doing the new fit to the elastic peak, then this elastic fit is subtracted before doing a new quasielastic fit, and so on. This particular procedure was iterated 500 times, with the resulting fits separating the peaks shown in the following Figures.

\begin{figure}[htb]
\centering
\includegraphics[width=0.8\textwidth]{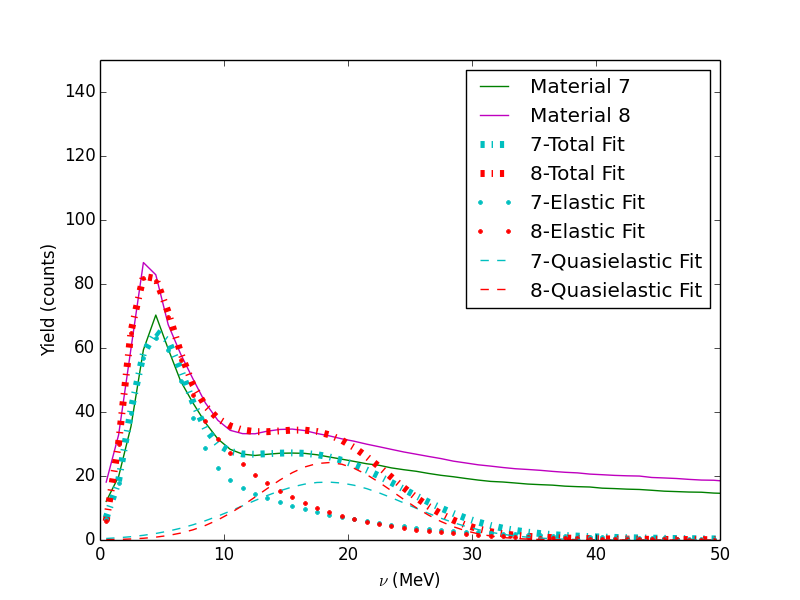}
\caption{2.5T 1.7 GeV Transverse Fits.}
\label{fig:gaussfit1}
\end{figure}

\begin{figure}[htb]
\centering
\includegraphics[width=0.8\textwidth]{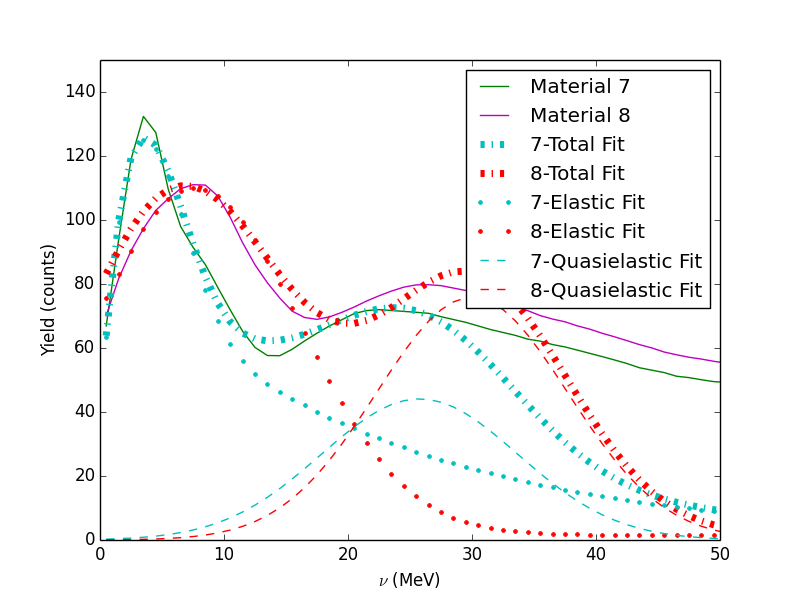}
\caption{2.5T 2.2 GeV Transverse Fits.}
\label{fig:gaussfit2}
\end{figure}

\begin{figure}[htb]
\centering
\includegraphics[width=0.8\textwidth]{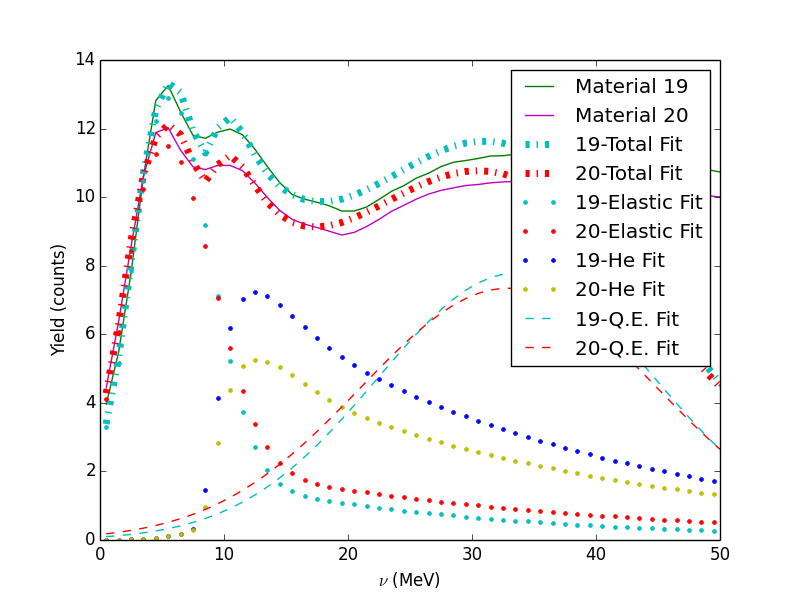}
\caption{5T 2.2 GeV Transverse Fits.}
\label{fig:gaussfit3}
\end{figure}

\begin{figure}[htb]
\centering
\includegraphics[width=0.8\textwidth]{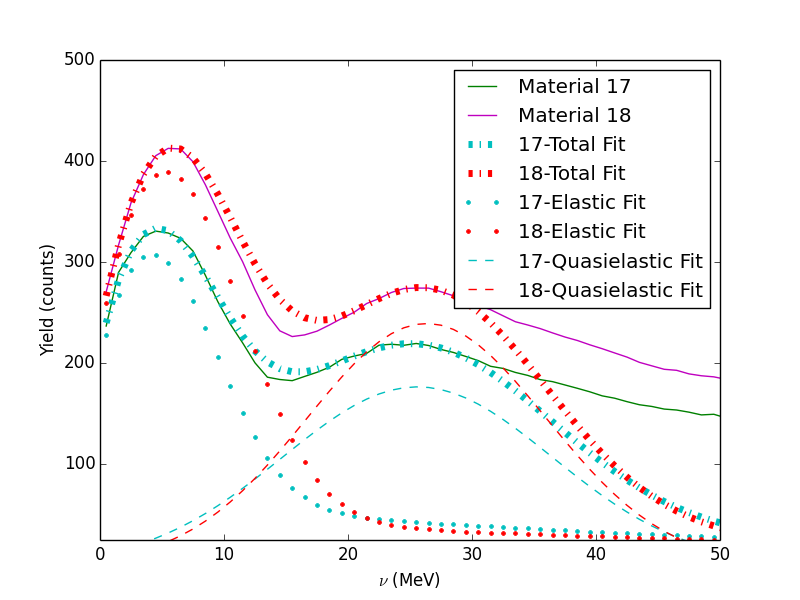}
\caption{5T 2.2 GeV Longitudinal Fits.}
\label{fig:gaussfit4}
\end{figure}
\clearpage

Once the combined quasielastic and proton elastic peak is subtracted we are left with, in theory, a pure nitrogen elastic peak, if our initial packing fraction guess was correct. However, the contributions from these subtracted peaks are the smallest near $\nu$ = 0, so we will focus on the superelastic region, or the side of the nitrogen elastic peak closest to $\nu$ = 0. In this region, we integrate the yields from any materials associated with the energy setting, as well as the yields produced with the simulation at packing fractions of 0.4 and 0.6. We can perform a linear fit through the simulated yields and then plug in the integrated nitrogen yields to find their packing fraction. For the 2.2 GeV, 2.5T, Transverse setting, the shape of the elastic peaks was very different between the two materials, due to the Septa configuration changing between the time of the relevant runs being taken. It was therefore necessary to run the simulation for each Septa configuration and produce two different linear fits to fit these materials. The comparison between data and simulation is shown below, followed by the linear fits produced for each setting. As only the elastic peak is generated for the simulations, the higher $\nu$ region displays nothing but noise that should be ignored. The comparison is performed in the superelastic region, ending at the peak of the elastic peak.

\begin{figure}[htb]
\centering
\includegraphics[width=0.8\textwidth]{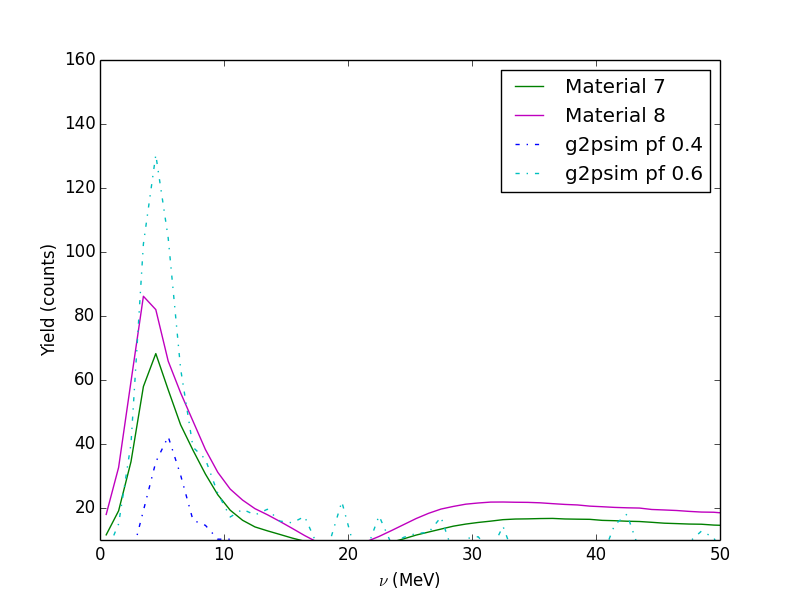}
\caption{2.5T 1.7 GeV Transverse Simulation \& Yields.}
\label{fig:pfsim1}
\end{figure}

\begin{figure}[htb]
\centering
\includegraphics[width=0.8\textwidth]{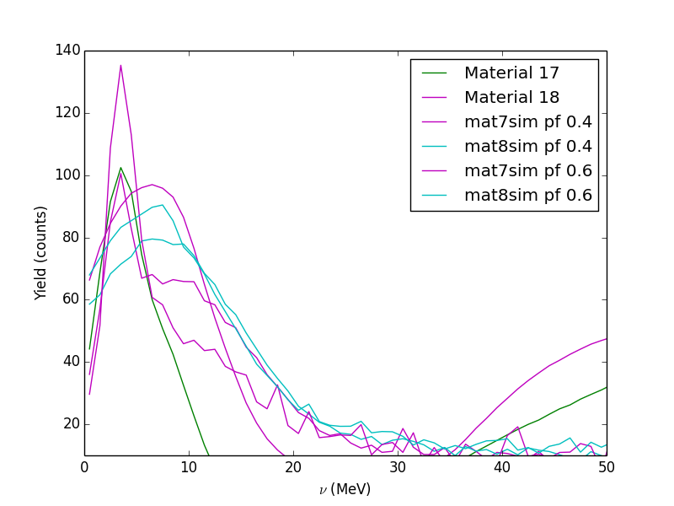}
\caption{2.5T 2.2 GeV Transverse Simulation \& Yields.}
\label{fig:pfsim2}
\end{figure}

\begin{figure}[htb]
\centering
\includegraphics[width=0.8\textwidth]{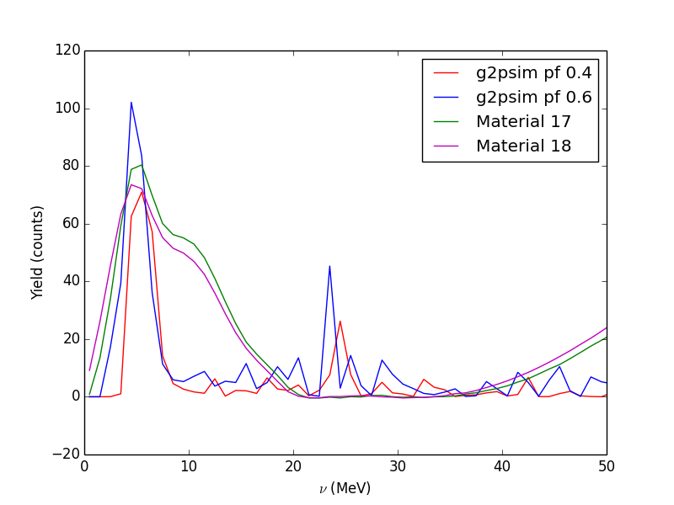}
\caption{5T 2.2 GeV Transverse Simulation \& Yields.}
\label{fig:pfsim3}
\end{figure}

\begin{figure}[htb]
\centering
\includegraphics[width=0.8\textwidth]{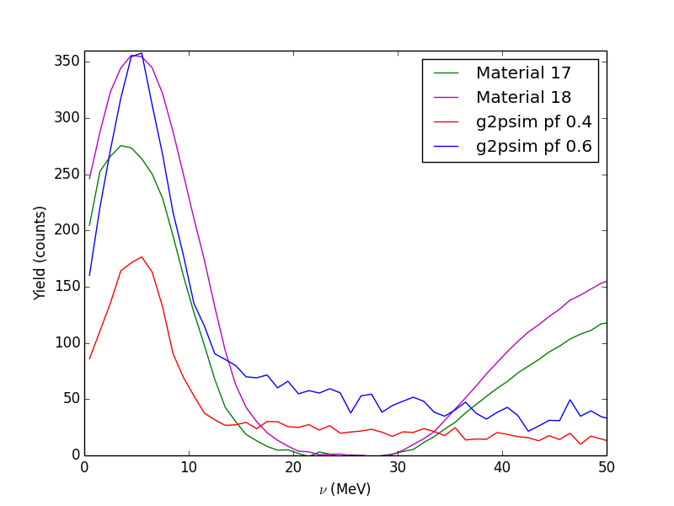}
\caption{5T 2.2 GeV Longitudinal Simulation \& Yields.}
\label{fig:pfsim4}
\end{figure}

\begin{figure}[htb]
\centering
\includegraphics[width=0.8\textwidth]{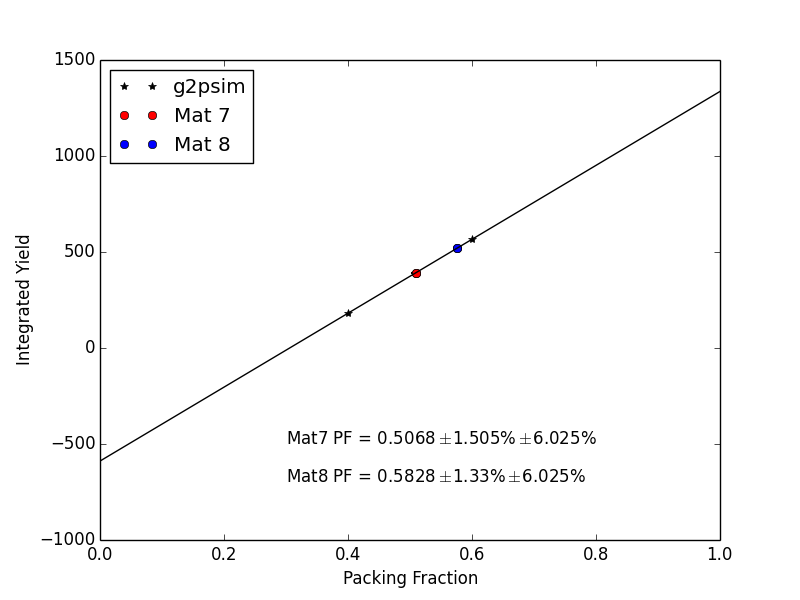}
\caption{2.5T 1.7 GeV Transverse Linear pf Fit.}
\label{fig:pf1}
\end{figure}

\begin{figure}[htb]
\centering
\includegraphics[width=0.8\textwidth]{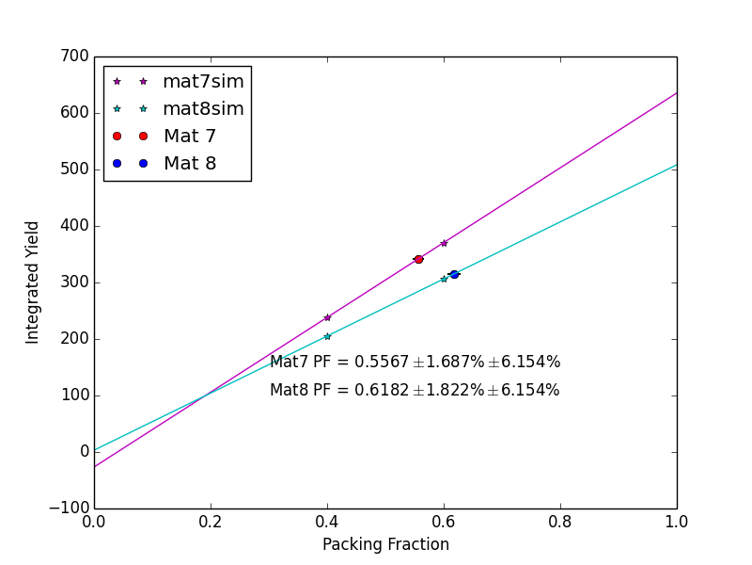}
\caption{2.5T 2.2 GeV Transverse Linear pf Fit.}
\label{fig:pf2}
\end{figure}

\begin{figure}[htb]
\centering
\includegraphics[width=0.8\textwidth]{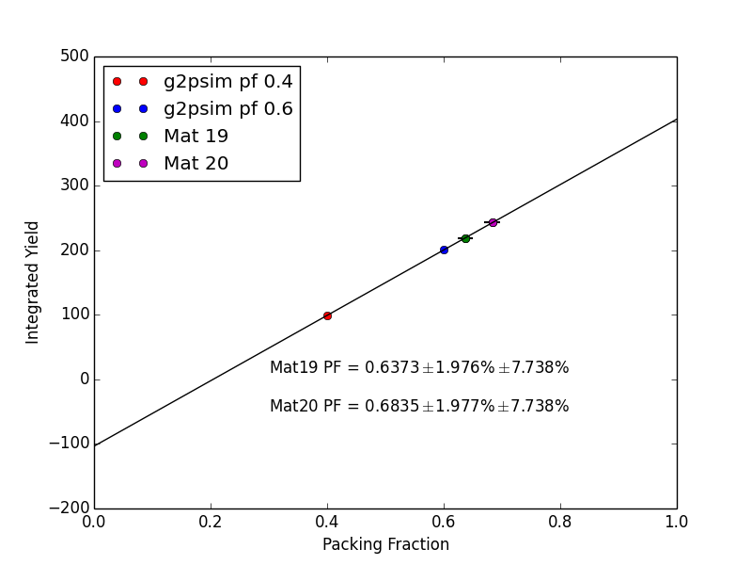}
\caption{5T 2.2 GeV Transverse Linear pf Fit.}
\label{fig:pf3}
\end{figure}

\begin{figure}[htb]
\centering
\includegraphics[width=0.8\textwidth]{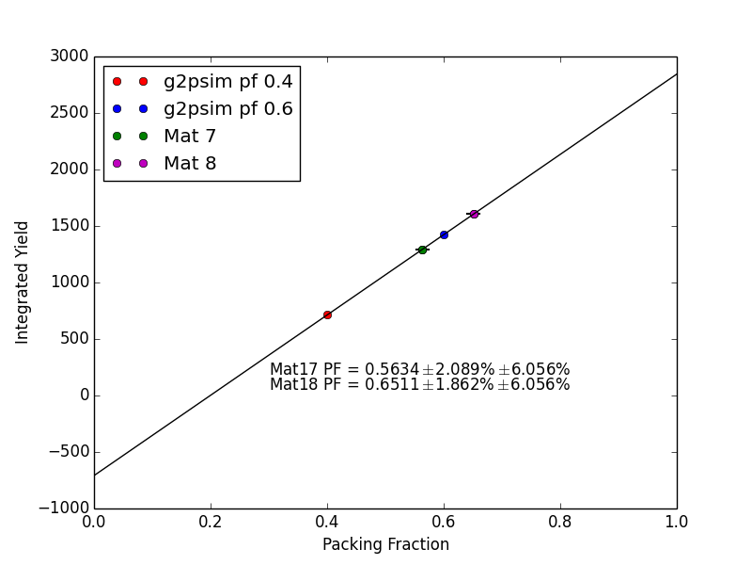}
\caption{5T 2.2 GeV Longitudinal Linear pf Fit.}
\label{fig:pf4}
\end{figure}

\clearpage

After iterating the above procedure until the initial packing fraction guess converges, the final packing fractions are presented in Table~\ref{table:packingfraction}.

\begin{table}[h!]
\centering
\begin{tabular}{|c|c|c|c|c|} 
 \hline
Setting & Material ID & Packing Fraction & Stat. Error (\%) & Syst. Error (\%) \\ [0.5ex] 
 \hline
 
 \multirow{2}{*}{2.5T 1.7GeV T} &  7& 0.51&  1.51 &  6.03 \\ 
 & 8 & 0.58 & 1.33 & 6.03 \\
 \hline
 \multirow{2}{*}{2.5T 2.2GeV T} &  7& 0.56&  1.69 &  6.15 \\ 
 & 8 & 0.62 & 1.82 & 6.15 \\
 \hline
 \multirow{2}{*}{5T 2.2GeV L} &  17& 0.56&  2.09 &  6.06 \\ 
 & 18 & 0.65 & 1.86 & 6.06 \\
 \hline
 \multirow{2}{*}{5T 2.2GeV T} &  19& 0.64&  1.98 &  7.74 \\ 
 & 20 & 0.68 & 1.98 & 7.74 \\
 \hline
 \multirow{2}{*}{5T 3.3GeV T} &  19& 0.64&  1.98 &  9.51 \\ 
 & 20 & 0.68 & 1.98 & 11.24 \\
 \hline
\end{tabular}
\caption{Final Packing Fractions for E08-027.}
\label{table:packingfraction}
\end{table}

\section{Dilutions}

To produce the experimental asymmetry we will need to calculate a dilution factor $f$. This factor is used to eliminate contributions from scattering off materials in the beam path other than the proton of interest. On a fundamental level, the dilution factor is defined as the ratio of the proton cross section to the full production cross section:

\begin{equation}
\label{eqn:dil_1}
f = \frac{\sigma_{proton}}{\sigma_{prod}}
\end{equation}

To determine this, we define the production cross section as being a linear sum of the contributions from each material it may scatter from. The leading contributors are nitrogen, due to the target material being NH$_3$, helium, due to the liquid helium used to cool the target, and aluminum, from the end-caps of the target cell. Other contributions, such as those from the NMR coils used to measure the target polarization, are small enough that they can be ignored due to the small amount of material present in the target. Taking $\sigma_{prod} = \sigma_{proton} + \sigma_N + \sigma_{He} + \sigma_{Al}$ we can rewrite the dilution factor as:

\begin{equation}
\label{eqn:dil_2}
f = 1 - \frac{\sigma_{N} + \sigma_{He} + \sigma_{Al}}{\sigma_{prod}}
\end{equation}

Because this is a ratio of cross sections, we can cancel shared factors in the cross section and instead simply compare the yields of (\ref{eqn:yield}). However, we now recall that instead of having individual yields for Nitrogen, Helium, and Aluminum, we have several types of dilution run collected during the experiment. We have Carbon runs, where the beam scatters off a carbon disk of known length. We have Empty runs, where the beam passes through an open target cell completely filled with liquid helium. And finally, we have Dummy runs, where the beam passes through a target cup identical to the production cup, but with no material in it; namely, a cup filled with helium and NMR coils, and covered by aluminum end-caps.

To express the nitrogen contribution, we use the Carbon dilution runs, since Carbon-14 and the Nitrogen-14 present in the ammonia have the same nucleon number, their cross sections will be very similar. We use the Bosted-Fersch model~\cite{N2Scale} to generate a carbon and nitrogen cross section in the correct kinematic region for each setting. The ratio of these cross sections at every given energy bin produces a scale factor we define as:
\begin{equation}
\label{eqn:cnscale}
S_{C \rightarrow N} = \frac{\sigma_N^{Bosted}}{\sigma_C^{Bosted}}
\end{equation}
In addition to this, since the cross section definition of (\ref{eqn:exp_xs}) contained a material-dependent factor of $\frac{1}{\rho \Delta Z}$, with $\rho$ being the number density of the material and $\Delta Z$ being the length of the scattering material, we will also need to include a ratio of these factors for carbon and nitrogen respectively. The length of the carbon disk is known, and the `length' of the nitrogen in the target cell can be defined as the total length of the target cell, times the packing fraction $pf$ determined in the previous section.

By using the Dummy and Empty yields together, we can actually express both the helium and aluminum contributions together. If we remove the contribution in the Dummy yield from any space that will be filled by ammonia in the production yield, we are left with a combined yield which contains all the contributions from helium scattering, aluminum scattering, and even any small contribution from the NMR coils. To do this, we can simply scale the pure helium yield represented by our Empty run by the packing fraction of the production run, to represent how much helium will be lost to the presence of the production material.

Putting all of this together, our dilution factor becomes:

\begin{equation}
\label{eqn:dil_2}
f = 1 - \frac{S_{C\rightarrow N}\frac{\rho_N (pf) Z_{tg}}{\rho_C Z_C}Y_C + \big(Y_{Dummy} - (pf) Y_{Empty}\big)}{Y_{prod}}
\end{equation}

Up until now, we have only considered scattering off of the volume of the target cell itself. But the cell is immersed in a bath of liquid helium exterior to the cell. To get a pure `nitrogen' yield, we will need to remove helium both in and out of the cell, as well as consider the total length of helium that is being measured when scaling our empty yield to subtract from our dummy yield. Defining $Z_{out}$ as the length of helium outside the cell and $Z_{tot}=Z_{tg}+Z_{out}$ as the total length of helium, we now write:

\begin{equation}
\label{eqn:dil_2}
f = 1 - \frac{S_{C\rightarrow N}\frac{\rho_N (pf) Z_{tg}}{\rho_C Z_C}\big[Y_C - (\frac{Z_{tot} - Z_C}{Z_{tot}})Y_{Empty}\big] + \big(Y_{Dummy} - pf(\frac{Z_{tg}}{Z_{tot}}) Y_{Empty}\big)}{Y_{prod}}
\end{equation}

One final length correction is now still needed. The radiation length seen by the beam is different for each type of dilution run. The radiation thicknesses for each type of run are listed in ~\cite{Zielinski:2017gwp}. In other words, a different amount of energy is lost before scattering depending on which types of materials the beam must pass through. To solve this, we employ the same Bosted-Fersch model to create two more scaling factors. $S_{He(E)\rightarrow He(C)}$ is a ratio of the model cross section for helium with the radiation length of the Empty run, versus the radiation length of the Carbon run. $S_{He(D)\rightarrow He(P)}$ is similar, but instead scales two helium cross sections for the radiation length of the dummy run and the production run. $S_{C\rightarrow N}$ also includes the appropriate radiation lengths for both cross sections such that both terms end with the effective radiation length of the production target. This gives us a final dilution factor of:

\begin{equation}
\label{eqn:dil_final}
f = 1 - \frac{S_{C\rightarrow N}\frac{\rho_N (pf) Z_{tg}}{\rho_C Z_C}\big[Y_C - (\frac{Z_{tot} - Z_C}{Z_{tot}})S_{He(E)\rightarrow He(C)}Y_{Empty}\big] + S_{He(D)\rightarrow He(P)}\big(Y_{Dummy} - pf(\frac{Z_{tg}}{Z_{tot}}) Y_{Empty}\big)}{Y_{prod}}
\end{equation}

The yields employed in the dilution are shown in the following figures. These yields had a number of issues which complicated the analysis. First and foremost, the momentum acceptance of the spectrometers resulted in data near the edge of each momentum setting being somewhat distorted. This distortion was internally consistent, meaning that for the asymmetries of the following section, this effect mostly cancels out entirely, since the asymmetry is a ratio of yields taken from the same momentum setting. However, each dilution run had slightly different momentum settings from each other, and from the production runs. This means that the distortion was actually in many cases magnified by only being present on one side of the ratio. 

To deal with this, the yields used in producing the dilution had an extremely tight momentum cut applied, limiting the measured momentum to a window of -3.2\%$<$dp$<$1.0\%. This eliminates most of the badly distorted data but also leaves large gaps in the momentum coverage of the yield. These gaps are equally troublesome as they appear in different places for each dilution run type. To deal with this, the Bosted-Christy model~\cite{Bosted3} was used to create yields for each material. These yields, without proper normalization, vary greatly from the data normalization. For each gap, the model result was used internal to the gap, but first, had a linear scaling applied to it that would fix the beginning of the model in the gap to the last data point before the gap, and the end of the model to the first data point after the gap. This was done simply by doing a linear fit to the points on either side of the gap, fixing the first point of the model to match the data at the same $Q^2$, and then applying the linear fit to the model inside the gap region.

Another issue that complicated the dilution analysis was the presence of yield drifts due to a BPM miscalibration~\cite{JieBPM}. These yield drifts are a known factor in the analysis, but they do not contribute to the systematic of the asymmetry because, as for the distortion above, they cancel out due to the ratio nature of the asymmetry. For the dilution analysis, the settings with a known BPM miscalibration discussed in \cite{JieBPM} were simply shifted linearly up or down until the first data point for each setting aligned with the last data point from nearby settings with no known BPM issues. Because this method is not necessarily trust-able and relies on the assumption that the real yield will be smooth, we apply an increased systematic to these yields of magnitude equal to the applied shift.

Additionally, for the 1.7 GeV, 2.5T setting, there is a mismatch between the Septa and dipole current caused by the Septa issues discussed previously. The ratio nature of the asymmetry again cancels this issue as discussed in \cite{Zielinski:2017gwp}, but like the others, it contributes strongly to the dilution analysis. Without these runs, the overall statistical error bar is more than doubled, but including them distorts one region of the dilution tremendously. To avoid this, the Bosted model is used to produce all components of the dilution in the distorted region, and the model dilution is employed for that section of the final result. A systematic contribution must be added from this, equal to the difference between the model and distorted dilution. A model dilution is also used for the 3.3 GeV setting in many places, as a significantly reduced range of dilution data was collected.

\begin{figure}[htb]

\centering
    \begin{subfigure}[t]{0.45\textwidth}
        \centering
        \includegraphics[width=\linewidth]{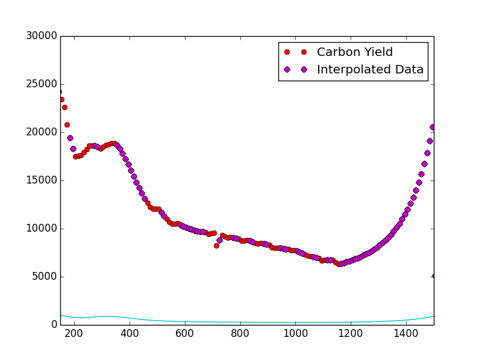} 
        \caption{Carbon} 
    \end{subfigure}
    \hspace{1.1em}
    \begin{subfigure}[t]{0.45\textwidth}
        \centering
        \includegraphics[width=\linewidth]{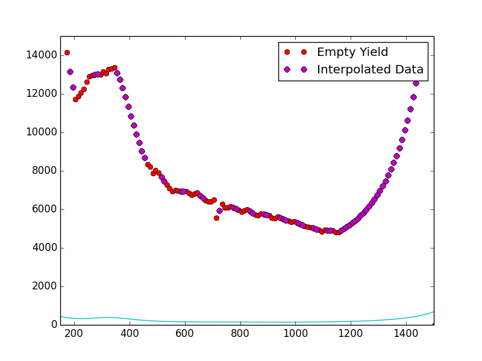} 
        \caption{Empty} 
    \end{subfigure}
    \begin{subfigure}[t]{0.45\textwidth}
        \centering
        \includegraphics[width=\linewidth]{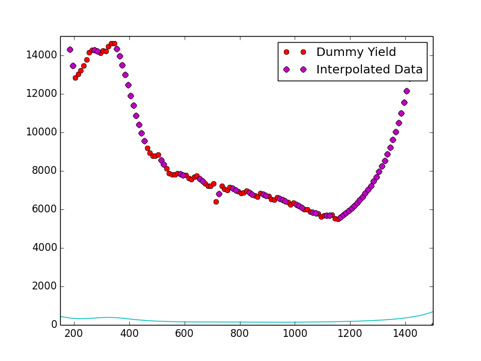} 
        \caption{Dummy} 
    \end{subfigure}
\caption{2.5T 1.7 GeV Transverse Dilution Yields.}
\end{figure}

\begin{figure}[htb]

\centering
    \begin{subfigure}[t]{0.45\textwidth}
        \centering
        \includegraphics[width=\linewidth]{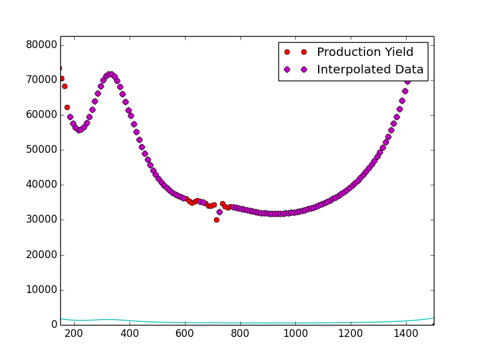} 
        \caption{Material 7} 
    \end{subfigure}
    \hspace{1.1em}
    \begin{subfigure}[t]{0.45\textwidth}
        \centering
        \includegraphics[width=\linewidth]{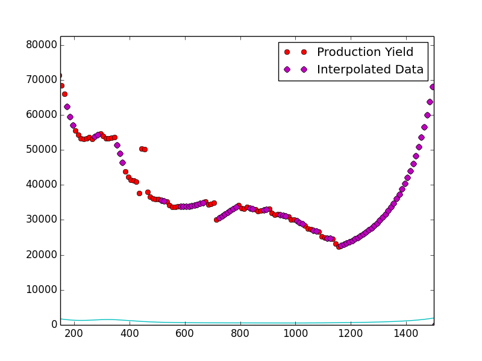} 
        \caption{Material 8} 
    \end{subfigure}
\caption{2.5T 1.7 GeV Transverse Production Yields.}
\end{figure}

\begin{figure}[htb]

\centering
    \begin{subfigure}[t]{0.45\textwidth}
        \centering
        \includegraphics[width=\linewidth]{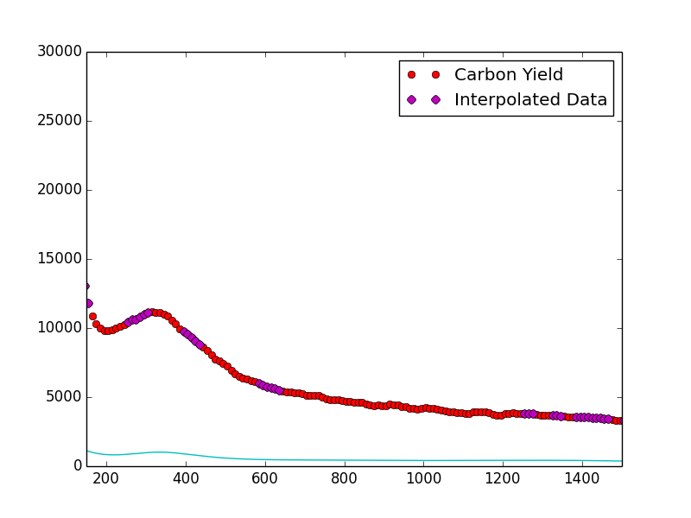} 
        \caption{Carbon} 
    \end{subfigure}
    \hspace{1.1em}
    \begin{subfigure}[t]{0.45\textwidth}
        \centering
        \includegraphics[width=\linewidth]{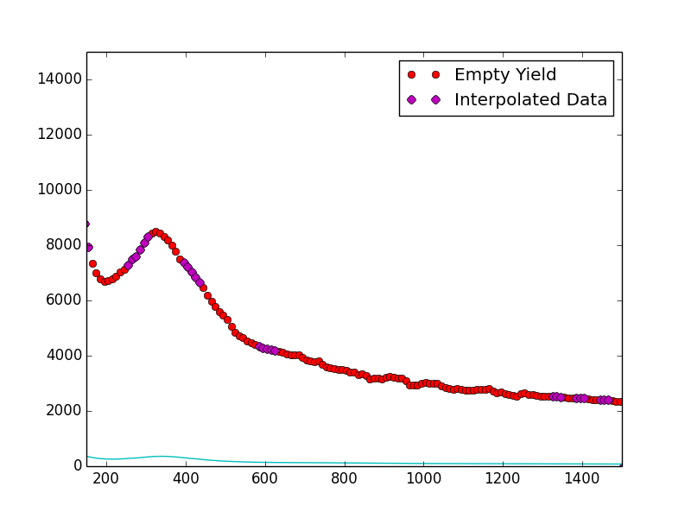} 
        \caption{Empty} 
    \end{subfigure}
    \begin{subfigure}[t]{0.45\textwidth}
        \centering
        \includegraphics[width=\linewidth]{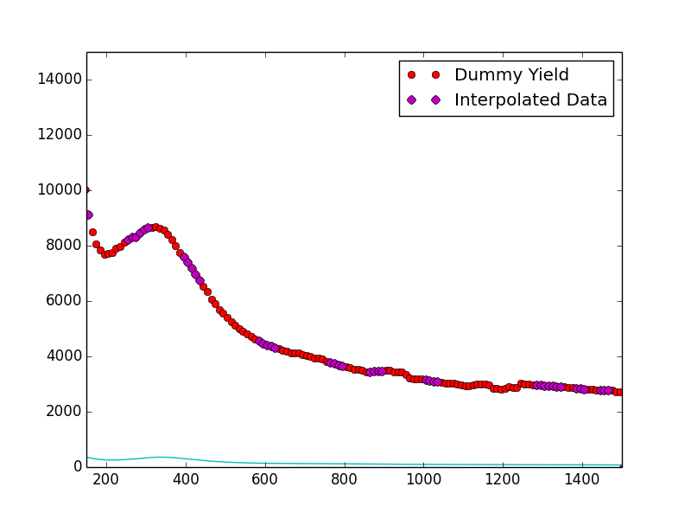} 
        \caption{Dummy} 
    \end{subfigure}
\caption{2.5T 2.2 GeV Transverse Dilution Yields.}
\end{figure}

\begin{figure}[htb]

\centering
    \begin{subfigure}[t]{0.45\textwidth}
        \centering
        \includegraphics[width=\linewidth]{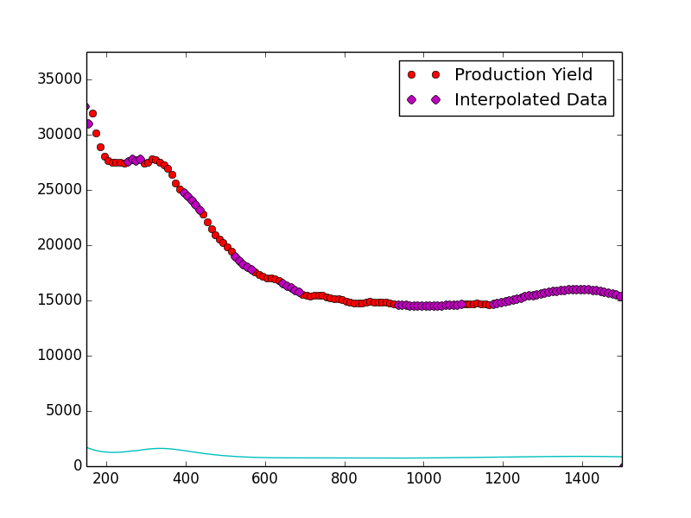} 
        \caption{Material 7} 
    \end{subfigure}
    \hspace{1.1em}
    \begin{subfigure}[t]{0.45\textwidth}
        \centering
        \includegraphics[width=\linewidth]{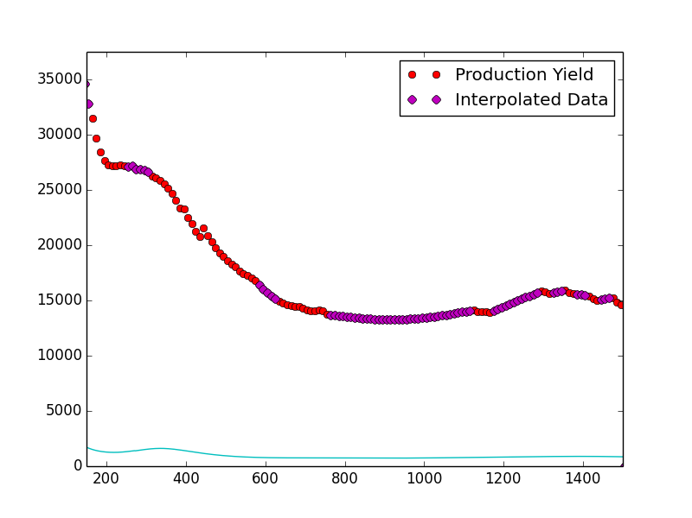} 
        \caption{Material 8} 
    \end{subfigure}
\caption{2.5T 2.2 GeV Transverse Production Yields.}
\end{figure}

\begin{figure}[htb]

\centering
    \begin{subfigure}[t]{0.45\textwidth}
        \centering
        \includegraphics[width=\linewidth]{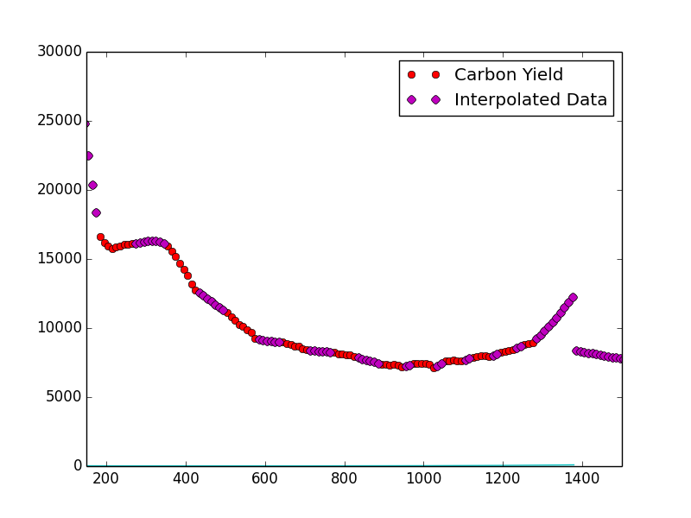} 
        \caption{Carbon} 
    \end{subfigure}
    \hspace{1.1em}
    \begin{subfigure}[t]{0.45\textwidth}
        \centering
        \includegraphics[width=\linewidth]{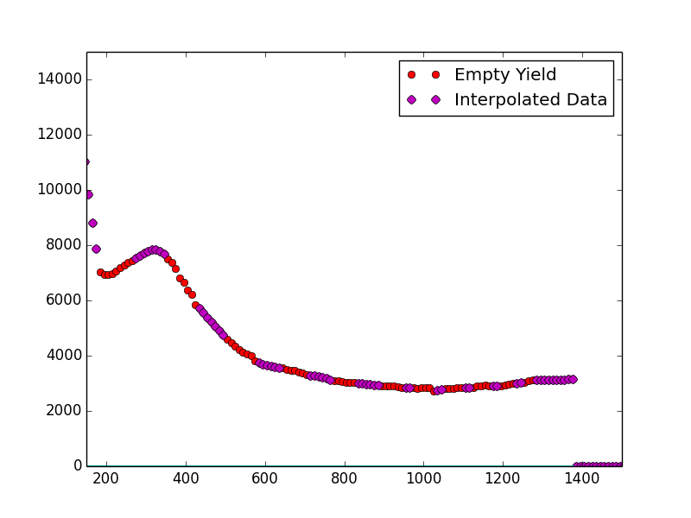} 
        \caption{Empty} 
    \end{subfigure}
    \begin{subfigure}[t]{0.45\textwidth}
        \centering
        \includegraphics[width=\linewidth]{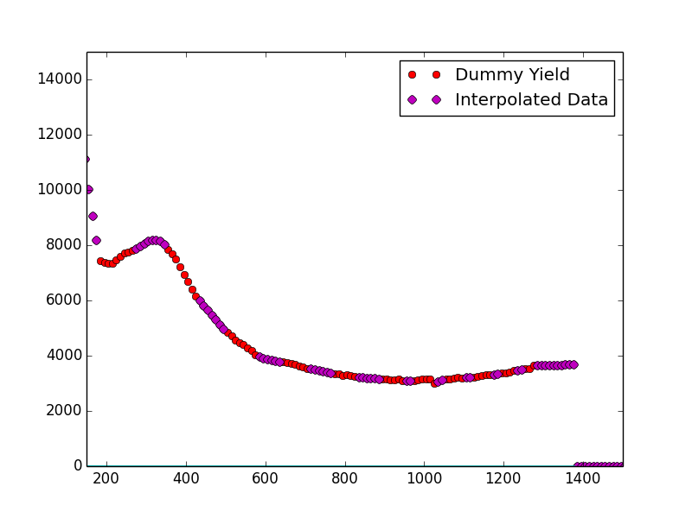} 
        \caption{Dummy} 
    \end{subfigure}
\caption{5T 2.2 GeV Longitudinal Dilution Yields.}
\end{figure}

\begin{figure}[htb]

\centering
    \begin{subfigure}[t]{0.45\textwidth}
        \centering
        \includegraphics[width=\linewidth]{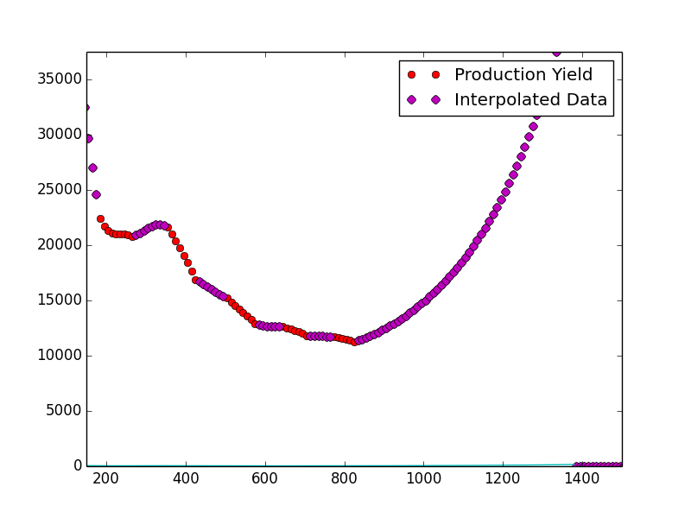} 
        \caption{Material 17} 
    \end{subfigure}
    \hspace{1.1em}
    \begin{subfigure}[t]{0.45\textwidth}
        \centering
        \includegraphics[width=\linewidth]{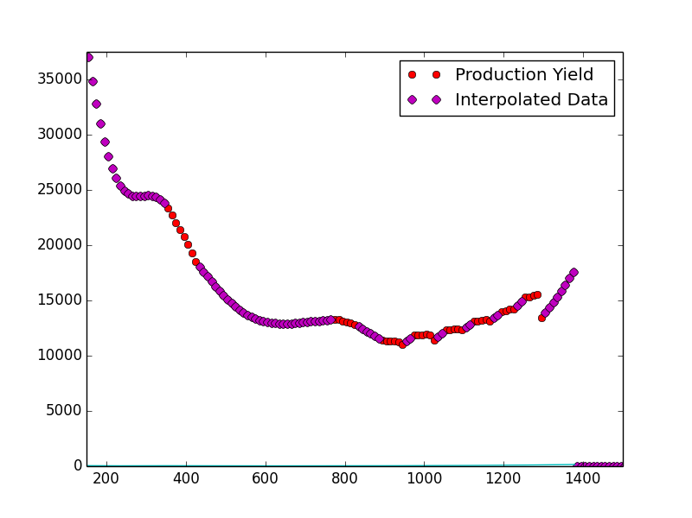} 
        \caption{Material 18} 
    \end{subfigure}
\caption{5T 2.2 GeV Longitudinal Production Yields.}
\end{figure}

\begin{figure}[htb]

\centering
    \begin{subfigure}[t]{0.45\textwidth}
        \centering
        \includegraphics[width=\linewidth]{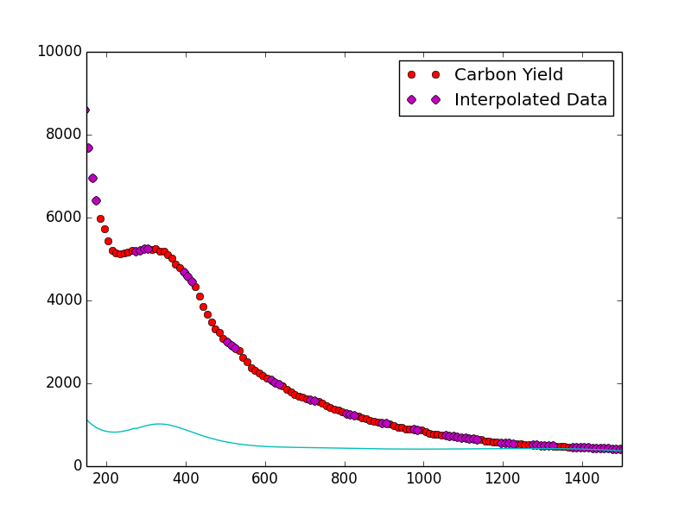} 
        \caption{Carbon} 
    \end{subfigure}
    \hspace{1.1em}
    \begin{subfigure}[t]{0.45\textwidth}
        \centering
        \includegraphics[width=\linewidth]{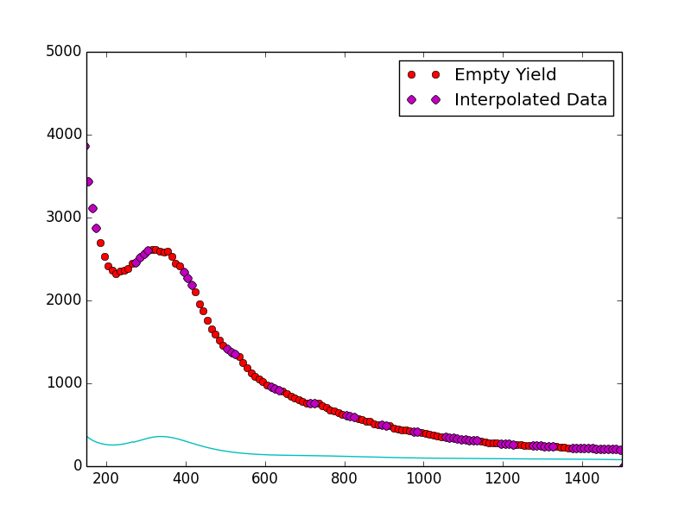} 
        \caption{Empty} 
    \end{subfigure}
    \begin{subfigure}[t]{0.45\textwidth}
        \centering
        \includegraphics[width=\linewidth]{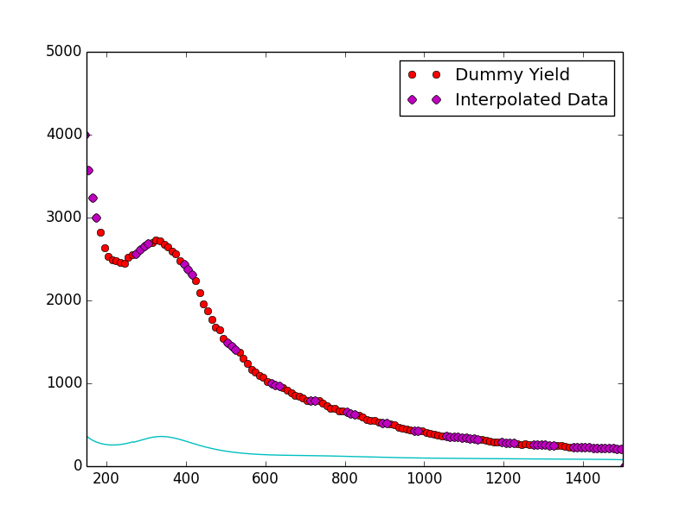} 
        \caption{Dummy} 
    \end{subfigure}
\caption{5T 2.2 GeV Transverse Dilution Yields.}
\end{figure}

\begin{figure}[htb]

\centering
    \begin{subfigure}[t]{0.45\textwidth}
        \centering
        \includegraphics[width=\linewidth]{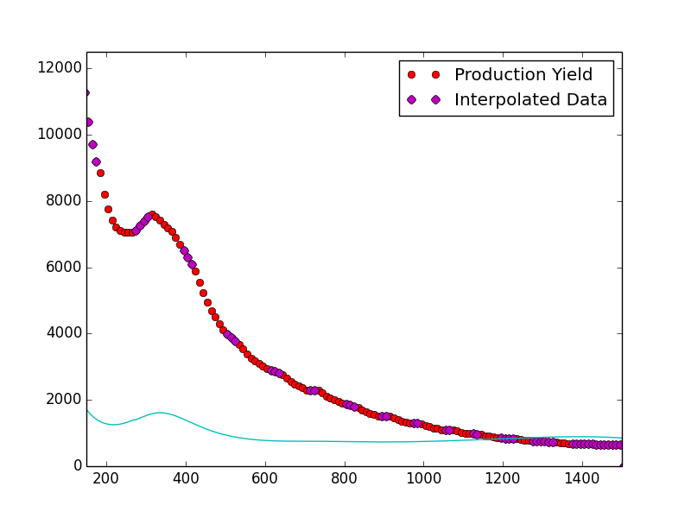} 
        \caption{Material 19} 
    \end{subfigure}
    \hspace{1.1em}
    \begin{subfigure}[t]{0.45\textwidth}
        \centering
        \includegraphics[width=\linewidth]{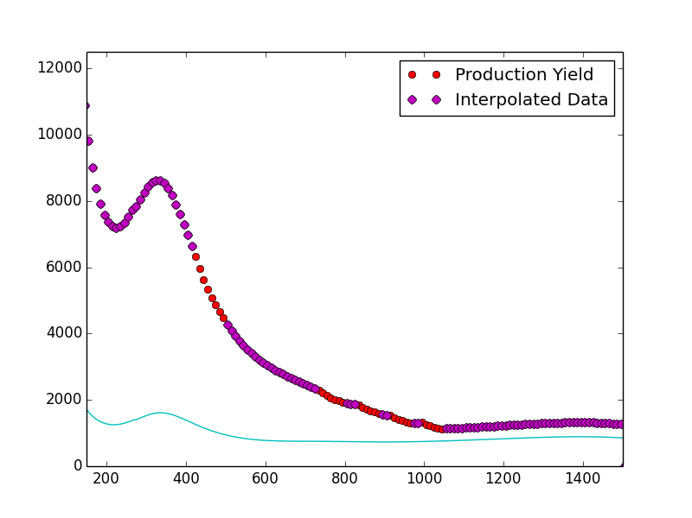} 
        \caption{Material 20} 
    \end{subfigure}
\caption{5T 2.2 GeV Transverse Production Yields.}
\end{figure}

\begin{figure}[htb]

\centering
    \begin{subfigure}[t]{0.45\textwidth}
        \centering
        \includegraphics[width=\linewidth]{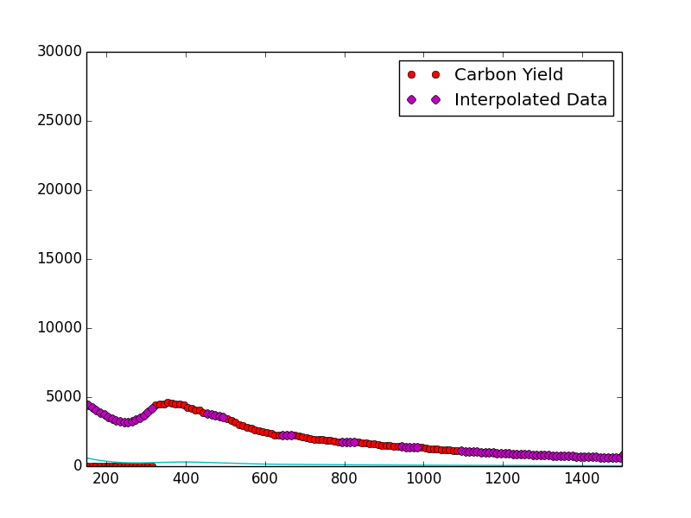} 
        \caption{Carbon} 
    \end{subfigure}
    \hspace{1.1em}
    \begin{subfigure}[t]{0.45\textwidth}
        \centering
        \includegraphics[width=\linewidth]{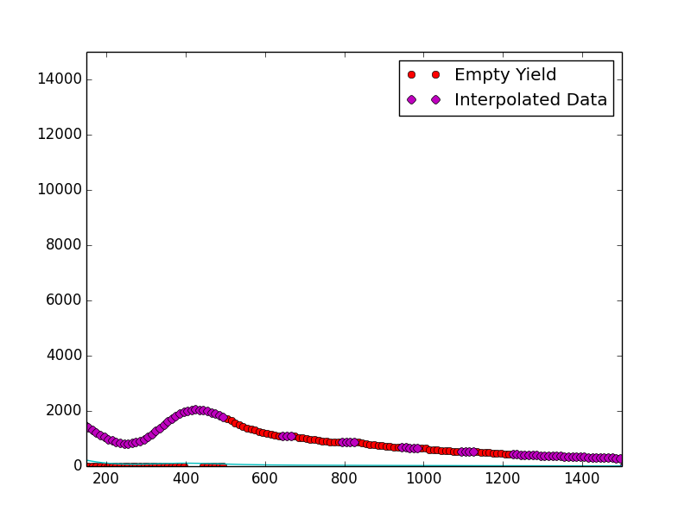} 
        \caption{Empty} 
    \end{subfigure}
    \begin{subfigure}[t]{0.45\textwidth}
        \centering
        \includegraphics[width=\linewidth]{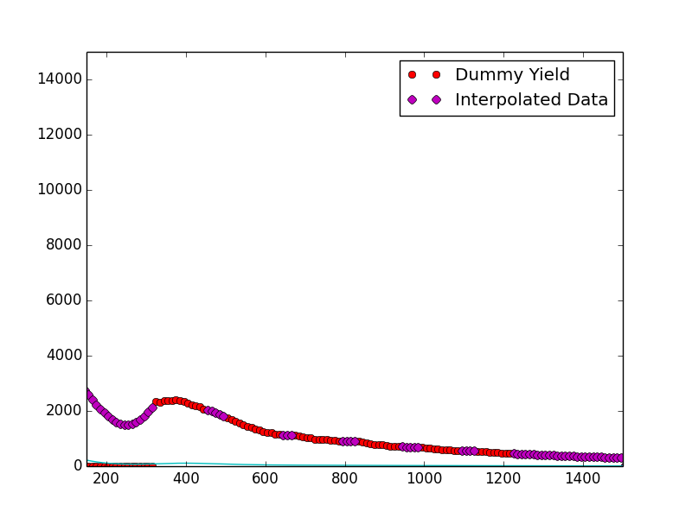} 
        \caption{Dummy} 
    \end{subfigure}
\caption{5T 3.3 GeV Transverse Dilution Yields.}
\end{figure}

\begin{figure}[htb]

\centering
    \begin{subfigure}[t]{0.45\textwidth}
        \centering
        \includegraphics[width=\linewidth]{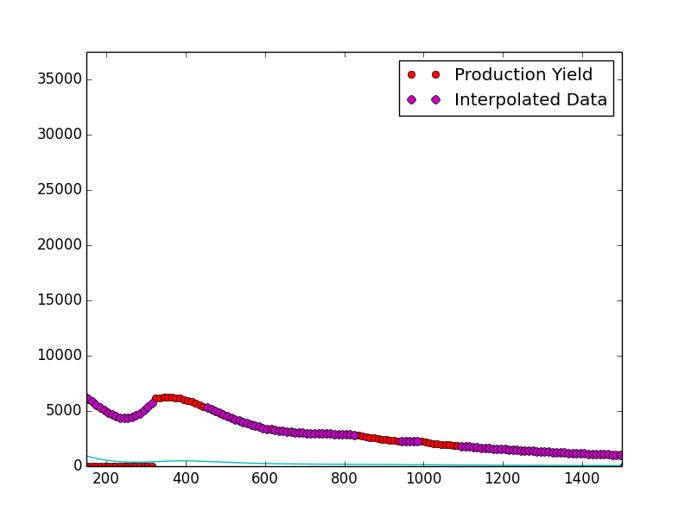} 
        \caption{Material 19} 
    \end{subfigure}
    \hspace{1.1em}
    \begin{subfigure}[t]{0.45\textwidth}
        \centering
        \includegraphics[width=\linewidth]{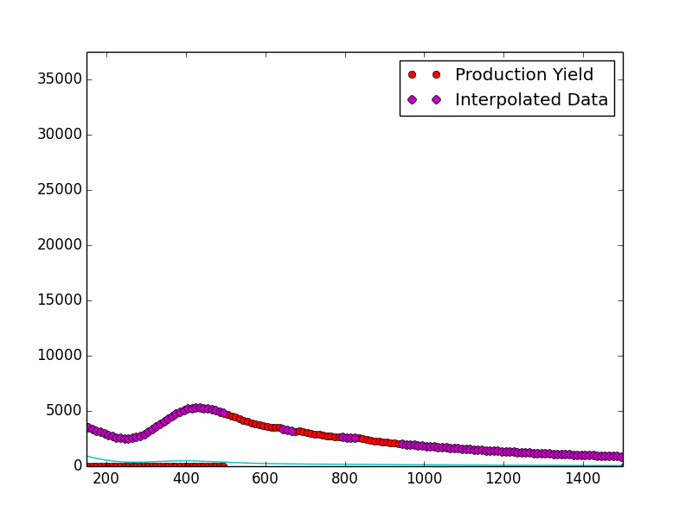} 
        \caption{Material 20} 
    \end{subfigure}
\caption{5T 3.3 GeV Transverse Production Yields.}
\end{figure}
\clearpage

Considering these corrected yields and dilution formalism established above, the final dilutions are presented in the following figures. Note that the final dilutions still suffer from visible jumps and edge effects, even with a significant amount of work to correct these. To fully correct these would require a fix for the BPM calibration issues and full and accurate modeling of the detector acceptance. Without a solution for these issues, the dilution must be used as-is, and consequently contributes one of the leading systematics to the final result. However, these effects become less notable in a larger binning, as they tend to cancel each other when enough data is averaged. A larger 30 MeV binning is used for the 5T asymmetries, and an even larger binning which scales up to 50 MeV is used for the 2.5T asymmetries.

\begin{figure}[htb]

\centering
    \begin{subfigure}[t]{0.7\textwidth}
        \centering
        \includegraphics[width=\linewidth]{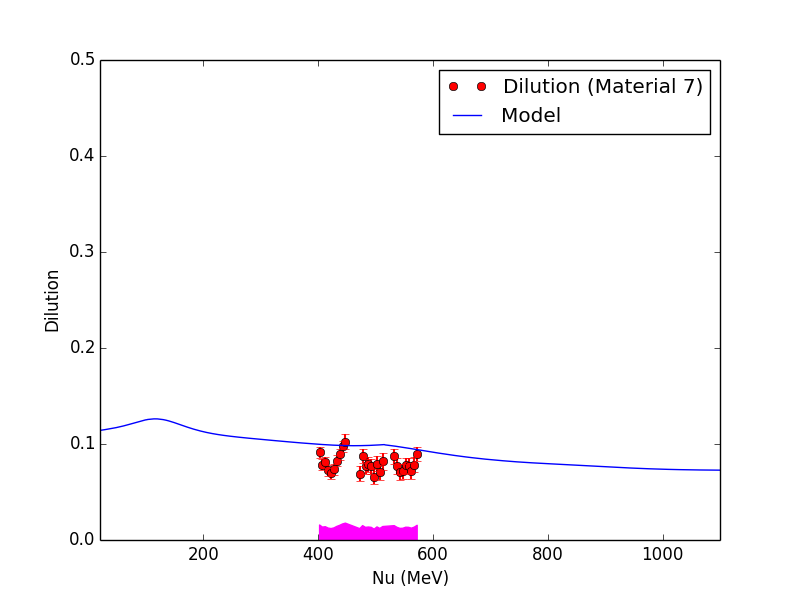} 
        \caption{Material 7} 
    \end{subfigure}
    \hspace{1.1em}
    \begin{subfigure}[t]{0.7\textwidth}
        \centering
        \includegraphics[width=\linewidth]{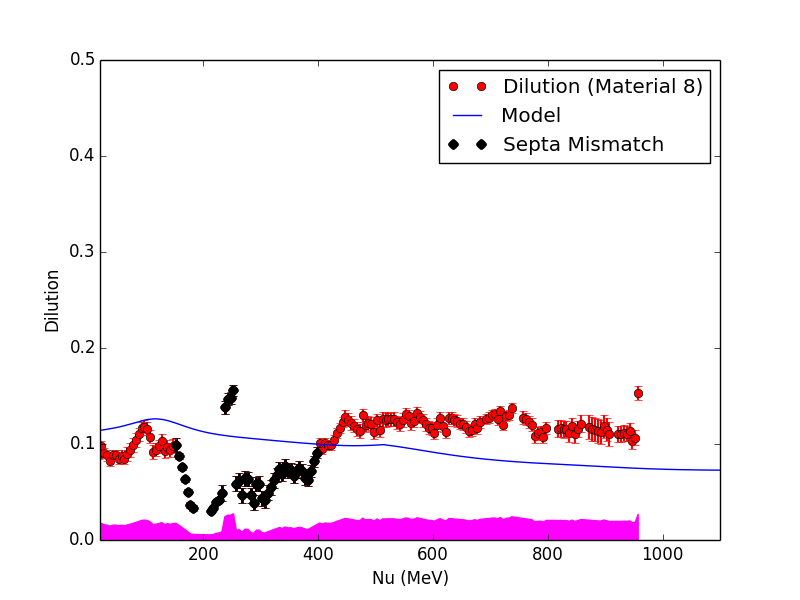} 
        \caption{Material 8} 
    \end{subfigure}
\caption{2.5T 1.7 GeV Transverse Dilutions.}
\end{figure}

\begin{figure}[htb]

\centering
    \begin{subfigure}[t]{0.7\textwidth}
        \centering
        \includegraphics[width=\linewidth]{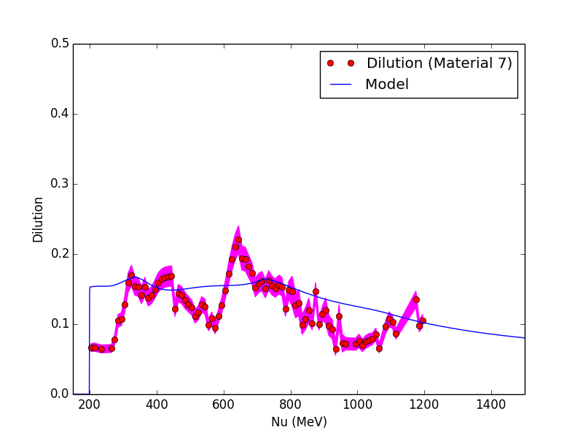} 
        \caption{Material 7} 
    \end{subfigure}
    \hspace{1.1em}
    \begin{subfigure}[t]{0.7\textwidth}
        \centering
        \includegraphics[width=\linewidth]{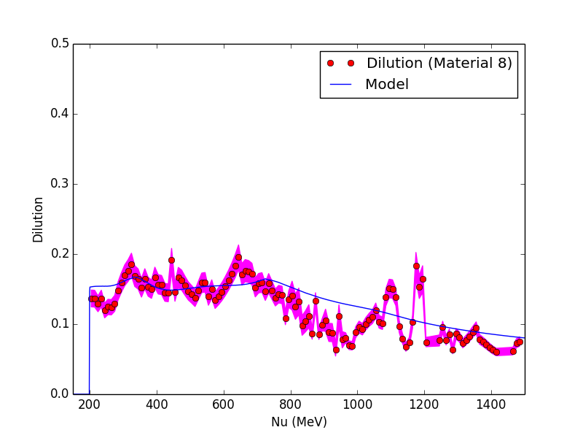} 
        \caption{Material 8} 
    \end{subfigure}
\caption{2.5T 2.2 GeV Transverse Dilutions.}
\end{figure}

\begin{figure}[htb]

\centering
    \begin{subfigure}[t]{0.7\textwidth}
        \centering
        \includegraphics[width=\linewidth]{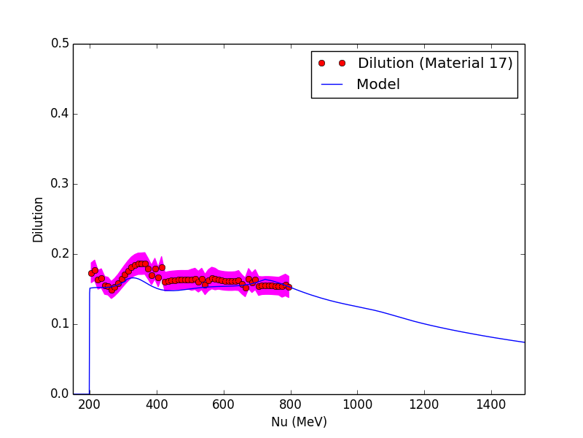} 
        \caption{Material 17} 
    \end{subfigure}
    \hspace{1.1em}
    \begin{subfigure}[t]{0.7\textwidth}
        \centering
        \includegraphics[width=\linewidth]{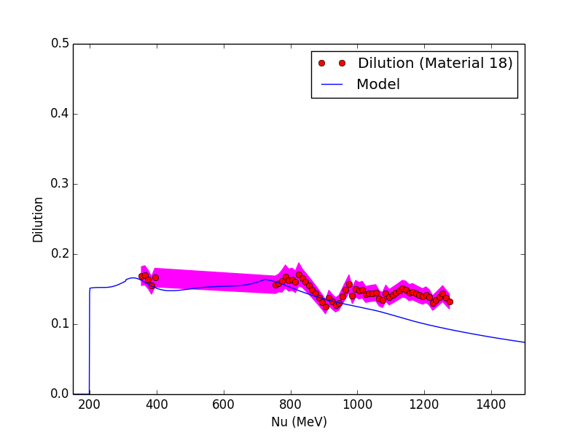} 
        \caption{Material 18} 
    \end{subfigure}
\caption{5T 2.2 GeV Longitudinal Dilutions.}
\end{figure}

\begin{figure}[htb]

\centering
    \begin{subfigure}[t]{0.7\textwidth}
        \centering
        \includegraphics[width=\linewidth]{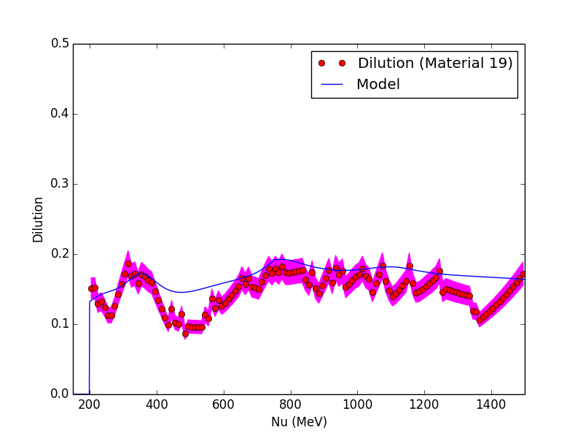} 
        \caption{Material 19} 
    \end{subfigure}
    \hspace{1.1em}
    \begin{subfigure}[t]{0.7\textwidth}
        \centering
        \includegraphics[width=\linewidth]{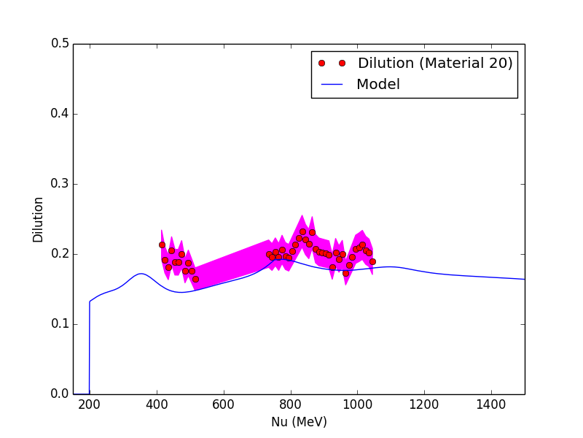} 
        \caption{Material 20} 
    \end{subfigure}
\caption{5T 2.2 GeV Transverse Dilutions.}
\end{figure}

\begin{figure}[htb]

\centering
    \begin{subfigure}[t]{0.7\textwidth}
        \centering
        \includegraphics[width=\linewidth]{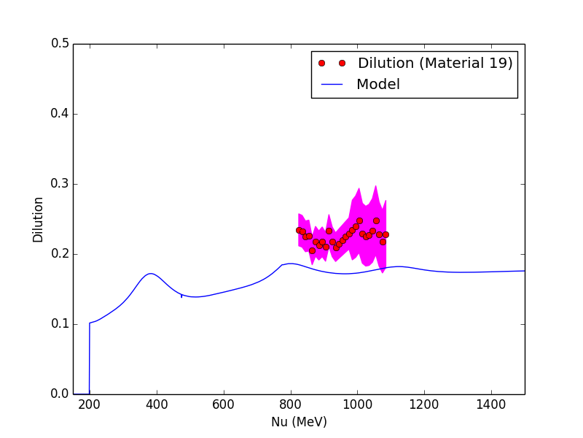} 
        \caption{Material 19} 
    \end{subfigure}
    \hspace{1.1em}
    \begin{subfigure}[t]{0.7\textwidth}
        \centering
        \includegraphics[width=\linewidth]{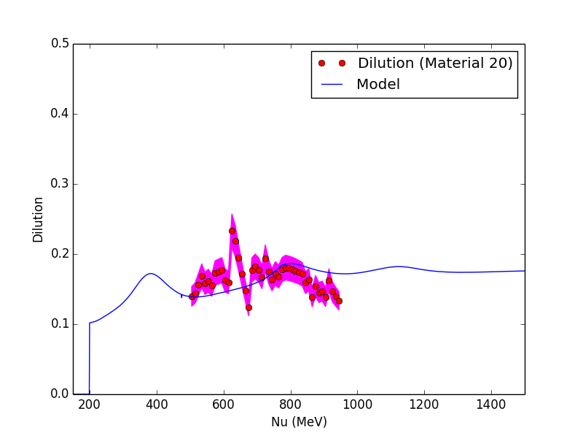} 
        \caption{Material 20} 
    \end{subfigure}
\caption{5T 3.3 GeV Transverse Dilutions.}
\end{figure}

\clearpage

\section{Asymmetries}

Now that we have the dilution, we can calculate the experimental asymmetry necessary to get the polarized cross section differences. As mentioned in the previous section's discussion, the measured asymmetry is defined as a ratio of cross sections, comparing measured electrons with forward (+) helicity, and those with backward (-) helicity:
\begin{equation}
\label{eqn:asym}
A = \frac{\sigma_+ - \sigma_-}{\sigma_+ + \sigma_-} = \frac{Y_+ - Y_-}{Y_+ + Y_-}
\end{equation}

where

\begin{equation}
\label{eqn:helyield}
Y_{\pm} = \frac{N_{\pm}}{LT_{\pm} Q_{\pm}}
\end{equation}

Where the yield is formed from the measured counts N, livetime L, and deposited charge Q for forward and backward helicity electrons.

The final measured asymmetry is generated by combining results at similar energies into a single energy `bin' through a statistically weighted average. The statistical error is given to first order by Poisson statistics, where
\begin{equation}
\label{eqn:poissonerror}
\delta N = \frac{1}{\sqrt{N}}
\end{equation}
with N equal to the total number of measured events. Because of the prescale factor mentioned earlier, not every event which occurs is collected in the final results, so it is necessary to introduce a second order correction factor S defined as~\cite{Zielinski:2017gwp}:
\begin{equation}
\label{eqn:poissoncorrection}
S = \sqrt{1 - LT\frac{N_{accepted}}{N_{total}}(1-\frac{1}{ps})}
\end{equation}
Here, LT is the livetime of the detector, ps is the prescale factor, and $N_{accepted}$ and $N_{total}$ are the total number of measured counts and the fraction accepted by the DAQ respectively. The final statistical error for each bin can then be written as:
\begin{equation}
\label{eqn:asymstat}
\delta A = \frac{1}{2}\sqrt{\frac{S_+^2}{N_+}+\frac{S_-^2}{N_-}}
\end{equation}

We combine the asymmetry from different runs in a statistically weighted average to find the asymmetry for each bin:

\begin{equation}
\label{eqn:asym_weighted_err}
\Delta A_{meas} = \sqrt{\frac{1}{\sum_i \frac{1}{\delta A^{2}_i}}}
\end{equation}
\begin{equation}
\label{eqn:asym_weighted}
A_{meas} = (\Delta A_{meas})^2 \sum_i \frac{A_i}{\delta A_i^2}
\end{equation}

The final sign of the asymmetry depends on whether the insertable half-waveplate is in (+) or out (-) of the beamline, whether the target has positive (+) or negative (-) polarization, and for the transverse settings, whether the data was measured in the left HRS (+) or right HRS(-)~\cite{Zielinski:2017gwp}.  Combining all of these gives the final sign of the asymmetry. 

A detailed description of the analysis cuts performed on the asymmetry can be found in~\cite{Zielinski:2017gwp}, nothing has changed from the description in that work for the cuts on the final analysis. Because the asymmetry is a ratio quantity, much of the necessary correction to the acceptance is canceled out, allowing the use of a wider cut in phase space and the inclusion of more data. In general, the employed cuts are:
\begin{itemize}
    \item -4\% $<$ dp $<$ 4\%
    \item -0.04 $<$ $\phi_{target}$ $<$ 0.04
    \item -0.04 $<$ $\theta_{target}^{\perp}$ $<$ 0.08
    \item -0.06 $<$ $\theta_{target}^{\parallel}$ $<$ 0.06
\end{itemize}
Where dp is the measured momentum in terms of the center of the current momentum setting.

To reduce the statistics, the asymmetry results from both the LHRS and RHRS are combined. To prove that they are statistically compatible, a reduced $\chi^2$ test is performed. For each HRS, the asymmetry is compared to a set of random fluctuations within the statistical error bar of the asymmetry, iterated 10,000 times to find a reduced $\chi^2$ of around 1.3. The LHRS And RHRS results are then compared to each other. For all settings, it was found that the resulting $\chi^2$ was less than the statistically random result of 1.3, tending to an average $\chi^2$ of around 1.1. This is taken to imply that the results are statistically compatible, and they are combined to form a final asymmetry.

By using (\ref{eqn:asymmetry_meas}) with the dilution factor calculated in the previous section, and the measured beam and target polarizations, we can get the experimental asymmetry from these total combined asymmetries. Due to the reduced 2.5T target polarization discussed earlier, the statistics on the 2.5T settings are much larger, and consequently, a larger binning is taken to obtain similar statistical error bars, including more data in each energy bin. The final asymmetries are shown in the following figures.

\begin{figure}[htb]
\centering
\includegraphics[width=0.8\textwidth]{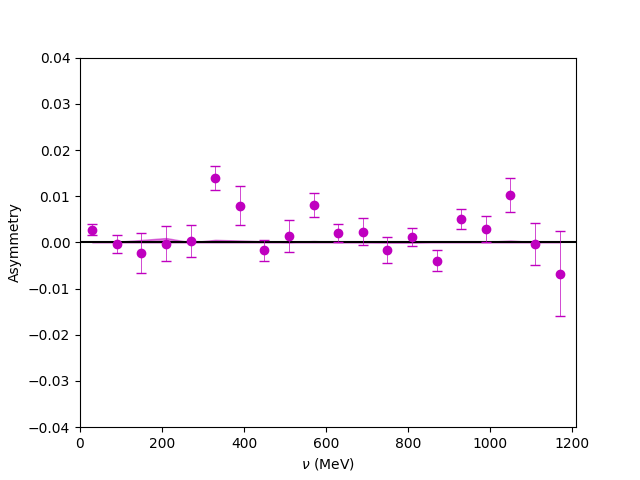}
\caption{2.5T 1.7 GeV Transverse Asymmetry.}
\label{fig:asym1}
\end{figure}
\begin{figure}[htb]
\centering
\includegraphics[width=0.8\textwidth]{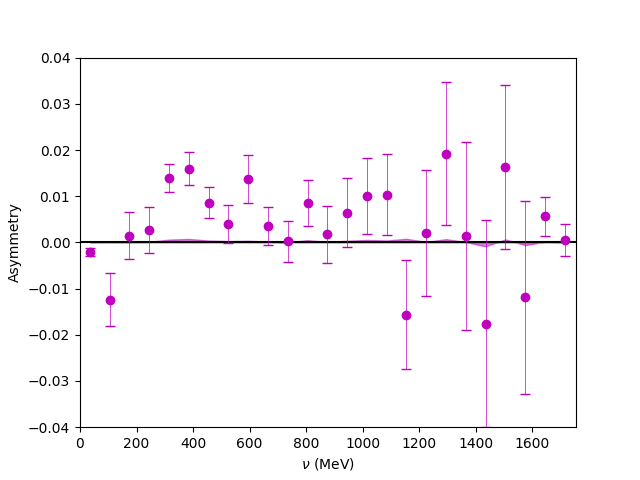}
\caption{2.5T 2.2 GeV Transverse Asymmetry.}
\label{fig:asym2}
\end{figure}

\begin{figure}[htb]
\centering
\includegraphics[width=0.8\textwidth]{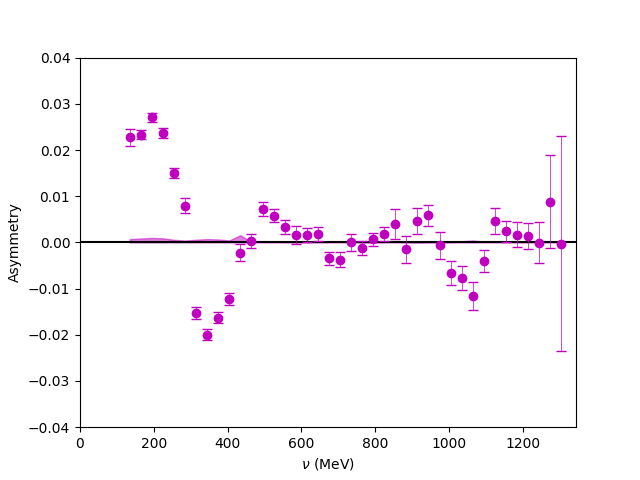}
\caption{5T 2.2 GeV Longitudinal Asymmetry.}
\label{fig:asym3}
\end{figure}

\begin{figure}[htb]
\centering
\includegraphics[width=0.8\textwidth]{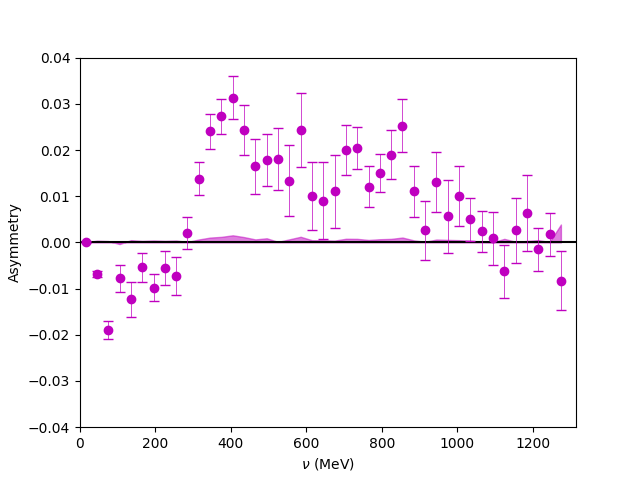}
\caption{5T 2.2 GeV Transverse Asymmetry.}
\label{fig:asym4}
\end{figure}

\begin{figure}[htb]
\centering
\includegraphics[width=0.8\textwidth]{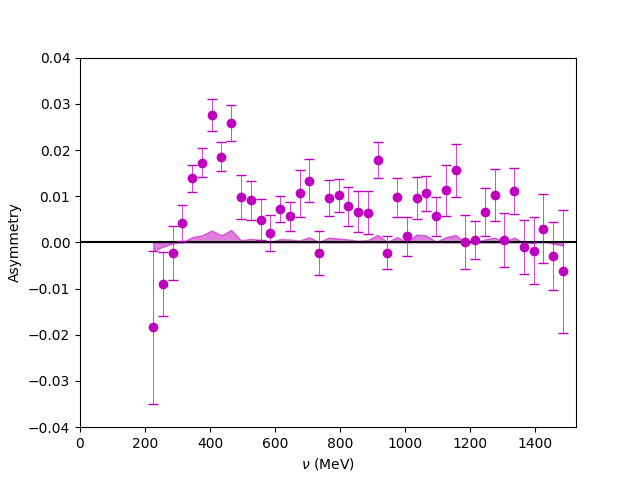}
\caption{5T 3.3 GeV Transverse Asymmetry.}
\label{fig:asym5}
\end{figure}

\clearpage

\section{Model Cross Section}

The asymmetry gives us half of what we need to form the polarized cross sections. To get the unpolarized cross section we will need to use~(\ref{eqn:exp_xs}) to form a cross section. However, the acceptance $\Delta \Omega$ has all the same issues that plagued the dilution analysis, so to form cross sections we would need to correct them. This is done by using the same g2psim simulation that was used for the packing fractions~\cite{Chao3} to reproduce and correct the acceptance. For the longitudinal setting, this process is well understood. But the transverse settings create a very complex acceptance which is extremely challenging to model and reproduce.

\begin{figure}[htb]
\centering
\includegraphics[width=0.8\textwidth]{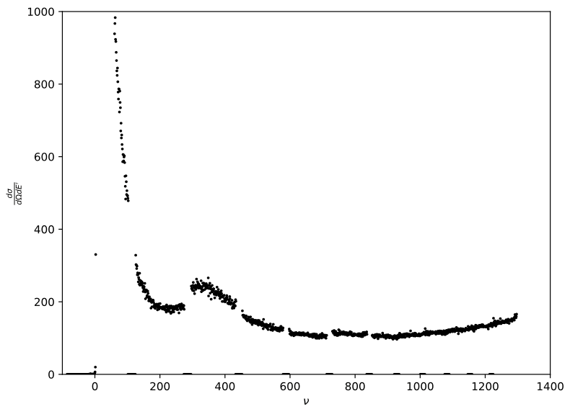}
\caption{5T 2.2 GeV Longitudinal Unpolarized Cross Section.}
\label{fig:longxs}
\end{figure}

The acceptance has consequently only been accurately modeled for the longitudinal setting to produce a smooth cross section, which is shown in Figure~\ref{fig:longxs}. Fortuitously, though the asymmetry represents a polarized part of the scattering which has never been measured before, unpolarized protons and their cross sections have been measured over a broad kinematic range. This large amount of world data means the models for unpolarized proton cross section are quite good, especially the Bosted-Christy model of~\cite{Bosted3}. This model is therefore an excellent cross section input for the settings where producing a cross section from the data is impossible.

\begin{figure}[htb]
\centering
\includegraphics[width=0.8\textwidth]{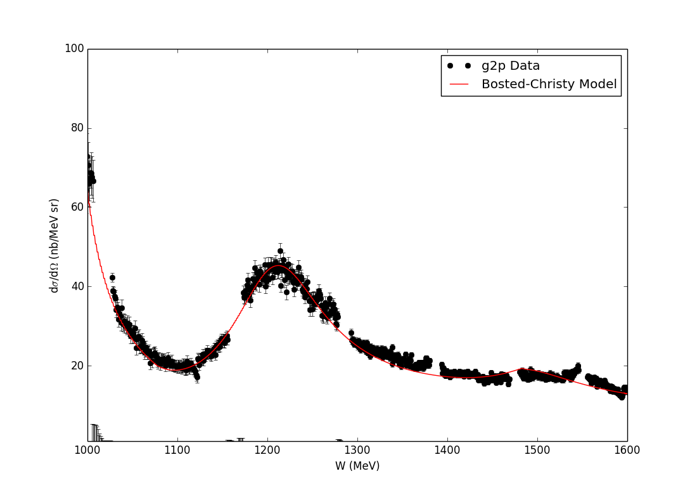}
\caption{5T 2.2 GeV Longitudinal Cross Section Comparison to Bosted-Christy Model~\cite{Bosted3}.}
\label{fig:longxs_compare}
\end{figure}

\begin{figure}[htb]
\centering
\includegraphics[width=0.8\textwidth]{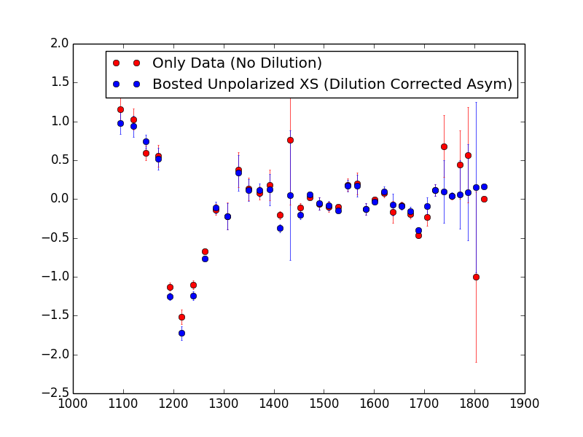}
\caption{5T 2.2 GeV Longitudinal Polarized Cross Section Difference Data Comparison to Bosted-Christy Model~\cite{Bosted3}.}
\label{fig:longpoldiff_compare}
\end{figure}

The Bosted-Christy model's validity for this purpose is tested by comparing it to the longitudinal setting's cross section. The model reproduces the structure of the data quite well, but the central value is lower by around a factor of 1.15. If we scale the model by 1.15, the agreement is very good, as is shown in the resonance region in Figure~\ref{fig:longxs_compare}. We can also check the use of this model by forming a longitudinal polarized cross section difference with (\ref{eqn:polxsdiff_par}), and comparing the result formed with the data cross section versus the model cross section. For the data cross section, it would normally be necessary to multiply it by the dilution factor $f$ to get a true experimental cross section. However, because the asymmetry scales as $\frac{1}{f}$, we can actually exclude the dilution entirely as it cancels at the level of the polarized cross section difference. This comparison is shown in Figure~\ref{fig:longpoldiff_compare}, and demonstrates very good agreement between both methods, acting as a sanity check on both the use of the model, and the construction of the dilution factor.

\begin{figure}[htb]
\centering
\includegraphics[width=0.8\textwidth]{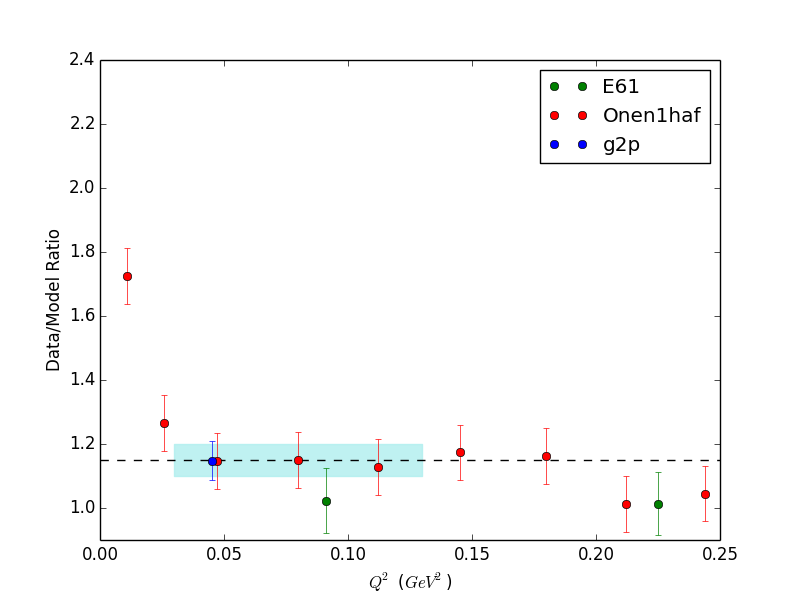}
\caption{Ratio of data to Bosted-Christy Model~\cite{Bosted3} for experiments in the low $Q^2$ region. The shaded region shows the approximate coverage of the transverse results of the E08-027 experiment.}
\label{fig:bostedscalecompare}
\end{figure}

Before finalizing the polarized cross section differences, however, the pressing question is whether the scaling factor of 1.15 needed for model agreement is reasonable, and how much systematic contribution it has. The Bosted-Christy model integrates a wide range of experimental data into its phenomenological fit, but the fit in the kinematic region relevant to the g2p experiment is driven by the results of the SLAC E61 experiment, consequently, the model fits the data of that experiment with no scaling whatsoever. However, another unpublished experiment in the same region, the SLAC ONEN1HAF experiment, requires a very similar scaling factor as the g2p data for the model to fit well. The comparison between the necessary scaling for all three experiments is shown in Figure~\ref{fig:bostedscalecompare}. This figure makes it clear that when statistical and systematic errors are considered, there is actually very little tension between the three of them, and the choice of a scaling factor of 1.15 is consistent with world data in this region.

It is now necessary to determine if this 15\% scaling on the model is equivalent to a 15\% difference, and consequently systematic uncertainty, on the final results. To do this, we use the model to form the structure functions following the procedure in the next section, both with an unscaled model, and a model scaled by 1.15. We then integrate the result to form moments, both second order and zero order:

\begin{equation}
\label{eqn:m0}
M_0 = \int_0^{x_{th}} g_1 (x,Q^2)dx
\end{equation}
\begin{equation}
\label{eqn:m0}
M_2 = \int_0^{x_{th}} x^2 g_1 (x,Q^2)dx
\end{equation}

Finally, we compare $M_0$ including the scaled model to $M_0$ with an unscaled model, and do the same for $M_2$. What we find is that the other terms of the structure function extraction lead $M_0$ to vary by a maximum of 6\%, and the kinematic scaling on the higher order moment causes the variation to be even lower. After considering the uncertainty on the longitudinal g2p cross section itself, the total systematic error of employing the model is approximately 9-10\%.

\section{Scattering Angle and Out of Plane Polarization Angle}

To employ (\ref{eqn:xs_poldiff_long2}) and (\ref{eqn:xs_poldiff_trans}) we will need to reconstruct two angles, the scattering angle from the target $\theta$, and the angle between the polarization and scattering planes, $\theta_{OoP}$.

The scattering angle $\theta$ can be reconstructed from the vectors corresponding to the electron beam before and after scattering. Before scattering, this vector is reconstructed from the angles $\theta_b$ and $\phi_b$ measured by the BPMs~\cite{Zielinski:2017gwp}:
\begin{equation}
\label{eqn:beamvector}
\vec{k} = \bigg(\sin\theta_b\cos\phi_b,\sin\theta_b\sin\phi_b,\cos\theta_b\bigg)
\end{equation}

The scattered electron position and angle are measured by the drift chambers in the high resolution spectrometers. Using the equations of motion for an electron in a magnetic field, as well as an experimentally measured field map of the target field~\cite{TECHNOTE:fieldmap}, it is possible to reconstruct angles $\phi_{rec}$ and $\theta_{rec}$ for the electron leaving the target relative to the central angle of the detectors $\theta_0$. With these angles, we can write the vector for the scattered electron as:
\begin{equation}
\label{eqn:beamvector2}
\vec{k'} = \bigg([\phi_{rec}\cos\theta_0 + \sin\theta_0],[-\theta_{rec}],[\cos\theta_0 - \phi_{rec}\sin\theta_0]\bigg)
\end{equation}

The scattering angle is defined as the angle between these two vectors~\cite{Zielinski:2017gwp}:
\begin{equation}
\label{eqn:thetascat}
\theta = \cos^{-1}\bigg(\frac{\vec{k}\cdot\vec{k'}}{|\vec{k}||\vec{k'}|}\bigg)
\end{equation}

The other angle we need is the out of plane polarization angle $\theta_{OoP}$, to form the transverse polarized cross section difference. This is defined as the angle between the normal vectors of the scattering plane and polarization planes. With the polarization vector for the proton:
\begin{equation}
\label{eqn:nplane}
\vec{n_s} = \vec{k} \times \vec{k'}
\end{equation}
\begin{equation}
\label{eqn:nplane2}
\vec{n_p} = \vec{k} \times \vec{s}
\end{equation}
and the final angle is:

\begin{equation}
\label{eqn:thetaoop}
\theta_{OoP} = \cos^{-1}\bigg(\frac{\vec{n_s}\cdot\vec{n_p}}{|\vec{n_s}||\vec{n_p}|}\bigg)
\end{equation}

\section{Polarized Cross Section Differences \& Radiative Corrections}

With an asymmetry from the g2p data and a scaled model cross section from the Bosted-Christy model, we can now employ~(\ref{eqn:polxsdiff_par}) and  (\ref{eqn:polxsdiff_perp}) to form polarized cross section differences. Before forming the final structure functions, it is necessary to radiatively correct these polarized cross section differences, as was discussed for the dilution factor. The Feynman diagram originally discussed in Figure~\ref{fig:epscattering} actually depicts the Born-level process. In actuality, there are many other diagrams which can take place, with the Born approximation being only a small fraction of the true interaction. Additional diagrams, radiation of brehmstrahhlung, and ionizing collisions cause the energy when an electron scatters off the target to differ from that measured with the Arc method, and the energy observed in the spectrometer to differ from the true scattered energy. The corrections for these complexities to restore the data to the Born approximation assumed for most of the formalism in Chapter 2 are collectively known as radiative corrections. This process is detailed at length in \cite{Zielinski:2017gwp}, and involves the calculation of a cross section which can be subtracted from our polarized cross section differences to obtain the Born cross section.

\begin{figure}[htb]
\centering
\includegraphics[width=0.8\textwidth]{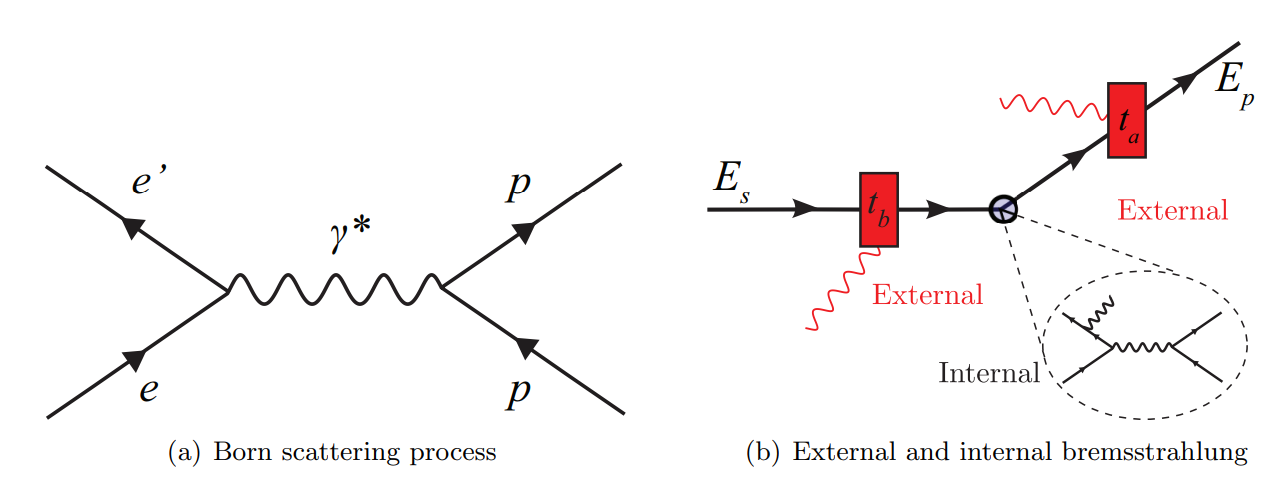}
\caption{Born Process and Bremstrahhlung corrections. Reproduced from~\cite{Zielinski:2017gwp}.}
\label{fig:bornradiations}
\end{figure}

The Born interaction and bremstrahhlung corrections are shown in Figure~\ref{fig:bornradiations}. This can be viewed as possible sources of energy loss for the scattering electrons. This includes the possibility that energy is radiated before or after scattering, due either to bremstrahhlung or ionizing collisions, as well as the possibility that a photon is radiated before or after scattering during the primary scattering diagram. To treat the Born level process as we did in Chapter 2, it is necessary to handle the external energy loss through estimates of the radiation thickness before and after the target, referred to as $t_b$ and $t_a$ respectively. It is also necessary to calculate the cross section for internal diagrams other than the one-photon exchange shown in Figure \ref{fig:bornradiations}(a), such as one-loop corrections, internal bremstrahhlung, and self-interaction of the electron.

In addition to being divided between external and internal corrections, the radiative corrections process is split between the calculation of a large elastic tail corresponding to energy loss in elastic collisions, and inelastic radiative corrections covering all other cases. The formalism is then further divided into unpolarized radiative corrections, and polarized radiative corrections, which must be considered for the spin-polarized beam of g2p. Each of these contributions is detailed in depth in ~\cite{Zielinski:2017gwp}, and the calculation of a final radiative corrections cross section there is ultimately subtracted from the polarized cross section differences calculated with the Bosted-Christy unpolarized cross section and the g2p asymmetry results.

After applying these radiative corrections, the final Born-level polarized cross section differences are shown in the following Figures.


\begin{figure}[htb]
\centering
\includegraphics[width=0.8\textwidth]{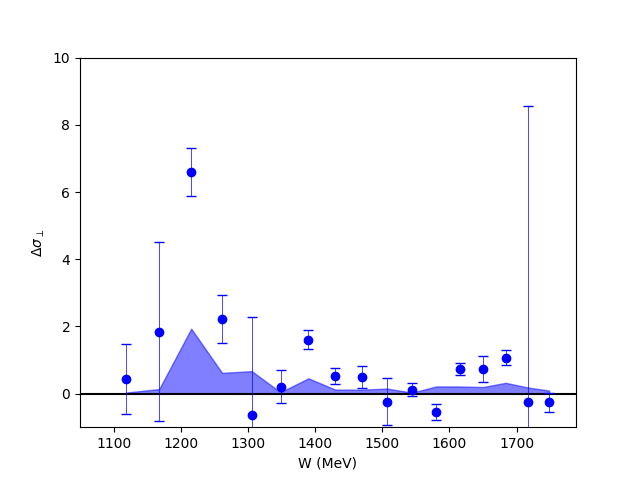}
\caption{2.5T 1.7 GeV Transverse Polarized Cross Section Difference.}
\label{fig:ds2}
\end{figure}
\begin{figure}[htb]
\centering
\includegraphics[width=0.8\textwidth]{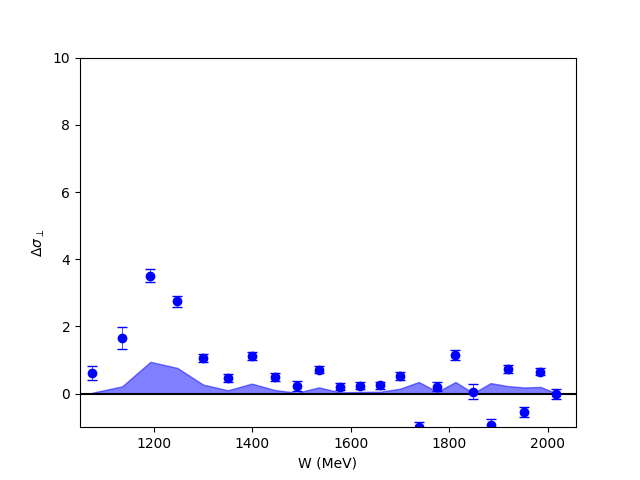}
\caption{2.5T 2.2 GeV Transverse Polarized Cross Section Difference.}
\label{fig:ds2}
\end{figure}

\begin{figure}[htb]
\centering
\includegraphics[width=0.8\textwidth]{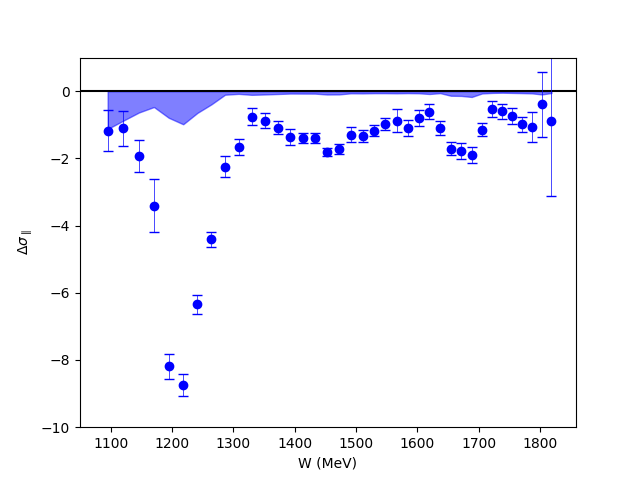}
\caption{5T 2.2 GeV Longitudinal Polarized Cross Section Difference.}
\label{fig:ds3}
\end{figure}

\begin{figure}[htb]
\centering
\includegraphics[width=0.8\textwidth]{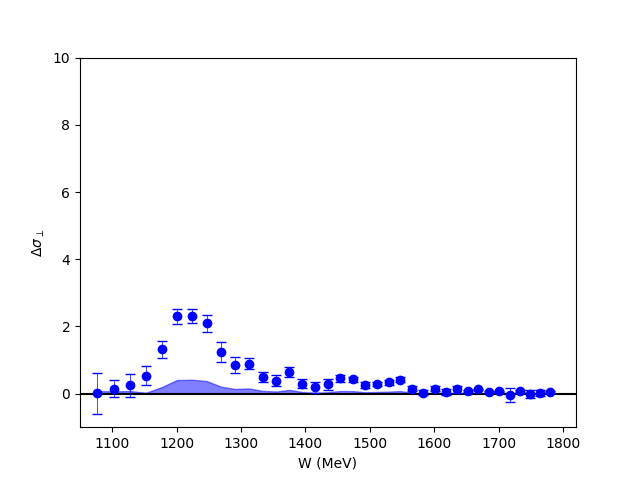}
\caption{5T 2.2 GeV Transverse Polarized Cross Section Difference.}
\label{fig:ds4}
\end{figure}

\begin{figure}[htb]
\centering
\includegraphics[width=0.8\textwidth]{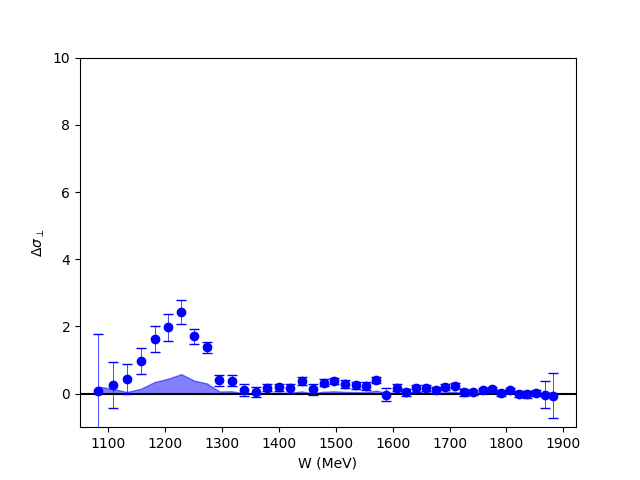}
\caption{5T 3.3 GeV Transverse Polarized Cross Section Difference.}
\label{fig:ds5}
\end{figure}

\clearpage


\section{Extracting the Structure Functions}

We know from (\ref{eqn:xs_poldiff_trans}) and (\ref{eqn:xs_poldiff_long2}) that the structure functions can be related to these polarized cross section differences. For a $Q^2$ of 0.045 GeV$^2$, we have both longitudinal and transverse data, so it is possible to solve these equations directly. But for the other settings, it is most convenient to reformulate the solution for the structure functions as:
\begin{equation}
\label{eqn:g1}
g_1 = K_1\bigg[\Delta \sigma_{\parallel}\big(1+\frac{1}{K_2}\tan\frac{\theta}{2}\big)\bigg] + \frac{2 g_2 (x,Q^2)}{K_2 y}\tan\frac{\theta}{2}
\end{equation}
\begin{equation}
\label{eqn:g2}
g_2 = \frac{K_1 y}{2}\bigg[\Delta \sigma_{\perp}(K_2+\tan\frac{\theta}{2}\big)\bigg] - \frac{g_1 (x,Q^2)y}{2}
\end{equation}
where $K_1$ and $K_2$ are defined by:
\begin{equation}
\label{eqn:k1}
K_1 = \frac{M_p Q^2}{4 \alpha}\frac{y}{(1-y)(2-y)}
\end{equation}
\begin{equation}
\label{eqn:k2}
K_2 = \frac{1+(1-y)\cos\theta}{(1-y)\sin\theta}
\end{equation}

This allows us to use the polarized cross section differences in combination with a smaller contribution from the other structure function to form each structure function. Since we have no $Q^2$ with only longitudinal data, this is primarily a concern for the formulation of $g_2$. In this case, we make use of the high precision $g_1$ proton data collected by other experiments, and the phenomenological models based on it, in this case, the CLAS Hall B model~\cite{EG1b2}. This $g_1$ term is responsible for around 30\% of the total $g_2$ structure function. Using a $g_1$ from this model, we are able to form $g_2$ for all settings without a relevant longitudinal measurement.

To form moments and obtain our final results, it is necessary to have our structure function measurements at a constant $Q^2$. Due to the variation in $E'$ across a given energy setting, each point of the structure function for that setting will land at a slightly different $Q^2$. We also employ the Hall B Model~\cite{EG1b2} to make this correction, by adding a correction factor comparing the model structure function result at the desired constant $Q^2$ and the model result at the $Q^2$ of the relevant data point:

\begin{equation}
\label{eqn:modelstruc_Correct}
g_{1,2}(x,Q^2_{const}) = g_{1,2}^{data}(x,Q^2) + \bigg(g_{1,2}^{model}(x,Q^2_{const}) -g_{1,2}^{model}(x,Q^2)\bigg)
\end{equation}

This method was also checked with a multiplicative correction factor and found almost no difference. This correction factor was also found to have an extremely small impact, on the order of 2\% or less, on the final structure function, meaning the model dependency introduced by this correction is not large.

\begin{figure}[htb]
\centering
\includegraphics[width=0.95\textwidth]{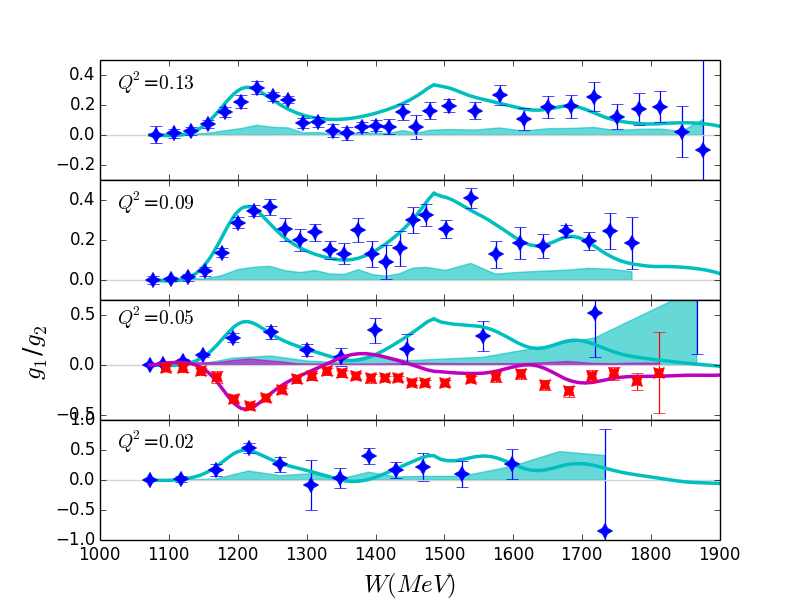}
\caption{Final structure functions results for all kinematic settings of the g2p experiment. The solid lines indicate the Clas Hall B Model~\cite{EG1b2}. Red points indicate $g_1$ results, while blue points show the $g_2$ results. The shaded regions show the respective systematic errors of each structure function.}
\label{fig:structure_functions}
\end{figure}

After these corrections, the final structure function results are shown in Figure~\ref{fig:structure_functions}.

\begin{figure}[htb]
\centering
\includegraphics[width=0.95\textwidth]{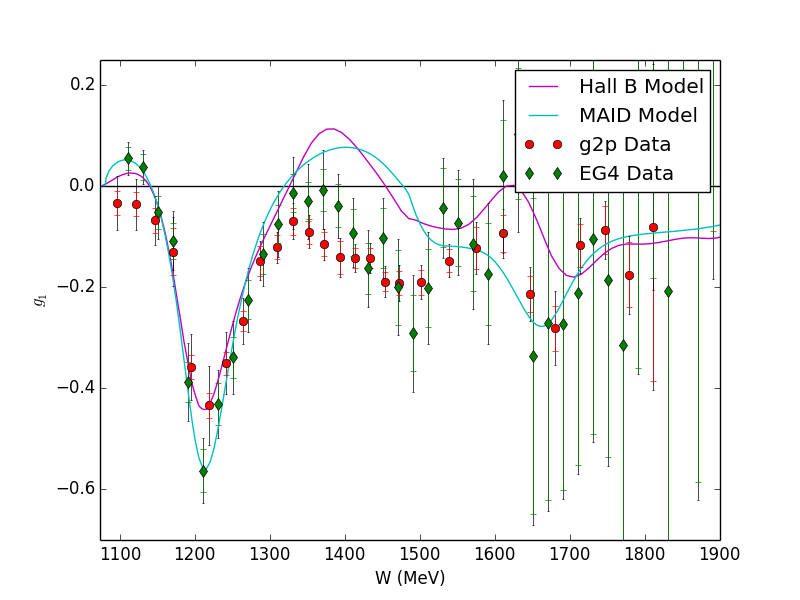}
\caption{Comparison between the $g_1$ data of the g2p~\cite{g2p_nature} and EG4~\cite{EG4_final} experiments at Jefferson Lab.}
\label{fig:g1_eg4_comparison}
\end{figure}

Though there is no other data in this kinematic region for the $g_2$ structure function, the $g_1$ results can be compared to recent $g_1^p$ results from the EG4 experiment at Jefferson Lab~\cite{EG4_final}. This comparison is shown in Figure~\ref{fig:g1_eg4_comparison}. For the most part, the g2p data agrees within error bars with the EG4 result, though showing a slightly smaller $\Delta$-resonance and very good statistics. The major point of interesting difference is near the pion production threshold at W=1073.2 MeV. Phenomenological models and EG4 results indicate a small positive section and a zero crossing, while g2p data stays negative. Though the g2p data is compatible with the positive result within error bars, the central value being negative has a powerful impact on the moments, especially the kinematically weighted moments where the data near threshold dominates the moment, in which case the sign of the structure function near threshold has a colossal impact on the moment.

\section{Systematic Contributions}

The final systematic uncertainty for the physics results is shown in Table~\ref{table:syst}. Here the longitudinal setting is denoted with an (L) and the other settings are all transverse.

\begin{table}[h!]
\centering
\begin{tabular}{|c|c|c|c|c|c|} 
 \hline
$Q^2$ (GeV$^2$) & 0.021 & 0.045 (L) & 0.045 & 0.09 & 0.13 \\ [0.5ex] 
 \hline
 Model Cross Section ($d\Omega dE'$) & 10\% & 10\% & 10\% & 10\% & 10\% \\
 \hline
 Dilution Factor($f$) \& Packing Fraction ($pf$)& 7\% & 4\% & 8\% & 7\% & 8\% \\
 \hline
 Beam Polarization($P_b$) \& Target Polarization ($P_t$)& 5\% & 5\% & 5\% & 5\% & 5\% \\
 \hline
 Dilution Factor($f$) \& Packing Fraction ($pf$)& 7\% & 4\% & 8\% & 7\% & 8\% \\
 \hline
 Inelastic Radiative Corrections ($\Delta \sigma_{RC}$)& 3\% & 3\% & 3\% & 3\% & 3\% \\
 \hline
 Elastic Radiative Corrections ($\Delta \sigma_{tail}$)& 3\% & 3\% & 3\% & 3\% & 3\% \\
 \hline
 $g_1$ Model Input& 2\% & 2\% & 1\% & 1\% & 1\% \\
 \hline
 Out-of-Plane Scattering Angle Correction($\theta_{OoP}$) & $<$1\% & $<$1\% & $<$1\% & $<$1\% & $<$1\% \\
 \hline
 Adjustment to Const $Q^2$ & $<$1\% & $<$1\% & $<$1\% & $<$1\% & $<$1\% \\
 \hline
 \textbf{Total} & \textbf{14\%} & \textbf{13\%} & \textbf{14\%} & \textbf{14\%} & \textbf{14\%} \\
 \hline
\end{tabular}
\caption{Final systematics for E08-027.}
\label{table:syst}
\end{table}

The dominating systematic is the model cross section contribution discussed above. Due to the uncertainty in the $g_2^p$ cross section and the issues scaling a phenomenological model to it, this uncertainty is no less than 10\%. The secondary uncertainty comes from the dilution factor and the packing fraction, as a direct consequence of the acceptance issues, BPM miscalibration, and septa/dipole current mismatch. The beam and target polarization together have around a 5\% systematic uncertainty, due to the uncertainty associated with the M\o{}ller polarimeter and Q-Meter NMR system. The elastic and inelastic radiative corrections both contribute around 3\% systematic error, and the use of the Hall B model for the $g_1$ part of the $g_2$ results contributes only 1-2\%. The out of plane correction to the scattering angle, introduced by the transverse target field, was investigated and found to contribute less than 1\%. Likewise, the adjustment to constant $Q^2$ with the Hall B model of the final structure functions was a very small systematic. These errors were analyzed thoroughly and found to be uncorrelated, so the final systematic error is all of the above errors added in quadrature, and comes out to around 14\%.

\chapter{Results}

The structure functions are a useful result on their own, because they provide information on how the proton's spin structure varies from pointlike behavior over a previously unmeasured kinematic range. As discussed previously, the proton's spin is a subject of tremendous uncertainty, and having a test of its spin structure at low $Q^2$ is valuable. But much of the impact of these results comes from the moments you can form with this structure functions. These quantities, derived in chapter 2, are kinematically weighted integrals of the structure functions. They are valuable in large part because they can be directly compared to various theoretical predictions, allowing a direct experimental test of effective theories of QCD in the low energy regime.

\section{Burkhardt-Cottingham Sum Rule ($\Gamma_2$)}

\begin{figure}[htb]
\centering
\includegraphics[width=0.9\textwidth]{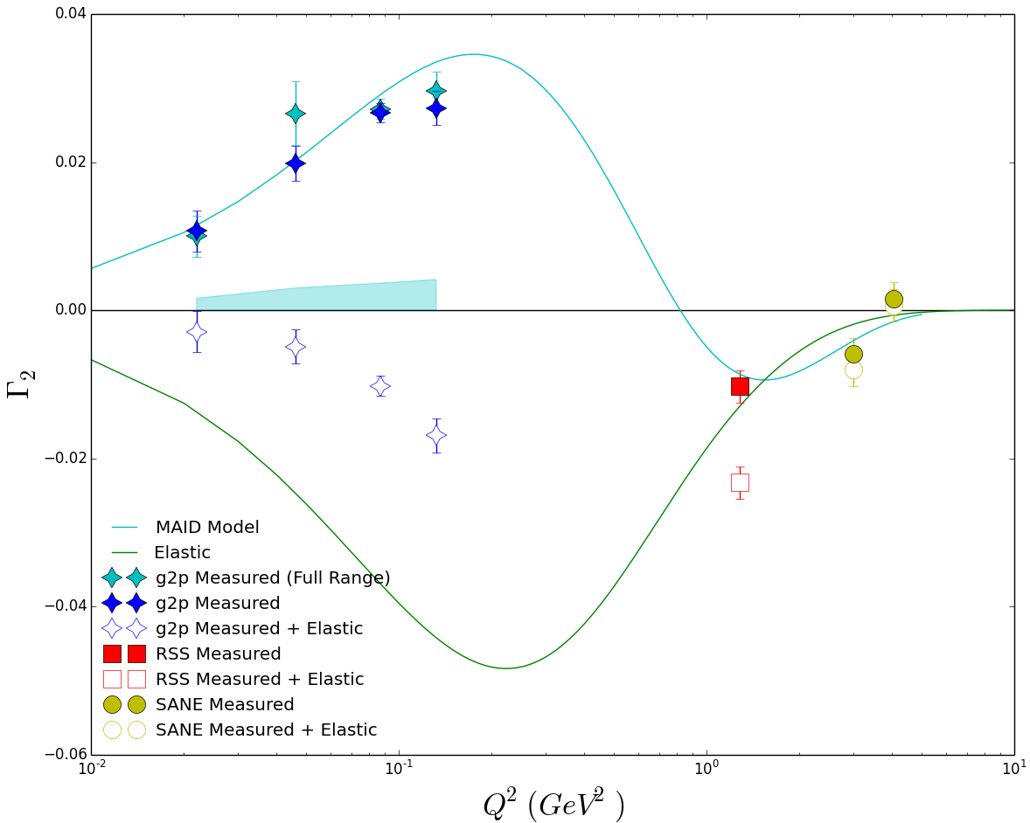}
\caption{The $\Gamma_2$ moment. Light blue points represent an integral of the full measured region of E08-027, while dark blue points are limited to a maximum W of 1725. The MAID phenomenological model~\cite{MAID2007} is also run over this limited W range, as are the other experiments' results, to provide an even comparison. Results of RSS~\cite{RSS} are shown in red, and results of SANE~\cite{warmstrong} are shown in yellow. The open symbols represent the sum of the measured and elastic parts of the integral, excluding only the unmeasured high-W part.}
\label{fig:gamma2}
\end{figure}

One of the easiest moments to form with the transverse results is the first moment $\Gamma_2$, as derived in (\ref{eqn:Gamma2}). This is simply an integral of $g_2$ over the full range in $x_{bj}$. The major prediction for this quantity is the Burkhardt-Cottingham Sum Rule, which states that the full $\Gamma_2$ integral must be equal to zero at every $Q^2$. This sum rule has so far been proven accurate by numerous measurements of the $g_2$ structure function at higher $Q^2$. To properly test this sum rule, it is necessary to handle the full integral of the moment, but the measured data is only in the resonance region. Consequently, we need a way to handle the inelastic, low-$x$ region, as well as the elastic $x = 1$ contribution. 

The $g_2$ elastic contribution can be calculated directly with~\cite{MELNITCHOUK_2001}:

\begin{equation}
\label{eqn:g2el}
g_2^{el}(x,Q^2) = \frac{\tau}{2}G_M(Q^2)\frac{G_E(Q^2) - G_M(Q^2)}{1+\tau}\delta(x-1)
\end{equation}

Here, $G_E$ and $G_M$ represent the Sachs electric and magnetic form factors~\cite{HalzenMartin}, and $\tau = \frac{Q^2}{4 M_p^2}$. The delta function makes integrating this to form a $\Gamma_2$ trivial:

\begin{equation}
\label{eqn:gamma2el}
\Gamma_2^{el}(Q^2) = \frac{\tau}{2}G_M(Q^2)\frac{G_E(Q^2) - G_M(Q^2)}{1+\tau}
\end{equation}

For this analysis, the Arrington fit to the Sachs form factors is used for $G_E$ and $G_M$~\cite{ArringtonFit}.

The inelastic, low-$x$ part is a bit trickier. To measure the full integral in this way at any $Q^2$ would require a measurement up to infinite W, so it is always necessary to estimate the contribution of the unmeasured region. For other experiments, the method employed is to use the Wandzura-Wilczek relation~\cite{Wandzura}. At high $Q^2$, we make the assumption that leading twist (twist-2) dominates. If we take our dispersion relation for $d_n$~(\ref{eqn:dn_disp}) and set $d_2$ to zero, a series of Mellin transforms leads to the relation:

\begin{equation}
\label{eqn:g2ww}
g_2^{WW}(x,Q^2) = \int_x^1 \frac{1}{y}g_1(y,Q^2)dy - g_1(x,Q^2)
\end{equation}

It is then possible to use $g_1$ data and models to calculate $g_2$ in the unmeasured region and consequently fill out the $\Gamma_2$ integral. However, in the very low $Q^2$ regime of the g2p experiment, the assumption that higher twist contributions are zero ceases to be valid. The higher twist contribution causes this $g_2^{WW}$ method to no longer be accurate, meaning that there is no known good way to estimate the unmeasured part.

The results for the g2p calculation of $\Gamma_2$ are shown in Figure~\ref{fig:gamma2}. Due to the lack of an unmeasured estimate, it is impossible to fully test the B.C. sum rule with this data, and the fact that the total sum of the integral shown in the open symbols does not lay along zero should not be viewed as a refutation of the Burkhardt-Cottingham sum rule. However, due to the low $Q^2$ of the experiment, the lowest measured $x$ becomes very low by the lowest point, down to $x = 0.008$. This means the unmeasured region grows very small, and consequently, the data comes very close to agreeing with the B.C. sum rule without any unmeasured estimate at all.

\section{Transverse-Longitudinal Spin Polarizability ($\delta_{LT}$) }

\begin{figure}[htb]
\centering
\includegraphics[width=0.9\textwidth]{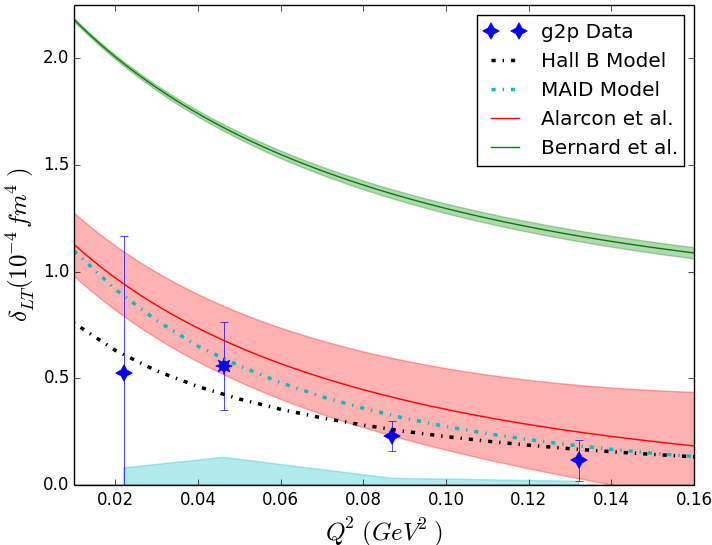}
\caption{The Transverse-Longitudinal Spin Polarizability $\delta_{LT}$. The 8-pointed marker indicates the $Q^2$ where both transverse and longitudinal data from E08-027 are available, while the other three points are formed with $g_1$ from the CLAS Hall B Model~\cite{EG1b2}. The points are compared to the MAID~\cite{MAID2007} and Hall B~\cite{EG1b2} phenomenological models, as well as $\chi$PT calculations from Alarcon et al.~\cite{Alarcon} and Bernard et al~\cite{Bernard_moments}. The blue shaded region indicates the systematic uncertainty.}
\label{fig:dlt}
\end{figure}

One of the most exciting results of the g2p experiment is the Transverse-Longitudinal Spin Polarizability $\delta_{LT}$, derived in ~(\ref{eqn:dlt}). Polarizabilities are a class of moment considered to be a fundamental property of the nucleons, which describes the nucleonic response to an external field. In this case, $\delta_{LT}$ described a spin-dependent response to an external electromagnetic field. It was considered to be a benchmark test of Chiral Perturbation Theory because it was expected to be insensitive to the contribution of the $\Delta$-resonance~\cite{Drechsel}. A vast disagreement in 2001 with Neutron data for this moment produced the `$\delta_{LT}$-puzzle' discussed in Chapter 1. For this moment, unlike $\Gamma_2$, it is not necessary to calculate an unmeasured part because the kinematic weighting of the integral heavily biases the result towards the data near the pion production threshold. This minimizes the contribution from low-$x$ data. The same is true for every other moment with a strong kinematic weighting, such as $d_2$ and $\gamma_0$.

The result shown in Figure~\ref{fig:dlt} represents the world's first experimental determination of $\delta_{LT}$ for the proton. The seemingly large error bar at the lowest point is a reflection of the strong $\frac{1}{Q^6}$ weighting on the moment, drastically inflating the statistical error of points at very low $Q^2$. The result agrees well with one calculation of $\chi$PT produced by the Alarcon et al.~\cite{Alarcon} group, showing no evidence of a `$\delta_{LT}$-puzzle' for the proton. However, the two cutting-edge calculations of $\chi$PT disagree with each other, and the g2p data shows a much larger discrepancy with the calculation of Bernard et al~\cite{Bernard_moments}.

The reason for this disagreement is not yet well understood, though there are several known differences between the two calculations. Firstly, the Bernard calculation makes the assumption that the higher order terms of the expansion are negligible, but recent results indicate that these terms may be important, and should be included in the expansion. Secondly, the two calculations use a different method of power-counting to expand the VVCS amplitudes. Bernard favors an `$\epsilon$-counting' method, where the assumption is made that the pion mass $m_\pi$ is of similar scale to the mass difference between the $\Delta$-resonance and the proton, which we will denote as $\varDelta$. This allows the two quantities to be expanded to the same order. Alarcon instead uses a `$\delta$-counting' method, where the underlying assumption is that the ratio between the pion mass and this mass difference $\frac{m_\pi}{\varDelta}$ is similar to the ratio between the mass difference and the nucleon mass $\frac{\varDelta}{M_p}$. Finally, the Alarcon et al. paper stresses their use of `consistent couplings' to the $\Delta$-field, where this consistency is not enforced in the Bernard result. Alarcon stresses that the use of these consistent couplings will likely not have an effect on the structure of the calculation, but may have an impact on the normalization at $Q^2=0$, which seems very similar to what we see for $\delta_{LT}$.

The g2p result for this polarizability represents a new benchmark which can help to discriminate between these calculations of chiral perturbation theory, and which other low-energy effective theories such as Lattice QCD must be able to reproduce.

\section{Twist-3 Local Operator ($d_2$)}

\begin{figure}[htb]
\centering
\includegraphics[width=0.9\textwidth]{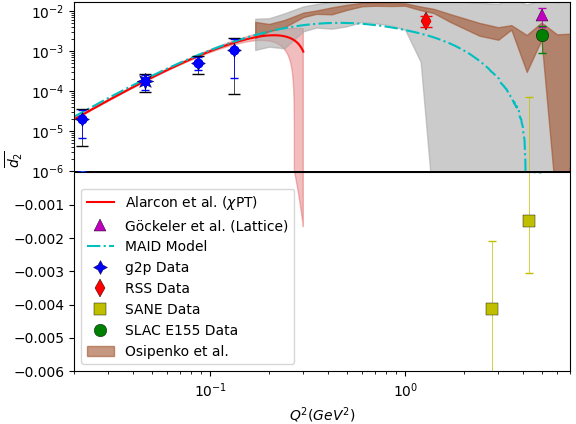}
\caption{The Twist-3 Local Operator $d_2$. The 8-pointed marker indicates the $Q^2$ where both transverse and longitudinal data from E08-027 are available, while the other three points are formed with $g_1$ from the CLAS Hall B Model~\cite{EG1b2}. The points are compared to the MAID~\cite{MAID2007} phenomenological model, as well as a $\chi$PT calculations from Alarcon et al.~\cite{Alarcon} and a higher $Q^2$ lattice QCD calculation~\cite{Gockeler}. Also shown are results from the RSS~\cite{RSS}, SANE~\cite{warmstrong} and SLAC E155~\cite{Anthony:2002hy} experiments.}
\label{fig:d2}
\end{figure}

Another polarizability of interest which is dominated by the $g_2$ structure function is the twist-3 local operator $\overline{d_2}$, defined in (\ref{eqn:d2bar}). At high $Q^2$, this moment is equated with a color polarizability, considered by some to quantify a `color Lorentz force'~\cite{Meziani:2004ne}. This means that it shows the nucleon's response to color electric and magnetic fields. However, at low $Q^2$, this definition drops out, and instead it can be interpreted as a `pure polarizability', which gives information about the quark-gluon correlations of the nucleon~\cite{Burkardtg2d2,Alarcon}.

The g2p results for $\overline{d_2}$ are shown in Figure~\ref{fig:d2}. There is only one chiral perturbation theory calculation in this regime~\cite{Alarcon}, but our data agrees very well with this calculation, as well as with the phenomenological MAID model~\cite{MAID2007}. This moment is interesting also because it must vanish at $Q^2 = 0$ and at high $Q^2$, where the twist-3 contribution to the nucleon's response becomes negligible. But in the strong-QCD regime and transition region, its structure reveals the interaction between nucleonic quarks and gluons at low energy.

At higher $Q^2$, the results of SANE~\cite{warmstrong} indicate that the behavior of $\overline{d_2}$ may be very complex, with a potential sign change in contrast to the expectation of the models. But at low $Q^2$ our new data shows no tension with the theoretical predictions or models. The area in between is a very interesting one, where a new measurement may be of great value in determining the maximum value of $\overline{d_2}$ and mapping out the transition between high and low $Q^2$.

\section{First Moment of $g_1$ ($\Gamma_1$)}

\begin{figure}[htb]
\centering
\includegraphics[width=0.9\textwidth]{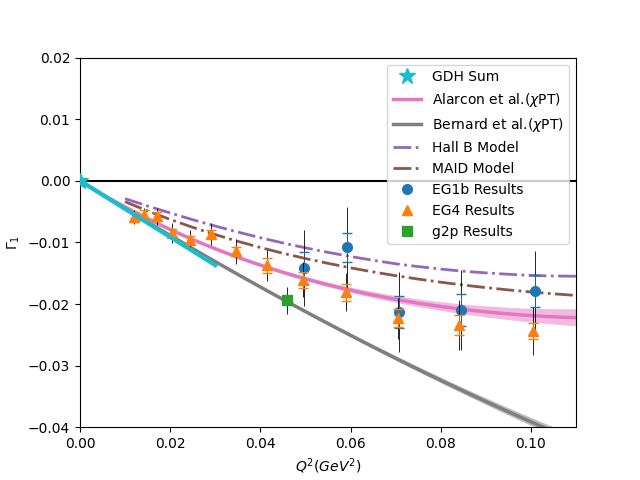}
\caption{The First Moment of $g_1$, $\Gamma_1$. The points are compared to the MAID~\cite{MAID2007} and Hall B~\cite{EG1b2} phenomenological model, as well as a $\chi$PT calculations from Alarcon et al.~\cite{Alarcon} and Bernard et al~\cite{Bernard_moments}. Also shown are results from the EG1B~\cite{EG1b2} experiment.}
\label{fig:gamma1}
\end{figure}

The first moment of $g_1$, similar to $\Gamma_2$, is simply an integral of the longitudinal structure function $g_1$ as defined in (\ref{eqn:Gamma1}). Data on this moment is useful for discriminating between $\chi$PT calculations as it was for the other moments, and for checking the well known GDH-Slope approaching $Q^2=0$. Similar to the first moment of $g_2$ we must add in an elastic contribution:
\begin{equation}
\label{eqn:gamma1el}
\Gamma_1^{el}(Q^2) = \frac{\tau}{2}G_M(Q^2)\frac{G_E(Q^2) + \tau G_M(Q^2)}{1+\tau}
\end{equation}
The new data point from g2p data is comparable to previously published data from the EG1b~\cite{EG1b2} experiment. However, it trends slightly lower and has significantly lower statistical uncertainty. Differences between the data sets can be explained by more than simple uncertainty, as this g2p point includes both transverse and longitudinal data used together, where the EG1b data uses only longitudinal data, with a model to reproduce the transverse contribution, similar to how g2p uses a model for the longitudinal part for the other $Q^2$ points. The g2p data also goes significantly closer to the pion production threshold, and its behavior in this region shows possible differences from the model used for this region by EG1b.

For this moment, the g2p data does not provide discriminating power between the two leading $\chi$PT calculations, as the error bars make it essentially compatible with either. However, the Bernard group is aware of the seeming discrepancy between their calculation and the data and model predictions at higher $Q^2$, believing this to be an issue of the missing next-to-next-to-leading-order diagrams discussed for the $\delta_{LT}$ moment.

\section{Generalized Forward Spin Polarizability ($\gamma_0$)}

\begin{figure}[htb]
\centering
\includegraphics[width=0.9\textwidth]{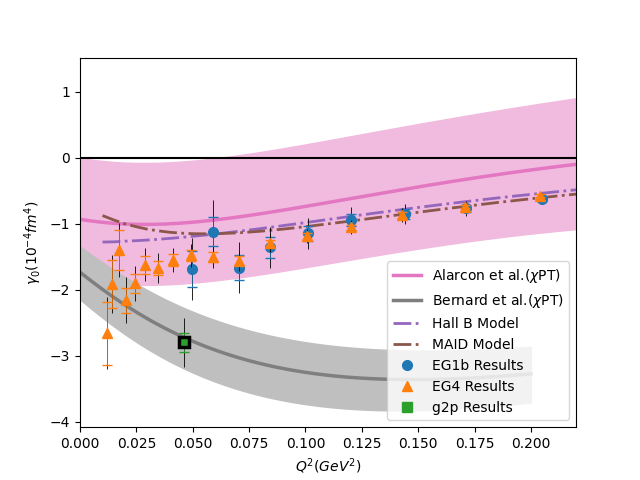}
\caption{The Generalized Forward Spin Polarizability, $\gamma_0$. The points are compared to the MAID~\cite{MAID2007} and Hall B~\cite{EG1b2} phenomenological model, as well as a $\chi$PT calculations from Alarcon et al.~\cite{Alarcon} and Bernard et al~\cite{Bernard_moments}. Also shown are results from the EG1B~\cite{EG1b2} and EG4~\cite{EG4_final} experiments.}
\label{fig:gamma0}
\end{figure}

Like $\delta_{LT}$, another important second moment spin polarizability is the complementary Generalized Forward Spin Polarizability $\gamma_0$, which is defined in (\ref{eqn:gamma0}). This polarizability tracks a longitudinally-dominated spin-dependent response to an electromagnetic field. Similar to the discussion for $\Gamma_1$, the g2p data here differs from the other existing data sets in the area of EG4~\cite{EG4_final} and EG1b~\cite{EG1b2} due to its collection of both transverse and longitudinal data for this $Q^2$ point, and the very close approach to the pion production threshold. For these moments with a strong $x^2$ dependence in the integral, the total result is heavily weighted towards the high $x$ data near the pion production threshold, where the g2p $g_1$ differs most from the other experiments. 

\section{Extended GDH Sum ($I_A$)}

\begin{figure}[htb]
\centering
\includegraphics[width=0.9\textwidth]{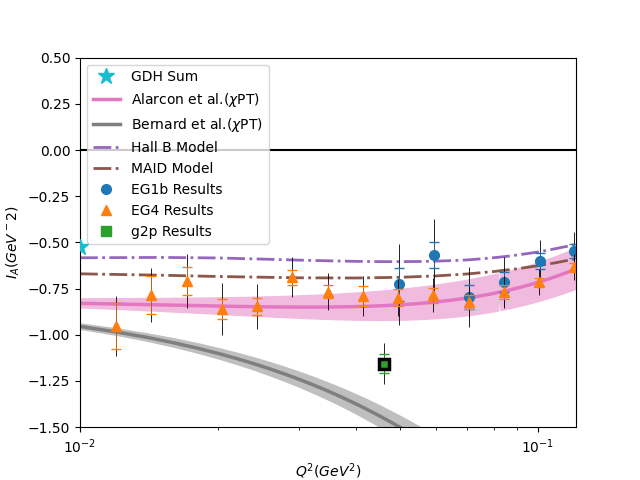}
\caption{The Extended GDH Sum, $I_A$. The points are compared to the MAID~\cite{MAID2007} and Hall B~\cite{EG1b2} phenomenological model, as well as a $\chi$PT calculations from Alarcon et al.~\cite{Alarcon} and Bernard et al~\cite{Bernard_moments}. Also shown are results from the EG1B~\cite{EG1b2} and EG4~\cite{EG4_final} experiments.}
\label{fig:gdh}
\end{figure}

The Extended GDH Sum, as defined in (\ref{eqn:gdh}). This moment too is dominated by the longitudinal data, and thus $Q^2$ points where we have only transverse data are not presented. Though the g2p point here falls between the two $\chi$PT calculations as it does for the other two longitudinally dominated moments, here it shows a slight preference for the Alarcon et al.~\cite{Alarcon} calculation. One of the most important theoretical comparisons for this moment is to the known value at the real photon point of $Q^2$ = 0. Though the g2p data is not at low enough $Q^2$ to compare directly with the real photon point, its result is compatible with the approach to this value.

\section{Hyperfine Structure}

For most purposes of scattering interactions, the energy scale associated with the electron orbitals of an atom differs greatly from the energy scale associated with its nucleus, meaning that these two sets of interactions can be treated with very different theoretical and experimental techniques. But as calculations and measurements become more precise, it becomes very desirable to bridge the gap between these two treatments, to understand the interaction between the atomic and nuclear regimes. The hydrogen atom is the simplest possible atom in which to understand this interaction.

In Hydrogen, the dipole moment of the proton that forms a hydrogen nucleus interacts with the dipole moment of its orbiting electron to create a splitting in the energy levels of the hydrogen atom, a shift known as `hyperfine splitting'. The magnitude of this shift has been measured with very high experimental precision, up to the level of $10^{-13}$ MHz uncertainty, to a value of $\Delta E_{hfs} = 1420.405751767(9)$ MHz~\cite{Zielinski:2017gwp,Hyperfine}. However, theoretical calculations of this value have an uncertainty more than a million times greater, on the level of $10^{-6}$. The leading uncertainty in this calculation is related to our theoretical understanding of the proton structure, and can be derived from the structure functions $g_1$ and $g_2$, making their measurement necessary for bridging the gap between theory and experiment for the hyperfine splitting.

The theoretical calculation of the hyperfine splitting is~\cite{Zielinski:2017gwp,Hyperfine}:
\begin{equation}
\label{eqn:hyperfine_theory}
\Delta E_{hfs} = (1 + \Delta^R + \Delta^{small} + \Delta^{QED} + \Delta_p^{struc})E_F^p
\end{equation}

Where $E_F^p = 1418.840$ MHz is the magnetic dipole interaction energy, $\Delta^R$ is a correction due to recoil effects, $\Delta^{small}$ is a correction associated with the muonic and vacuum polarizations and the weak interaction, $\Delta^{QED}$ is a correction associated with radiative quantum electrodynamics, and $\Delta_p^{struc}$ is the correction associated with the proton structure. To fully quantify this last correction, experimental measurements are necessary. This correction can be broken up into a ground state term and an excited state term~\cite{Zielinski:2017gwp}:
\begin{equation}
\label{eqn:hyperfine_deltastruc}
\Delta_p^{struc} = \Delta^Z + \Delta^{pol}
\end{equation}

The excited state term can be further broken up into two additional terms:

\begin{equation}
\label{eqn:hyperfine_deltapol}
\Delta^{pol} = \frac{\alpha m_e}{\pi g_p M_p}(\Delta_1 + \Delta_2)
\end{equation}

Where $\alpha$ is the fine structure constant and $g_p$ is the proton g-factor. The first term can be defined as:

\begin{equation}
\label{eqn:hyperfine_delta1}
\Delta_1 = \frac{9}{4}\int_0^\infty \frac{dQ^2}{Q^2}\bigg[\bigg(\frac{G_M(Q^2) + G_E^2(Q^2)}{1+\tau}\bigg)^2+\frac{8 M_p^2}{Q^2}B_1(Q^2)\bigg]
\end{equation}
\begin{equation}
\label{eqn:hyperfine_b1}
B_1(Q^2) = \int_0^{x_{pp}}\beta_1(\tau)g_1(x,Q^2)dx
\end{equation}
\begin{equation}
\label{eqn:hyperfine_beta1}
\beta_1(Q^2) = \frac{4}{9}\bigg(-3\tau + 2 \tau^2 + 2(2-\tau)\sqrt{\tau(\tau+1)}\bigg)
\end{equation}

\begin{figure}[htb]
\centering
\includegraphics[width=0.9\textwidth]{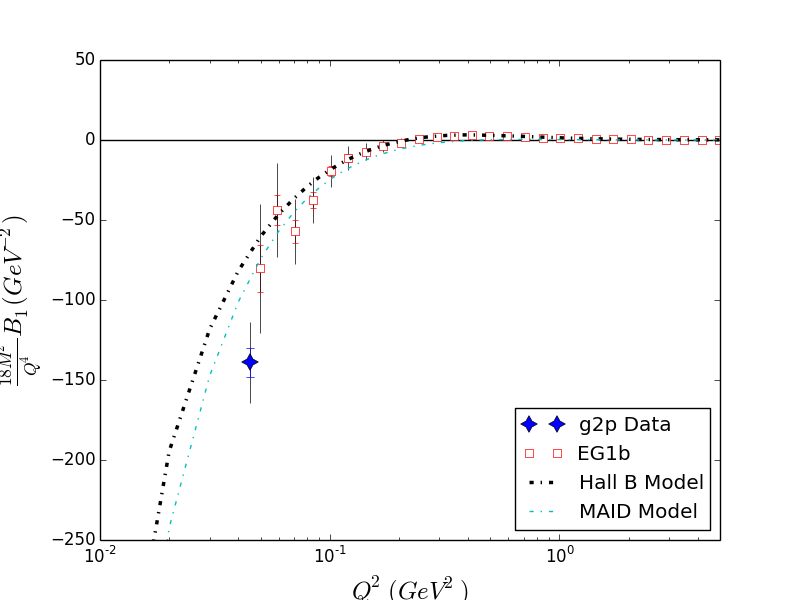}
\caption{The $B_1(Q^2)$ part of the $\Delta_1$ integral. The points are compared to the MAID~\cite{MAID2007} and Hall B~\cite{EG1b2} phenomenological model. Also shown are calculated results from internal data of the EG1B~\cite{EG1b2} experiment.}
\label{fig:b1}
\end{figure}

A full calculation of $\Delta_1$ therefore requires measurements of $g_1$ over a broad range in $Q^2$. The results for the $B_1$ part of the integral from g2p longitudinal data are shown in Figure~\ref{fig:b1}. To perform the full integral would require an integration over all $Q^2$, requiring either additional data or a model contribution. The $B_1$ result is in fairly good agreement with phenomenological models and with the prior data from the CLAS EG1b experiment~\cite{EG1b2}, but when combined with the lowest data point of that experiment, is perhaps suggestive of a slightly steeper descent than the models expect, which would result in a slightly larger contribution to the value of $\Delta_1$

A similar equation for the other part of $\Delta^{pol}$ gives:

\begin{equation}
\label{eqn:hyperfine_delta2}
\Delta_2 = -24 M_p^2 \int_0^\infty \frac{dQ^2}{Q^4} B_2(Q^2)
\end{equation}
\begin{equation}
\label{eqn:hyperfine_b2}
B_2(Q^2) = \int_0^{x_{pp}}\beta_2(\tau)g_2(x,Q^2)dx
\end{equation}
\begin{equation}
\label{eqn:hyperfine_beta2}
\beta_2(Q^2) = 1 + 2 \tau - 2 \sqrt{\tau(\tau+1)}
\end{equation}

\begin{figure}[htb]
\centering
\includegraphics[width=0.9\textwidth]{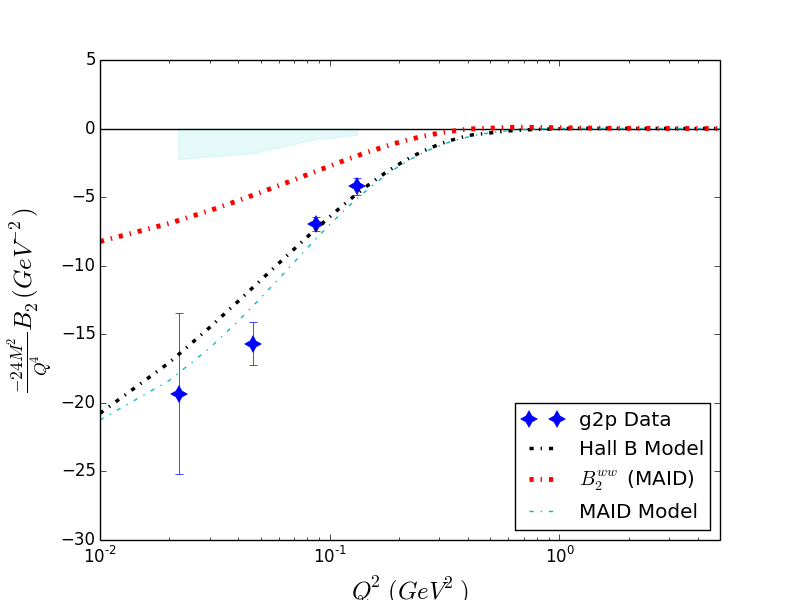}
\caption{The $B_2(Q^2)$ part of the $\Delta_2$ integral. The points are compared to the MAID~\cite{MAID2007} and Hall B~\cite{EG1b2} phenomenological model.}
\label{fig:b2}
\end{figure}

The results for the integrand of $\Delta_2$ are shown in Figure~\ref{fig:b2}. This represents the first experimental determination of this contribution. Though it is possible to form a $B_2^{WW}$ using the Wandzura-Wilczek $g_2$ defined in (\ref{eqn:g2ww}), as the plot shows, this $B_2$ has very different and incorrect behavior at low $Q^2$ when leading twist assumptions fail. The g2p data is otherwise in relatively good agreement with the phenomenological models. Similar to $B_1$, to perform the full integral of $B_2$ would require additional data over a broader $Q^2$ range, or to use a model for most of the range of the integral.

The uncertainty in current theoretical calculations of hyperfine splitting is driven almost entirely by the proton structure term, and consequently, by the lack of data for constructing $\Delta_2$~\cite{Hyperfine}. This data therefore helps to shrink the current largest uncertainty in understanding the dipole interaction in atomic hydrogen. 

\section{P-N and P+N Channels}

One outstanding question from these results is how they can help us to understand the discrepancies between Neutron data and calculations and Proton data and calculations. This is especially relevant considering the "$\delta_{LT}$ Puzzle" in the Small-Angle GDH~\cite{saGDH} neutron results, and the discrepancy between Bernard~\cite{Bernard_moments} and Alarcon~\cite{Alarcon} calculations for moments such as proton $\delta_{LT}$, proton $\gamma_0$, and neutron $\gamma_0$. One way to begin investigating these differences is by examining the difference of the proton and neutron moment results (the P-N channel) and comparing it to the sum of the respective nucleon results (the P+N channel.)

These channels are shown in the following plots for several moments where $\chi$PT calculations are available for both the proton and the neutron. For the following plots, data points shown are a result of subtracting the Small-Angle GDH Neutron results from the g2p Proton results shown in this chapter. A small correction is applied with the Hall B model to correct for any difference in the central $Q^2$ of the relevant data points.
\begin{figure}[htb]
\centering
\includegraphics[width=0.8\textwidth]{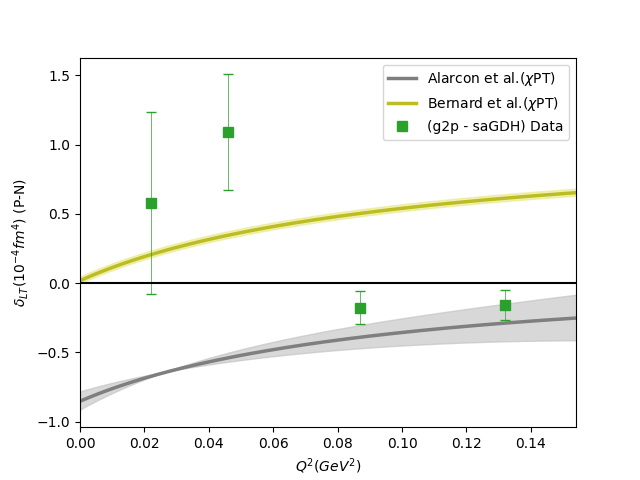}
\caption{The Transverse Longitudinal Spin Polarizability $\delta_{LT}$ in the P-N Channel. Chiral Perturbation Theory calculations are shown from Alarcon et al.~\cite{Alarcon} and Bernard et al.~\cite{Bernard_moments}. Data is combined from the g2p~\cite{g2p_nature} proton results and Small-Angle GDH~\cite{saGDH} neutron results.}
\label{fig:dltp-n}
\end{figure}
\begin{figure}[htb]
\centering
\includegraphics[width=0.8\textwidth]{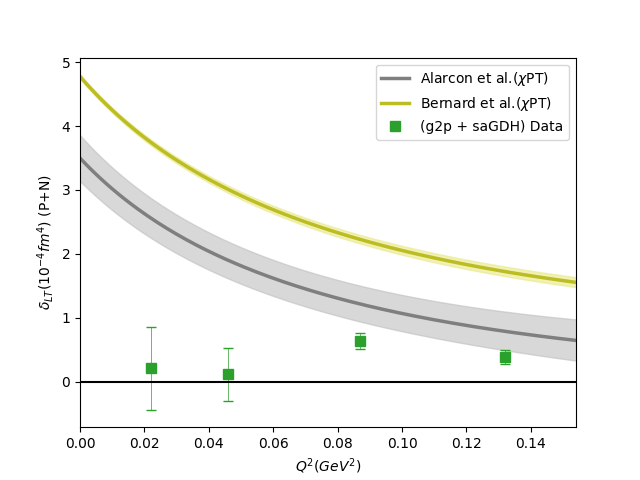}
\caption{The Transverse Longitudinal Spin Polarizability $\delta_{LT}$ in the P+N Channel. Chiral Perturbation Theory calculations are shown from Alarcon et al.~\cite{Alarcon} and Bernard et al.~\cite{Bernard_moments}. Data is combined from the g2p~\cite{g2p_nature} proton results and Small-Angle GDH~\cite{saGDH} neutron results.}
\label{fig:dltp+n}
\end{figure}

\begin{figure}[htb]
\centering
\includegraphics[width=0.8\textwidth]{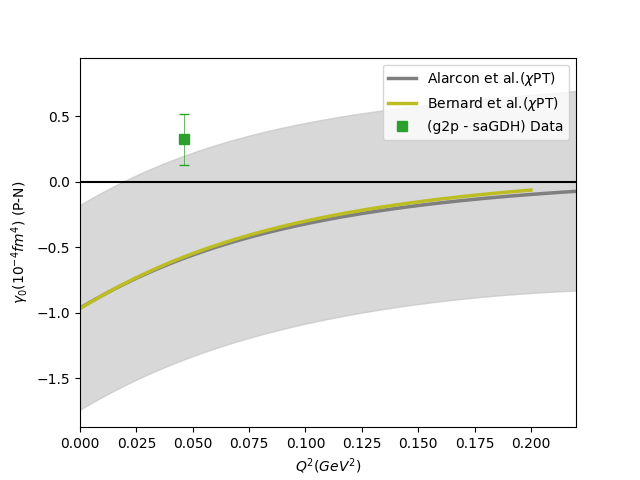}
\caption{The Generalized Forward Spin Polarizability $\gamma_0$ in the P-N Channel. Chiral Perturbation Theory calculations are shown from Alarcon et al.~\cite{Alarcon} and Bernard et al.~\cite{Bernard_moments}. Data is combined from the g2p~\cite{g2p_nature} proton results and Small-Angle GDH~\cite{saGDH} neutron results.}
\label{fig:gamma0p-n}
\end{figure}
\begin{figure}[htb]
\centering
\includegraphics[width=0.8\textwidth]{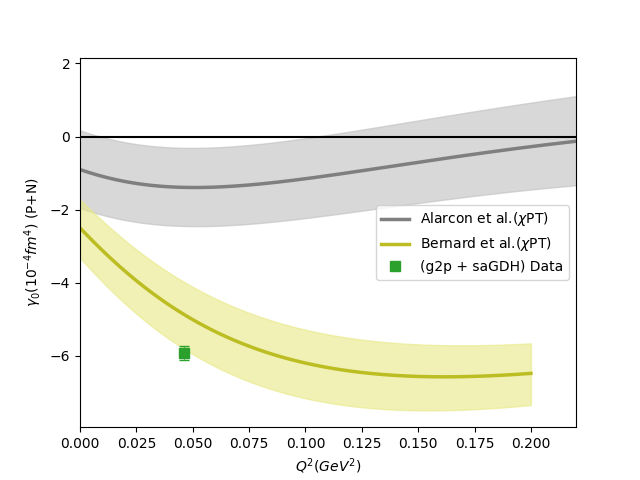}
\caption{The Generalized Forward Spin Polarizability $\gamma_0$ in the P+N Channel. Chiral Perturbation Theory calculations are shown from Alarcon et al.~\cite{Alarcon} and Bernard et al.~\cite{Bernard_moments}. Data is combined from the g2p~\cite{g2p_nature} proton results and Small-Angle GDH~\cite{saGDH} neutron results.}
\label{fig:gamma0p+n}
\end{figure}

\begin{figure}[htb]
\centering
\includegraphics[width=0.8\textwidth]{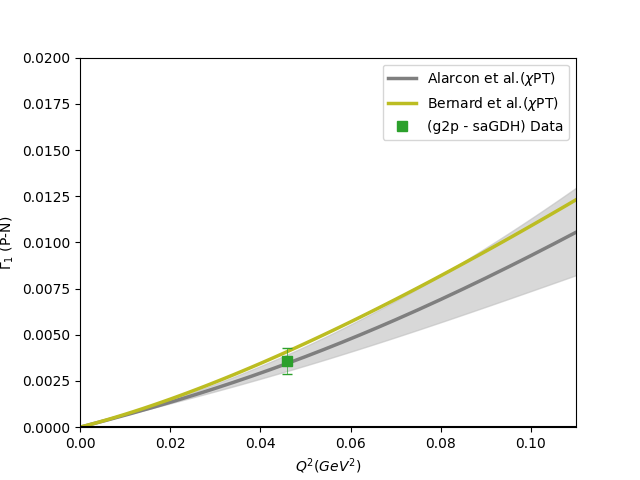}
\caption{The $\Gamma_1$ moment in the P-N Channel. Chiral Perturbation Theory calculations are shown from Alarcon et al.~\cite{Alarcon} and Bernard et al.~\cite{Bernard_moments}. Data is combined from the g2p~\cite{g2p_nature} proton results and Small-Angle GDH~\cite{saGDH} neutron results.}
\label{fig:Gamma1p-n}
\end{figure}
\begin{figure}[htb]
\centering
\includegraphics[width=0.8\textwidth]{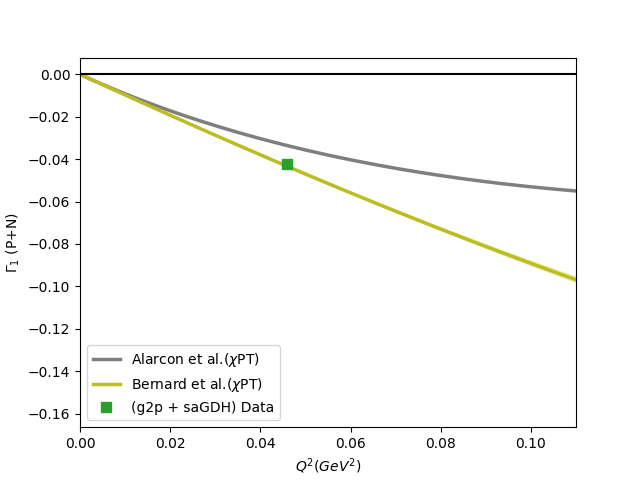}
\caption{The $\Gamma_1$ moment in the P+N Channel. Chiral Perturbation Theory calculations are shown from Alarcon et al.~\cite{Alarcon} and Bernard et al.~\cite{Bernard_moments}. Data is combined from the g2p~\cite{g2p_nature} proton results and Small-Angle GDH~\cite{saGDH} neutron results.}
\label{fig:Gamma1p+n}
\end{figure}

An interesting potential clue is found in the P-N channel, where for the $\Gamma_1$ and $\gamma_0$ moments, the discrepancy between $\chi$PT calculations nearly vanishes. This seems to indicate that the difference between the calculations is constant between the two nucleons. However, the situation is different for $\delta_{LT}$, where the calculations \textit{already} agree well for the neutron (although both fail to satisfy the recent experimental results.) Consequently, the P-N channel for $\delta_{LT}$ shows a remaining discrepancy not present in the other moments. It is possible that examining these differences provides important information on the strange issues with $\delta_{LT}$, and on the underlying disagreement between calculations present in several moments and polarizabilities.

\chapter{Physics Conclusions}

The g2p experiment was the first ever measurement of a fundamental proton observable in the region where the perturbative definition of QCD breaks down and our understanding becomes more limited. The experiment was challenging, due to the low energy and necessary transverse field in the polarized target, but obtained high precision data in a very difficult regime to measure. The results clarify our understanding of the proton's spin structure, as well as our understanding of low $Q^2$ QCD, by acting as a benchmark for several important moments and polarizabilities which theoretical calculations must be able to reproduce. This result is transformative and was recently published in the journal Nature Physics under the title, ``Proton spin structure and generalized polarizabilities in the strong quantum chromodynamics regime'' in October 2022.


In general, the g2p proton results show good agreement with the predictions of chiral perturbation theory, giving no indication of a proton `$\delta_{LT}$ Puzzle', and providing evidence in favor of current cutting edge effective theories of QCD. However, which calculation the results favor is highly moment dependent. The moments which are dominated by $g_2$ such as $\delta_{LT}$ tend to show very good agreement with the calculation of Alarcon et al.~\cite{Alarcon}. However, the g2p data for the moments dominated by $g_1$ is either unable to decisively favor one calculation over the other, or in the case of $\gamma_0$, actively favors the calculation of Bernard et al.~\cite{Bernard_moments}. This is driven almost entirely by the behavior of the $g_1$ result near the pion production threshold, where the g2p result stays negative, while CLAS data and phenomenological models indicate a small positive crossing.

The results presented above imply a number of possibilities for future work to follow up. Primarily, the major result for $\delta_{LT}$ and other moments shows that to fully understand the low $Q^2$ regime of QCD, and the structure of the particle which comprises most of the mass of the visible universe, the proton, it is necessary for additional theory work to occur, to reconcile the competing calculations of chiral perturbation theory for the polarizabilities, and to accurately estimate the unmeasured part of the integral at low $Q^2$ for other moments. This theory work must be able to reproduce the newly published g2p data.

Several of the results also present an important motivation for a further measurement of both structure functions $g_1$ and $g_2$ over a transition range from low to medium $Q^2$ for the proton. $d_2$ seems to have a maximum value in a region which has not yet been measured, finding this maximum and understanding it may have theoretical implications for the connection between the low $Q^2$ realm of quark-gluon correlations and the high $Q^2$ interpretation of color polarizability. The behavior of $g_2$ in this unmeasured range is also interesting. As can be seen most prominently in Figure~\ref{fig:gamma2}, the sign of $g_2$ in the resonance region inverts from the high $Q^2$ RSS and SANE data to the low $Q^2$ g2p data. Understanding exactly why this sign flip happens and where it occurs may be extremely interesting for understanding the spin structure of the proton. For both of these reasons, as well as the prospect of computing a number of these moments in the regime of the upper limit of $\chi$PT calculations, I believe the g2p results call for another measurement of the proton structure functions at slightly higher $Q^2$, with a large number of energy settings for the highest possible $Q^2$ resolution. In an ideal world, the best possible measurement would ensure that longitudinal data and transverse data are taken at similar $Q^2$ for all kinematic points, to reduce the model dependency of the final result as was possible for the $Q^2 = 0.045$ point in g2p.

\singlespacing
\bibliographystyle{unsrtnat}

\bibliography{sources}

\doublespacing
\titleformat{\chapter}[display]{\normalfont\bfseries}{\centering APPENDIX \thechapter}{0pt}{\centering}
\appendix


\end{document}